\documentclass[symmetry,article,submit,moreauthors,pdftex]{Definitions/mdpi} 

\preto{\abstractkeywords}{\nolinenumbers}

\firstpage{1} 
\pubvolume{1}
\issuenum{1}
\articlenumber{0}
\pubyear{2022}
\copyrightyear{2020}
\datereceived{14 May 2022} 
\dateaccepted{12 June 2022} 
\datepublished{} 
\hreflink{https://doi.org/} 

\def\p{\partial}
\def\msun{M$_{\odot}$}
\def\Msun{M$_{\odot}$ }
\def\be{\begin{equation}}
\def\ee{\end{equation}}
\def\bi{\begin{itemize}}
\def\i{\item}
\def\ei{\end{itemize}}
\def\ben{\begin{enumerate}}
\def\een{\end{enumerate}}
\def\bea{\begin{eqnarray}}
\def\eea{\end{eqnarray}}
\def\bt{\begin{tabbing}}
\def\et{\end{tabbing}}

\def\edo{\end{document}}

\newcommand{\enth}{\mathcal{E}}

\newcommand{\tlg}{\tilde{\gamma}}
\newcommand{\tlG}{\tilde{\Gamma}}
\newcommand{\emfp}{e^{-4\phi}}
\newcommand{\tlA}{\tilde{A}}
\newcommand{\tlR}{\tilde{R}}

\newcommand{\dt}[1]{\partial_t #1}
\newcommand{\pdu}[3]{\partial_{#3} #1^{#2}}
\newcommand{\pdl}[3]{\partial_{#3} #1_{#2}}
\newcommand{\pdpdu}[4]{\partial_{#3}\partial_{#4} #1^{#2}}
\newcommand{\pdpdl}[4]{\partial_{#3}\partial_{#4} #1_{#2}}
\newcommand{\upwindu}[3]{\beta^{#3}\bar{\partial}_{#3} #1^{#2}}
\newcommand{\upwindl}[3]{\beta^{#3}\bar{\partial}_{#3} #1_{#2}}
\newcommand{\tlGn}{\tilde{\Gamma}_{\mathrm{(n)}}}
\newcommand{\tlGmixed}[2]{\tlG_{#1}^{\;\;\;#2}}

\def\x{{\bf x}}

\def\M4{M$_4$}
\def\M6{M$_6$}
\def\lcm6{LCM$_6$}
\def\qcm6{QCM$_6$}

\def\wh3{W$_{\rm H,3}$}
\def\wh4{W$_{\rm H,4}$}
\def\wh5{W$_{\rm H,5}$}
\def\wh6{W$_{\rm H,6}$}
\def\wh7{W$_{\rm H,7}$}

\newcommand{\SpB}{\texttt{SPHINCS\_BSSN}\; }
\newcommand{\spB}{\texttt{SPHINCS\_BSSN}}
\newcommand{\spid}{\texttt{SPHINCS\_ID}}
\newcommand{\spiD}{\texttt{SPHINCS\_ID}\; }

\newcommand{\Lo}{\texttt{LORENE}\;}
\newcommand{\lo}{\texttt{LORENE}}

\newcommand{\McL}{\texttt{McLachlan}\;}

\def\ent{\mathcal{E}} 

\Title{Thinking outside the box: Numerical Relativity with particles}

\TitleCitation{Numerical Relativity with particles}

\Author{Stephan Rosswog$^{1,*}$\orcidA{}, Peter Diener $^{2}$ and Francesco Torsello $^{3}$}

\AuthorNames{Firstname Lastname, Firstname Lastname and Firstname Lastname}

\AuthorCitation{Rosswog, S.; Diener, P.; Torsello, F.}

\address{%
$^{1}$ \quad Department of Astronomy and Oskar Klein Centre, Stockholm University; stephan.rosswog@astro.su.se\\
$^{2}$ \quad Center for Computation \& Technology and Department of Physics \& Astronomy, Louisiana State University, Baton Rouge, LA 70803, United States of America; diener@cct.lsu.edu\\
$^{3}$ \quad Department of Astronomy and Oskar Klein Centre, Stockholm University; francesco.torsello@astro.su.se}

\corres{Correspondence: stephan.rosswog@astro.su.se}

\abstract{
The observation of gravitational waves from compact objects  has now become an active part of observational astronomy. 
For a sound interpretation,  one needs to compare such observations against detailed Numerical Relativity simulations,
which are essential tools to explore the dynamics and physics of compact binary mergers.
To date, essentially all simulation codes that solve the full set of Einstein's equations are performed in the framework
of  Eulerian hydrodynamics. The exception is our recently developed Numerical Relativity code \SpB which 
solves the commonly used BSSN formulation of the Einstein equations on a structured mesh and
the matter equations via Lagrangian particles. We show here, for the first time, \SpB neutron star merger 
simulations with piecewise polytropic approximations to four nuclear matter equations of state. In this set of neutron
star merger simulations we focus on perfectly symmetric binary systems that are irrotational and have 1.3 $M_\odot$ masses.
We introduce some further
methodological refinements (a new way of steering dissipation, an improved particle-mesh mapping) and we explore
the impact of the exponent that enters in the calculation of the thermal pressure contribution. We find that it leaves
a noticeable imprint on the gravitational wave amplitude (calculated via both quadrupole approximation and the $\Psi_4$-formalism)
and has a noticeable impact on the amount of dynamic ejecta. Consistent with earlier findings, we only find a few times $10^{-3}$ \Msun
as dynamic ejecta in the studied equal mass binary systems, with softer equations of state (which are more prone to shock formation) 
ejecting larger amounts of matter.  In all of the cases, we see a credible high-velocity ($\sim0.5 .. 0.7c$) ejecta component 
of $\sim 10^{-4}$ \Msun that is launched at contact from the interface between the two neutron stars. 
Such a high-velocity component has been suggested to produce an early, blue precursor to the main kilonova emission
and it could also potentially cause a kilonova afterglow.
}


\begin{document}

\section{Introduction}

Detections of gravitational waves emitted during the violent collisions of compact objects
are now routinely observed by ground-based gravitational wave detectors \cite{abbott21b}.
Especially if more than one 
messenger can be detected, such events offer unprecedented insights into the physics under
the most extreme conditions: the dynamics of the last inspiral and subsequent merger stages
are governed  by the interplay of strong-field gravity and the equation of state at supra-nuclear 
densities \cite{baiotti19}, weak interactions determine the neutrino emission and ejecta composition, see e.g. 
\cite{ruffert97a,rosswog03a,sekiguchi11,perego14b,just15,fujibayashi20,foucart21a,just22,radice22},
and magnetic fields that can rise to beyond magnetar field strengths \cite{price06,kiuchi15,palenzuela15}
likely cause additional outflows and electromagnetic emission.\\
These violent multi-physics events can only be explored in a realistic way via full-blown numerical
simulations. Thus, Numerical Relativity simulation codes have become precious (but unfortunately rather
complicated) exploration tools. To date, there exists a  "mono-culture" in the sense that practically
all Numerical Relativity codes that solve the full set of Einstein equations make use of Eulerian hydrodynamics
methods. We have recently developed an alternative methodology \cite{rosswog21a,diener22a} where we follow the 
conventional and well tested approach to simulate the spacetime evolution via the  BSSN formulation
\cite{shibata95,baumgarte99}, but we evolve the fluid via freely moving Lagrangian particles. While the
spacetime part is very similar to Eulerian approaches, the fluid part has additional advantages, in particular
in following the material that becomes ejected in a neutron star merger. This material contains a small mass of
only $\sim 1\%$ of the binary system \cite{rosswog99,oechslin07b,bauswein13a,hotokezaka13a,radice18a}, but 
is responsible for all the electromagnetic emission and therefore
of utmost importance for understanding the multi-messenger signals. Tracing these small amounts of matter
with conventional Eulerian approaches is difficult, since a) the accuracy of the advection of matter is 
resolution-dependent and the resolution away from the central collision site usually degrades substantially
and b) these methods employ artificial "atmospheres" to model vacuum. The ejecta are expanding within
these atmospheres and this makes it hard to trace their properties with the desired accuracy. Within
our particle approach these challenges are absent since vacuum simply corresponds to the absence of computational
particles. Advection is exact in our approach:  a property such as the electron fraction is 
attached to a particle, and it is therefore simply  "carried around" as the particle moves, without any loss of information.
Equally important is that in our case the neutron star surface remains perfectly well-behaved and does not
need any special treatment, while it is a continuous source of concern in Eulerian neutron star simulations.
There, the sharp transition between high-density and vacuum, on a numerically difficult to resolve length scale,
can easily lead to failures in recovering the primitive variables and to an effective reduction of the 
convergence order \cite{schoepe18}.\\
We have recently presented our new \SpB code \cite{rosswog21a} and we have shown that it can accurately 
reproduce a number of challenging benchmark tests. These tests included standard hydrodynamics tests
such as relativistic shock tubes, the oscillations of relativistic neutron stars, both in fixed ("Cowling approximation")
and dynamically evolving spacetimes, and the transition of an unstable neutron star either to a stable neutron
star or to a black hole. In a recent paper \cite{diener22a} we have introduced a number of
new methodological elements in \spB, such as fixed mesh refinement, a new algorithm to transfer
particle properties to the spacetime mesh and we have laid out in detail how we set up our fluid particles
according to the "Artificial Pressure Method" (APM) \cite{rosswog20a} while making use of the 
library \Lo \cite{lor}, using our new code \spiD \cite{diener22a}. While we, for simplicity, used polytropic 
equations of state in the first two papers \cite{rosswog21a,diener22a}, we here adapt \SpB to use 
piecewise polytropic approximations to nuclear equations of state \cite{read09}. \\
The paper is structured as follows. In Sec.~\ref{sec:SPHINCS_BSSN} we describe the methodology that we use,
including the relativistic hydrodynamics, how we apply dissipation, evolve the spacetime, couple the spacetime
and the matter, the equations of state that we use and we also summarize how our initial conditions are constructed.
In Sec.~\ref{sec:results} we present our results, starting with standard shock tube tests to demonstrate where
our new  steering algorithm triggers dissipation. We then show the dynamical evolution of our binary systems, their 
gravitational wave emission and their dynamic mass ejection. In Sec.~\ref{sec:summary} we summarize the new 
methodological elements and discuss the main findings. Details of our recovery scheme for piecewise polytropic equations
of state and some resolution experiments are shown in two appendices. 
\section{SPHINCS\_BSSN}
\label{sec:SPHINCS_BSSN}
\SpB ("Smooth Particle Hydrodynamics lN Curved Spacetime using BSSN" ) \cite{rosswog21a,diener22a} has become a rather complex code with a large number of methodological ingredients.
Those parts where the current status has been described already in detail in the first two papers will only be briefly
summarized here and we refer to the original publications for more information. In this paper we include, for the first time
in  \spB, piecewise polytropic approximations to nuclear equations of state and this requires a modification of our 
recovery algorithm, see Appendix \ref{sec:recovery}. We also explore the impact of the thermal
component (more specifically of the thermal exponent $\gamma_{\rm th}$, see Appendix \ref{sec:recovery}) and we 
use a new way to steer dissipation in \SpB that is based on \cite{cullen10,rosswog15b}.

\subsection{Hydrodynamic evolution}
The hydrodynamic evolution equations  are in \SpB modelled  via a high-accuracy version of the Smooth Particle Hydrodynamics 
(SPH) method.
The basics of the relativistic SPH-equations have been derived very explicitly in Sec. 4.2 of \cite{rosswog09b} and we will
refer to this text for many of the derivations and only present the final equations here. Many new, substantially 
accuracy-enhancing elements (kernels, gradient estimators, dissipation steering) have been explored in \cite{rosswog15b,rosswog20a,rosswog20b} and most of them are also implemented in \spB.\\
We follow the conventions $G=c=1$, use metric signature (-,+,+,+)  and we measure all energies in units of $m_0 c^2$, where $m_0$ is the average baryon mass\footnote{This quantity depends on the actual nuclear composition, but simply using the atomic mass unit $m_u$ gives
a precision of better than 1\%. We therefore use  the approximation $m_0 \approx m_u$ in the following.}. 
Greek indices run from  0 to 3 and latin indices from 1 to 3. Contravariant spatial indices of a vector quantity $\vec{w}$ at particle $a$ are 
denoted as $w_a^i$, while covariant ones will be written as $(w_i)_a$.\\
To discretize our fluid equations we choose a "computing frame" in which the computations are performed. Quantities in this frame usually differ
from those calculated in the local fluid rest frame. 
The line element in a 3+1 split of spacetime reads
\be
ds^2= -\alpha^2 dt^2 + \gamma_{ij} (dx^i + \beta^i dt) (dx^j + \beta^j dt),
\ee
where $\alpha$ is the lapse function, $\beta^i$ the shift vector and $\gamma_{ij}$ the spatial 3-metric.
We use a generalized Lorentz factor 
\be
\Theta\equiv \frac{1}{\sqrt{-g_{\mu\nu} v^\mu v^\nu}} \quad {\rm with} \quad v^\alpha=\frac{dx^\alpha}{dt},
\label{eq:theta_def}
\ee
the coordinate velocities $v^\alpha$ are related to the four-velocities, normalized as $U_\mu U^\mu=-1$, by
\be
v^\mu= \frac{dx^\mu}{dt}= \frac{U^\mu}{\Theta}= \frac{U^\mu}{U^0}.
\label{eq:v_mu}
\ee
We choose the computing frame baryon number density $N$ as density variable, which is related to the
baryon number density as measured in the local fluid rest frame, $n$, by 
\be
N= \sqrt{-g}\, \Theta\, n.
\label{eq:N_def}
\ee
Here $g$ is the determinant of the spacetime metric.  Note that this density variable is  very similar to 
what is used in Eulerian approaches \cite{alcubierre08,baumgarte10,rezzolla13a,shibata16}.
We keep the baryon number of each SPH particle, $\nu_b$, constant so that exact numerical baryon 
number conservation is guaranteed. At every time step, the computing frame baryon number density
at the position of a particle $a$ is  calculated via a weighted sum (actually  very similar to 
Newtonian SPH)
\be
N_a= \sum_b \nu_b\, W(|\vec{r_a} - \vec{r}_b|,h_a),
\label{eq:N_sum}
\ee
where the smoothing length $h_a$ characterizes the support size 
of the SPH smoothing kernel $W$. As momentum variable, 
we choose the canonical momentum per baryon 
\be
(S_i)_a = (\Theta \mathcal{E} v_i)_a,
\label{eq:can_mom}
\ee
where $\mathcal{E}= 1 + u + P/n$ is the relativistic enthalpy per baryon with $u$ being the internal energy per
baryon and $P$ the gas pressure. This quantity evolves according to
\be
\frac{d(S_i)_a}{dt}  =  \left(\frac{d(S_i)_a}{dt}\right)_{\rm hyd} +  \left(\frac{d(S_i)_a}{dt}\right)_{\rm met},
\label{eq:dSdt_full}
\ee
where the hydrodynamic part is
\be
\left(\frac{d(S_i)_a}{dt}\right)_{\rm hyd}= -\sum_b \nu_b \left\{ \frac{P_a}{N_a^2}  D^a_i  +  
\frac{P_b}{N_b^2} D^b_i \right\}
\label{eq:dSdt_hydro}
\ee
and the gravitational part
\be 
\left(\frac{d(S_i)_a}{dt}\right)_{\rm met}= \left(\frac{\sqrt{-g}}{2N} T^{\mu \nu} \frac{\p g_{\mu \nu}}{\p x^i}\right)_a
\label{eq:dSdt_metric}.
\ee
Here we have used the abbreviations
\be
D^a_i \equiv   \sqrt{-g_a} \;  \frac{\p W_{ab}(h_a)}{\p x_a^i} \quad {\rm and} \quad 
D^b_i \equiv    \sqrt{-g_b} \; \frac{\p W_{ab}(h_b)}{\p x_a^i}
\label{eq:kernel_grad}
\ee
and $W_{ab}(h_k)$ is a shorthand for $W(|\vec{r}_a - \vec{r}_b|/h_k)$.
As energy variable we use the canonical energy per baryon 
\be
e_a= \left(S_i v^i + \frac{1 + u}{\Theta}\right)_a = \left(\Theta \mathcal{E} v_i v^i + \frac{1 + u}{\Theta}\right)_a,
\label{eq:can_en}
\ee
which is evolved according to
\be
\frac{d e_a}{dt}= \left(\frac{d e_a}{dt}\right)_{\rm hyd}  + \left(\frac{de_a}{dt}\right)_{\rm met},
\label{eq:energy_equation}
\ee
with
\be
\left(\frac{d e_a}{dt}\right)_{\rm hyd} = -\sum_b \nu_b \left\{ \frac{P_a}{N_a^2}  \;  v_b^i   \; D^a_i +  
\frac{P_b}{N_b^2} \;  v_a^i \; D^b_i \right\}
\label{eq:dedt_hydro}
\ee
and
\be
\left(\frac{de_a}{dt}\right)_{\rm met}= -\left(\frac{\sqrt{-g}}{2N} T^{\mu \nu} \frac{\p g_{\mu \nu}}{\p t}\right)_a.
\label{eq:dedt_metric}
\ee
Note that the hydrodynamics part of the momentum equation, Eq.~(\ref{eq:dSdt_hydro}), looks very similar to the Newtonian momentum equation,
and possesses in particular the same symmetries in the particle labelling indices $a$ and $b$: exchanging these indices, 
$a \leftrightarrow b$, leads to sign change due to the anti-symmetry of the kernel gradient, $\nabla_a W(|\vec{r}_a-\vec{r}_b|,h_a)=
-\nabla_b W(|\vec{r}_a-\vec{r}_b|,h_b)$ which, in the Newtonian and fixed-metric case ensures in a straight-forward way the 
exact conservation of momentum, see Sec. 2.4 in \cite{rosswog09b} for a detailed discussion of the anti-symmetry of
kernel gradients in relation to exact numerical conservation. The hydrodynamic part of the energy equation, Eq.~(\ref{eq:dedt_hydro}),
in turn, looks very similar to the Newtonian evolution equation of the "thermo-kinetic" energy, $u+v^2/2$, see Eq.~(34) in \cite{rosswog09b}.
For the involved kernel function we use the  Wendland C6-smooth kernel \cite{wendland95}
\be
W(q)= \frac{\sigma}{h^3} (1-q)^8_+ (32 q^3 + 25 q^2 + 8 q + 1),
\ee
where the normalization is $\sigma= 1365/(64 \pi)$ in 3D and the symbol $(.)_+$ denotes the cutoff function max$(.,0)$.
Our kernel choice is based on ample previous experiments \cite{rosswog15b,rosswog20a} and we choose 
(at every Runge-Kutta substep) the smoothing length so that every particle $a$ has exactly 300 neighbours in 
its own support of radius $2h_a$. Technically this is achieved via a very fast tree method \cite{gafton11}, more 
details on how this is done in practice can be found in \cite{rosswog20a}.\\ 
Note that this version of the relativistic SPH equations is very similar to the Newtonian case and it has
the additional advantage over older relativistic  SPH formulations \cite{laguna93a} that it does not involve 
numerically inconvenient terms such as time derivatives of Lorentz factors. The quantities that we evolve 
numerically, however, are not the physical quantities that we are interested in and we therefore have to 
recover the physical quantities $n$, $u$, $v^i$, $P$ from  
$N$, $S_i$, $e$ at every integration (sub-)step. This ``recovery step" is done in a very similar way as in Eulerian approaches,
for the case of polytropic equations of state our strategy is described in detail in Sec.~2.2.4 of \cite{rosswog21a}.
One of the new elements of \SpB that we introduce here is the use of piecewise polytropic equations of state that
also contain a thermal pressure contribution. This requires a modified approach for the recovery that we describe in 
detail in  Appendix \ref{sec:recovery}.

\subsubsection{Dissipative terms}
\label{sec:AV}
To  deal with shocks we have to include dissipative terms. We follow the approach originally suggested by 
von Neumann and Richtmyer \cite{vonNeumann50} which simply consists in everywhere replacing  the 
physical pressure $P$ with $P+Q$, where $Q$ is a suitable viscous pressure. Our prescription used here
does not differ substantially from the original paper \cite{rosswog21a}, but since we carefully gauge the
involved parameters, we will briefly summarize the basic equations for the ease of the subsequent discussion. \\
For the form of the viscous pressure, we follow \cite{liptai19} and use
\bea
Q_a&=& -\frac{1}{2} \alpha_{\rm AV} N_a v_{{\rm s},a} \enth_a \left(\Gamma_a^\ast V^\ast_a -  \Gamma_b^\ast V^\ast_b \right)
\label{eq:Qa}
\\
Q_b&=& -\frac{1}{2} \alpha_{\rm AV} N_b v_{{\rm s},b} \enth_b \left(\Gamma_a^\ast V^\ast_a -  \Gamma_b^\ast V^\ast_b \right).
\label{eq:Qb}
\eea
The $V^\ast$ are the fluid velocities seen by an Eulerian observer, $V^i= \frac{v^i + \beta^i}{\alpha}$, projected 
onto the line connecting two particles $a$ and $b$, $\hat{e}_{ab}$,
\be
V^\ast_a= \eta_{ij} \hat{e}^j_{ab} V_a^i \quad{\rm and} \quad \Gamma_a^\ast= \frac{1}{\sqrt{1 - V_a^{\ast 2}}},
\label{eq:projection}
\ee
the corresponding expression with indices $a$ and $b$ interchanged applies for $V^\ast_b$ and the parameter $\alpha_{\rm AV}$ determines the amount of dissipation. 
For the signal speeds we use
\be
v_{\rm s,a}= \frac{c_{\rm s,a} + |V^\ast_{ab}|}{1 + c_{\rm s,a}  |V^\ast_{ab}|},
\ee
where $c_{\rm s}= \sqrt{(\gamma-1)(\mathcal{E}-1)/\mathcal{E}}$ is the relativistic sound speed, $\gamma$ being the polytropic exponent, and
\be
V^\ast_{ab}= \frac{V^\ast_a - V^\ast_b}{1 - V^\ast_a V^\ast_b}.
\ee
We also include a small amount of artificial conductivity  by adding the following term to equation~(\ref{eq:dedt_hydro})
\be
\left(\frac{de}{dt}\right)^{\rm c}= \frac{\alpha_c}{2}  \sum_b \nu_b \; \xi^c_{ab}  \; v_{{\rm s},ab}^{\rm c} \left( \frac{\alpha_a u_a}{\Gamma_a} - 
\frac{\alpha_b u_b}{\Gamma_b}\right) \left\{ \frac{D^a_i }{N_a} +   \frac{D^b_i }{N_b}\right\} \hat{e}_{ab}^i,
\label{eq:cond}
\ee
where the $\alpha_a/\alpha_b$ are the lapse functions at the particle positions 
and $ \Gamma= (1 - V_i V^i)^{-1/2}$.  Apart from the limiter $\xi^c_{ab}$, see below, this
expression is the same as in  \cite{liptai19}. For the conductivity signal  velocity we use \cite{liptai19}
\be
v_{{\rm s},ab}^{\rm c}= {\rm min} \left(1, \sqrt{\frac{2|P_a-P_b|}{\ent_a n_a + \ent_b n_b}}\right)
\ee
for cases when the metric is known (i.e. cases where no consistent hydrostatic equilibrium needs to be maintained)
and $v_{{\rm s},ab}^c= |V^\ast_{ab}|$ otherwise. The prefactor has been chosen after extensive experiments with
shock tubes and single neutron star and we find good results for $\alpha_c= 0.3$. The role of the limiter
 $\xi^c_{ab}$ is to restrict the application of conductivity to regions that have large second derivatives $\partial_i \partial_j u$
 and to suppress it elsewhere. In \cite{rosswog21a} we had designed a simple dimensionless trigger aimed
 at quantifying the size of second-derivative effects
\begin{equation}
T_{u, ab}= \frac{h_{ab}}{u_{ab}} | (\nabla u)_a - (\nabla u)_b|,
\end{equation}
where $u_{ab}= (u_a+u_b)/2$ and $h_{ab}= (h_a+h_b)/2$, and the final
conductivity limiter reads 
\begin{equation}
\xi^c_{ab}= \frac{T_{u, ab}}{T_{u, ab} + 0.2}.
\end{equation}
The reference value 0.2 has been chosen after experiments
in both Sod-type shock tubes and self-gravitating neutron stars.

\subsubsection{Slope-limited reconstruction in the dissipative terms}
\label{sec:AV_reconst}
As demonstrated in a Newtonian context \cite{christensen90,frontiere17,rosswog20a}, slope-limited reconstruction
can be very successfully applied within an artificial viscosity approach and we also follow such a strategy here. In our earlier experiments 
\cite{rosswog20a} reconstruction has massively suppressed unwanted effects of artificial dissipation. Instead of
using $(\Gamma_a^\ast V_a^\ast - \Gamma_b^\ast V_b^\ast)$ in Eqs.~(\ref{eq:Qa}) and ~(\ref{eq:Qb}), we reconstruct
the velocities seen by an Eulerian observer  of both particle $a$ and $b$ to their common mid-point, $\bar{r}_{ab}^i= (r_a^i+r_b^i)/2$:
\be
\tilde{V}_a^i= V_a^i - \frac{1}{2} {\rm SL}(\partial_j V_a^i,\partial_j V_b^i)(r_a^j - r_b^j) \quad {\rm and}  \quad
\tilde{V}_b^i= V_b^i + \frac{1}{2} {\rm SL}(\partial_j V_a^i,\partial_j V_b^i)(r_a^j - r_b^j) ,
\ee
project them onto the line joining particle $a$ and $b$ as in Eq.~(\ref{eq:projection}), calculate the corresponding Lorentz factors, 
and use products based on the reconstructed values instead of  $(\Gamma_a^\ast V_a^\ast - \Gamma_b^\ast V_b^\ast)$.
For the slope limiter ${\rm SL}$ we use a modification of the {\tt minmod} limiter
\be
{\rm SL}=  {\rm SL}^{\rm mm} \times  \left\{
 \begin{array}{ll}
e^{-\left(\frac{\chi_{ab} - \chi_{\rm crit}}{\chi_{\rm fo}}  \right)^2} \quad {\rm for \;} \chi_{ab} < \chi_{\rm crit}\\
1 \hspace*{2.1cm}{\rm else,}\\
\end{array}
\right.
\label{eq:SL}
\ee
where ${\rm SL}^{mm}$ is the original {\tt minmod} limiter
\be
{\rm SL}^{\rm mm} (a,b)=    \left\{
    \begin{array}{ll}
    \; \; \;{\rm min}(|a|,|b|) & \rm{if \; }  a> 0 {\rm \; and \;} b > 0\\
   -  {\rm min}(|a|,|b|) & {\rm if \;} a < 0 {\rm \; and \;} b < 0\\
    \; \; \;0 & {\rm otherwise.}
   \end{array}
 \right.
 \ee
 The quantity $\chi$ in the exponential suppression factor is given by $\chi_{ab}= {\rm{min}}(r_{ab}/h_a,r_{ab}/h_b)$ 
 with $r_{ab}= \sqrt{\sum_j (r_a^j - r_b^j)^2}$.
 Our aim is to have a close-to-uniform particle distribution within the kernel support, and the 
 purpose of the exponential  factor in Eq.~(\ref{eq:SL}) is to suppress reconstruction for particles
 that get closer than a (de-dimensionalized) critical separation $\chi_{\rm crit}$. For this separation
 we  choose the typical separation of a uniform distribution of $\rm nei_{des}$ particles inside a sphere
 with the volume of the support, $4 \pi (2h)^3/3$, which yields
 \be
 \chi_{\rm crit}= \left( \frac{32 \pi}{3 \rm nei_{des}} \right)^{1/3}.
 \ee
 Particles that come closer than $\chi_{\rm crit}$ have their reconstruction suppressed (i.e. SL going to zero)
 and therefore more dissipation which has an ordering effect. For the fall-off scale, $\chi_{\rm fo}$, we follow
 \cite{frontiere17} and use their suggested value of 0.2. In practice, the exponential suppression factor has 
 only a very small, though welcome, effect.
 
\subsubsection{Steering the dissipation parameter $\alpha_{\rm AV}$}
\label{sec:AV_steering}
We apply here a dissipation steering strategy that is different from before \cite{rosswog21a,diener22a}.
Apart from enjoying the exploration of a new strategy, the main reason for this is that it liberates us from the need to 
define a suitable entropy variable. While this is easy for simple equations of state and delivers very good results \cite{rosswog20b}, 
it leads to an unnecessary dependence
between our dissipation scheme and the --in principle- completely unrelated equation of state/microphysics modules. A
decoupling of different parts of the code is good coding practice and in particular desirable with regard to our future 
development plans that include the implementation of different, microphysical equations of state: with an independent
steering mechanism no modifications are needed when the microphysics is updated.\footnote{We had previously compared
an entropy-based steering mechanism with one that is based on a $d(\nabla\cdot\vec{v})/dt$-steering similar to the one we 
use here \cite{rosswog20b}. Overall we found very good agreement between both approaches, but slight advantages for 
the entropy-steering.}\\
We want to apply dissipation only where it is needed and to do so we follow a variant of the strategies
suggested in \cite{cullen10} and \cite{rosswog15b}. We calculate at each time step and for each particle
a desired value $\alpha_a^{\rm des}$ for the dissipation parameter and if the  current value 
at a particle $a$, $\alpha_{{\rm AV},a}$, is larger than  $\alpha^{\rm des}_a$, we let it decay 
exponentially according to 
\be
\frac{d\alpha_{{\rm AV},a}}{dt}= - \frac{\alpha_{{\rm AV},a} - \alpha_0}{20 \tau_a},
\label{eq:AV_decay}
\ee
where $\tau_a= h_a/c_{s,a}$ is the particle's dynamical time scale and $\alpha_0$ is a low floor value. 
Otherwise, if $\alpha^{\rm des}_a > \alpha_{{\rm AV},a}$, 
the value of $\alpha_{{\rm AV},a}$  is instantaneously raised to $\alpha^{\rm des}_a$.\\
In determining  the desired dissipation value $\alpha^{\rm des}_a$ we apply two criteria, one for detecting shocks 
yielding $\alpha^{\rm des,S}$, and another one for detecting noise yielding $\alpha^{\rm des,N}$ and we use
$\alpha^{\rm des}= {\rm max}(\alpha^{\rm des,S}, \alpha^{\rm des,N})$. We  monitor compressions
that increase in time to detect shocks \cite{cullen10} (for ease of notation we are omitting a particle labelling index)
\be
\alpha^{\rm des,S}= \alpha^{\rm max} \; \frac{A}{0.1 (\frac{c_s}{h})^2 + A},
\label{eq:CD_trigger}
\ee
where 
\be
A= {\rm max}\left[\frac{-d(\nabla \cdot \vec{v})}{dt},0\right]
\ee
and $\alpha^{\rm max}$ is the maximally reachable dissipation parameter.
Note that in Eq.~(\ref{eq:CD_trigger}), we conservatively use a smaller prefactor (= 0.1) than what is suggested 
in \cite{cullen10} (= 0.25), so dissipation is increased more aggressively if an increasing compression is
detected. While this works very well, it may mean that in the current set of simulations we are applying 
somewhat more dissipation than is actually needed. This issue will be explored in future work, where we will
try to reduce the dissipation further. Our second criterion is based on
fluctuations in the sign of $\nabla \cdot \vec{v}$ to detect local noise (as suggested in the special relativistic
version {\tt SPHINCS\_SR}, \cite{rosswog15b})
\be
\alpha^{\rm des,N}= \frac{\mathcal{N}}{0.2 (\frac{c_s}{h}) + \mathcal{N}},
\ee
where the noise trigger is
\be
\mathcal{N}= \sqrt{\mathcal{S}^+ \mathcal{S}^-}
\ee
and
\be
\mathcal{S}^+ =   \frac{1}{N^+} \sum_{b,\nabla \cdot \vec{v}_b > 0}^{N^+} \nabla \cdot \vec{v}_b \quad {\rm and}  \quad 
\mathcal{S}^- =    \frac{1}{N^-} \sum_{b,\nabla \cdot \vec{v}_b < 0}^{N^-} \nabla \cdot \vec{v}_b
\ee
and $N^+/N^-$ are the number of positive/negative $\nabla \cdot \vec{v}$ contributions.
The noise trigger $\mathcal{N}$ is the product of two quantities, so if there are sign fluctuations, but they are small compared to 
$c_s/h$, then $\alpha^{\rm des,N}$ is close to zero. If instead we have a uniform expansion or compression, either $\mathcal{S}^+$ 
or $\mathcal{S}^-$ will be zero and therefore also the noise trigger.
So only for sign fluctuations and significantly large compressions/expansions will the product have a substantial value
and thus trigger a dissipation increase.\\
In all of the simulations shown in this paper we use a floor value $\alpha_0= 0.2$ and a maximum dissipation parameter
$\alpha^{\rm max}= 1.5$ which yields good results in shock tube tests, see below.
As a side remark we mention that \cite{cullen10} advertise to use a vanishing floor value $\alpha_0$, we will 
explore a possible reduction of this parameter in future work.

\subsection{Spacetime evolution}
\label{sec:spacetime_evolution}
We evolve the spacetime according to the (``$\Phi$-version'' of the) BSSN
equations \citep{shibata95,baumgarte99}. For this we have written a wrapper around code extracted
from the  \McL thorn~\citep{Brown:2008sb}  of the Einstein 
Toolkit~\citep{ETK:web,Loffler:2011ay}.
The variables  used in this method are related to the
Arnowitt-Deser-Misner (ADM) variables $\gamma_{ij}$ (3-metric), $K_{ij}$ (extrinsic curvature),
$\alpha$ (lapse function) and $\beta^{i}$ (shift vector) and they read
\begin{eqnarray}
  \phi & = & \frac{1}{12} \log(\gamma), \\
  \tlg_{ij} & = & \emfp \gamma_{ij}, \\
  K & = & \gamma^{ij} K_{ij}, \\
  \tlG^{i} & = & \tlg^{jk} \tlG^{i}_{jk}, \\
  \tlA_{ij} & = & \emfp\left ( K_{ij}-\frac{1}{3}\gamma_{ij} K\right ),
\end{eqnarray}
where $\gamma = \mathrm{Det}(\gamma_{ij})$,  $\tlG^{i}_{jk}$ are the
Christoffel symbols related to the conformal metric $\tlg_{ij}$ and $ \tlA_{ij}$ is
the conformally rescaled, traceless part of the extrinsic curvature. The corresponding
evolution equations read
\begin{eqnarray}
  \dt{\phi} & = & -\frac{1}{6} \left ( \alpha K - \pdu{\beta}{i}{i} \right) + \upwindu{\phi}{}{i}, \\
 \dt{\tlg_{ij}} & = & -2\alpha \tlA_{ij} + \tlg_{ik} \pdu{\beta}{k}{j}
                   + \tlg_{jk} \pdu{\beta}{k}{i}
                    -\frac{2}{3} \tlg_{ij} \pdu{\beta}{k}{k}
                         + \upwindl{\tlg}{ij}{k}, \\
  \dt{K} & = & -\emfp \left ( \tlg^{ij} \left [ \pdpdu{\alpha}{}{i}{j}
               +2\pdu{\phi}{}{i}\pdu{\alpha}{}{j} \right ]
               - \tlGn^{i}\pdu{\alpha}{}{i} \right ) \nonumber \\
         &   & + \alpha \left ( \tlA^{i}_{j} \tlA^{j}_{i} +\frac{1}{3} K^2
               \right ) + \upwindu{K}{}{i} + 4 \pi \alpha ( \rho + s ), \\
  \dt{\tlG^{i}} & = & -2 \tlA^{ij} \pdu{\alpha}{}{j} + 2 \alpha \left (
                    \tlG^{i}_{jk} \tlA^{jk} - \frac{2}{3} \tlg^{ij}
                    \pdu{K}{}{j} +
                  6 \tlA^{ij} \pdu{\phi}{}{j}\right )
                 \nonumber\\
                  & &                     +\tlg^{jk} \pdpdu{\beta}{i}{j}{k} + \frac{1}{3}
                    \tlg^{ij} \pdpdu{\beta}{k}{j}{k} -\tlGn^{j}\pdu{\beta}{i}{j}
                    + \frac{2}{3} \tlGn^{i}\pdu{\beta}{j}{j} \nonumber \\
                    &   & + \upwindu{\tlG}{i}{j} -16 \pi \alpha \tlg^{ij} s_j, \\
  \dt{\tlA_{ij}} & = & \emfp \left [ -\pdpdu{\alpha}{}{i}{j} + \tlG^{k}_{ij}
                       \pdu{\alpha}{}{k} + 2 \left ( \pdu{\alpha}{}{i}
                       \pdu{\phi}{}{j}+\pdu{\alpha}{}{j} \pdu{\phi}{}{i}\right )
                       +\alpha R_{ij} \right ]^{\mathrm{TF}} \nonumber\\
                 & & +\alpha ( K \tlA_{ij}- 2 \tlA_{ik} \tlA^{k}_{j} )
                       + \tlA_{ik} \pdu{\beta}{k}{j}
                       + \tlA_{jk} \pdu{\beta}{k}{i}
                       - \frac{2}{3} \tlA_{ij} \pdu{\beta}{k}{k} \nonumber \\
                 &   & +\upwindl{\tlA}{ij}{k} - \emfp \alpha 8 \pi
                       \left (T_{ij}-\frac{1}{3} \gamma_{ij} s\right ),
\end{eqnarray}
where
\begin{eqnarray}
  \rho & = & \frac{1}{\alpha^2} ( T_{00} - 2 \beta^{i} T_{0i} +
             \beta^{i}\beta^{j} T_{ij} ),\label{eq:BSSN_rho} \\
  s & = & \gamma^{ij} T_{ij}, \\
  s_{i} & = & -\frac{1}{\alpha} ( T_{0i} - \beta^{j} T_{ij}),\label{eq:BSSN_Si}
\end{eqnarray}
and $\upwindu{}{}{i}$ denote partial derivatives that are upwinded based on the
shift vector. The superscript "TF" in the evolution equation of $\tlA_{ij}$ denotes 
the trace-free part of the bracketed term.
Finally $R_{ij} = \tlR_{ij} + R^{\phi}_{ij}$, where
\begin{eqnarray}
  \tlG_{ijk} & = & \frac{1}{2}\left ( \pdl{\tlg}{ij}{k} + \pdl{\tlg}{ik}{j}
               - \pdl{\tlg}{jk}{i} \right ), \\
  \tlGmixed{ij}{k} & = & \tlg^{kl} \tlG_{ijl}, \\
  \tlG^{i}_{jk} & = & \tlg^{il}\tlG_{ljk}, \\
  \tlGn^{i} & = & \tlg^{jk} \tlG^{i}_{jk} \\
  \tlR_{ij} & = & -\frac{1}{2} \tlg^{kl} \pdpdl{\tlg}{ij}{k}{l}
                  +\tlg_{k(i} \pdu{\tlG}{k}{j)}
                  +\tlGn^{k} \tlG_{(ij)k} \nonumber \\
            &   & +\tlG^{k}_{il} \tlGmixed{jk}{l}
                  +\tlG^{k}_{jl} \tlGmixed{ik}{l}
                  +\tlG^{k}_{il} \tlGmixed{kj}{l}, \\
  R^{\phi}_{ij} & = & -2\left (\pdpdu{\phi}{}{i}{j}
                 -\tlG^{k}_{ij}\pdu{\phi}{}{k}\right )
                 -2\tlg_{ij} \tlg^{kl} \nonumber\\
            & &     \left ( \pdpdu{\phi}{}{k}{l}
                 -\tlG^{m}_{kl}\pdu{\phi}{}{m}\right )
                + 4\pdu{\phi}{}{i}\pdu{\phi}{}{j} \nonumber\\
             & &    - 4\tlg_{ij}\tlg^{kl}\pdu{\phi}{}{k}\pdu{\phi}{}{l}.
\end{eqnarray}
The derivatives on the right hand side of the BSSN equations are evaluated via standard Finite Differencing
techniques and, unless mentioned otherwise, we use sixth order differencing as a default. We have recently implemented
a fixed mesh refinement for the spacetime evolution which is described in detail in \cite{diener22a}, to which
we refer the interested reader. For the gauge choices we use a variant of 1+log-slicing, where the lapse
is evolved according to
\be
\partial_t \alpha= -2 \alpha K
\ee
and a variant of the $\Gamma$-driver shift evolution with
\be
\partial_t \beta^i= \frac{3}{4}(\tilde{\Gamma}^i-\beta^i).
\ee

\subsection{Coupling between fluid and spacetime}
\label{sec:PM_coupling}
The hydrodynamic equations need the spacetime metric and their
derivatives (known on the mesh; see Eqs.~(\ref{eq:dSdt_metric})
and (\ref{eq:dedt_metric})) and the BSSN equations
need the energy momentum tensor (known on the particles). We therefore need an accurate and efficient mapping
between particles and mesh. In the "mesh-to-particle" step we use a quintic Hermite interpolation which is described
in detail in Sec.~2.4 of \cite{rosswog21a}. For the "particle-to-mesh" step we use a hierarchy of sophisticated kernels
that have been developed in the context of "vortex methods" \cite{cottet00,cottet14}. For relatively uniformly distributed
particles, these kernels deliver results of excellent accuracy, but since they are (contrary to SPH kernels) {\em not} positive
definite, they can deliver unphysical results, if, for example, they are applied across a sharp edge such as a neutron star
surface.\\ 
For this reason we have developed  a "multi-dimensional optimal order detection" (MOOD) strategy, where we calculate
the results with three kernels of different accuracy, $\Lambda_{4,4}, \Lambda_{2,2}$  \cite{cottet14} and $M_4$ \cite{cottet00}.
For (close to) uniform particle distributions, $\Lambda_{4,4}$ is most and $M_4$ is least accurate. 
Out of those kernels only $M_4$ is positive definite and it serves as a robust "parachute" for the cases that the more
accurate kernels should deliver unacceptable results (e.g. because they are applied across a sharp neutron star surface). 
The main idea is to choose the most accurate kernel that is still "admissible". In our previous
paper \cite{diener22a}, we considered a mapping as "admissible" if all of the $T_{\mu \nu}$-components at a grid point $g$, 
$T_{\mu \nu,g}$, are bracketed by the $T_{\mu \nu}$ values of  the contributing particles, see Sec. 2.1.3 of \cite{diener22a}. 
Here, we refine the admissibility criterion further. At each grid point $g$, we calculate for each mapping option $K$
(i.e.  $\Lambda_{4,4}, \Lambda_{2,2}$ or $M_4$) how well the $T_{00,g}^K$ value agrees with the ones
of the nearby SPH particles (labelled by $b$).  More concretely, we choose the mapping option $K$ that minimizes the expression
 \be
 \epsilon^K_g \equiv \sum_b \left(1 - \frac{T_{00,g}^K}{T_{00,b}}\right)^2 M_4\left(\frac{|x_b-x_g|}{l_b}\right) 
 M_4 \left(\frac{|y_b-y_g|}{l_b}\right)  M_4\left(\frac{|z_b-z_g|}{l_b}\right),
 \ee
 where $b$ runs over all particles in the kernel support.
 The first term is just the squared relative energy density error, where the particle $T_{00}$-value is used for the normalization since it
 is guaranteed to be non-zero. The quantity $l_b$ in the tensor product weight is given by $l_b= \left( \nu_b/N_b \right)^{1/3}$.
 The mapping option with the smallest value of $\epsilon^K_g$  is then selected.\\
Clearly, this error measure is not unique and we have run many neutron star merger experiments with alternative
error measures (e.g. using different kernels for the weight; using spherical kernels rather than tensor products etc.). 
While the differences were overall only moderate, the above error measure provided the smoothest results. Compared 
to our previous "bracketing criterion", the new admissibility criterion  based on the above error measure delivers overall 
similar, but somewhat smoother results.

\subsection{Equations of state}
\label{sec:EOS}
In our first exploration of neutron star mergers  \cite{diener22a} we had restricted ourselves to polytropic equations of state (EOSs),
here we take a first step towards more realistic EOSs. We use piecewise polytropic EOSs to approximate microscopic 
models of cold nuclear matter \cite{read09} and we add a thermal, ideal gas-type contribution to both pressure and 
specific internal energy, a common practice in Numerical Relativity simulations. The explicit form of the equation of state 
and the recovery algorithm are explained in detail in Appendix \ref{sec:recovery}. \\
In this first \SpB study with piecewise polytropic equations of state we restrict ourselves to the following equations of state
\bi
\i  SLy \cite{SLY_eos}: with a  maximum TOV mass  $M_{\rm TOV}^{\rm max}= 2.05$ \msun, tidal deformability of a 1.4 \Msun star 
$\Lambda_{1.4}= 297$
\i APR3 \cite{akmal98}: $M_{\rm TOV}^{\rm max}= 2.39$ \msun, $\Lambda_{1.4}= 390$
\i MPA1 \cite{MPA1_eos}: $M_{\rm TOV}^{\rm max}= 2.46$ \msun, $\Lambda_{1.4}= 487$
\i MS1b \cite{MS1_EOS}: $M_{\rm TOV}^{\rm max}= 2.78$ \msun, $\Lambda_{1.4}= 1250$.
\ei
For the tidal deformabilities we have quoted the numbers from Tab.1 of \cite{pacilio22}.
For all cases we use the piecewise polytropic fit according to Tab. III in \cite{read09} and we use a  thermal polytropic  exponent
$\gamma_{\rm th}= 1.75$ as a default, but for one EOS (MPA1) we also use values of 1.5 and 2.0 to explore its impact on the evolution.\\
Given the observed  mass of 2.08$^{+0.07}_{-0.07}$ \Msun for J0740+6620 \cite{cromartie20} the SLy EOS is 
still above the $2\sigma$ lower bound of 1.94 \msun, but probably too soft and we consider it as a limiting case. 
Concerning the currently "best guess" of the maximum neutron star mass, a number of
indirect arguments point to  values of $\sim 2.2-2.4$  \Msun \cite{fryer15,margalit17,bauswein17,shibata17c,rezzolla18}, 
and a recent Bayesian study \cite{biswas22}  suggests a maximum TOV mass of 2.52$^{+0.33}_{-0.29}$ \msun, close to the 
values of APR3 and MPA1. We therefore consider these two EOSs as the most realistic ones in our selection which is 
also consistent with the findings of \cite{pacilio22}. The MS1b
EOS with its very high maximum mass of 2.78 \Msun brackets our selection on the stiffer end.
While its maximum TOV mass is already close to the often quoted upper limit of $\sim 3$ \Msun \cite{rhoades74,kalogera96}, 
it is worth keeping in mind that the upper limit (from causality constraints alone and assuming that nuclear matter is only known close to saturation density) could be as large as 4.2 \Msun \cite{schaffner_bielich20}. But since its tidal deformability is disfavored by
the observation GW170817 \cite{abbott17b} we consider MS1b as a limiting case.

\subsection{Constructing initial data for binary neutron stars} 
To perform simulations of neutron star mergers, we need  initial data (ID) that both satisfy the constraint equations and 
accurately describe the binary systems that we are interested in. Constructing ID for \SpB consists of two parts: a)
solving the general relativistic constraint equations for the standard 3+1 variables
 using the library \Lo \cite{lor} and b) placing
our SPH particles so that they represent the matter distribution found by \lo. This second step is actually non-trivial
since the particle setup should --apart from representing the \Lo matter distribution-- fulfil  a number of additional requirements:
a) the particles should ideally have the same masses since large differences can lead to numerical noise, b) the particles
should be locally very ordered so that they provide a good interpolation accuracy, c) but the particles should {\em not} be on a lattice
with preferred directions (as most simple lattices have) since this can lead to artefacts, e.g. for shocks travelling along those 
axes.\\
All these issues are addressed in the "Artificial Pressure Method" (APM) \cite{rosswog20a}. The main idea of this method
is to place equal mass particles in an initial guess and then let the particles themselves find the position where they best
approximate a given density profile (here provided by \lo). This is achieved in an iterative process where at each step
the current density is measured and compared to the desired profile density. We use the relative error between both  to define
an "artificial pressure" which is then applied in a position update prescription that is derived from an SPH momentum equation.
In other words, at each step each particle measures its own, current density error and then moves into a direction which reduces it.
The method was initially suggested in a Newtonian context  \cite{rosswog20a}, then translated to a General
Relativistic context  to accurately construct neutron stars \cite{rosswog21a}. More recently \cite{diener22a}, it has been adapted 
to model binary neutron star systems where it has delivered accurate General Relativistic SPH initial conditions. 
The construction of APM particle distributions for binary neutron stars has been implemented in our code \spid, already used in 
\cite{diener22a}. By now, \spiD has been improved so that it can also easily be extended to produce BSSN and SPH 
initial data for other astrophysical systems.

\subsection{Summary of the new elements}
\noindent In summary, the new elements described in this study are
\bi
\i We enhance the slope limiter used in the reconstruction ({\tt minmod}) by an exponential suppression term that 
    enhances the dissipation for those rare cases where particles should get too close to each other, see Sec.~\ref{sec:AV_reconst}.
    The effect of this change is only tiny, but we mention it for completeness.
\i We trigger dissipation based on a shock indicator similar to \cite{cullen10} and a noise indicator suggested for the {\tt SPHINCS\_SR} code \cite{rosswog15b}, see Sec.~\ref{sec:AV_steering}.
\i As in our previous study \cite{diener22a}, we use a MOOD approach to decide which kernel to use in the mapping, but here
   we use a more sophisticated acceptance measure, see our Sec.~\ref{sec:PM_coupling}.
\i We use, for the first time in \spB, piecewise polytropic approximations to nuclear equations of state. These fits to cold nuclear matter
   equations of state are enhanced by thermal pressure contributions, see Appendix  \ref{sec:recovery} for a detailed description.
\ei

\section{Results}
\label{sec:results}
After our initial code papers \cite{rosswog21a,diener22a}, we take here our next step towards more 
realistic simulations of neutron star mergers: we use piecewise polytropic approximations to nuclear matter
equations of state. These first \SpB simulations of such mergers are the main topic, but since we also have 
implemented a new dissipation steering, we demonstrate how it works in a first "Shock tube" section.
\begin{figure}[H]	
\widefigure
\centerline{\includegraphics[width=13 cm]{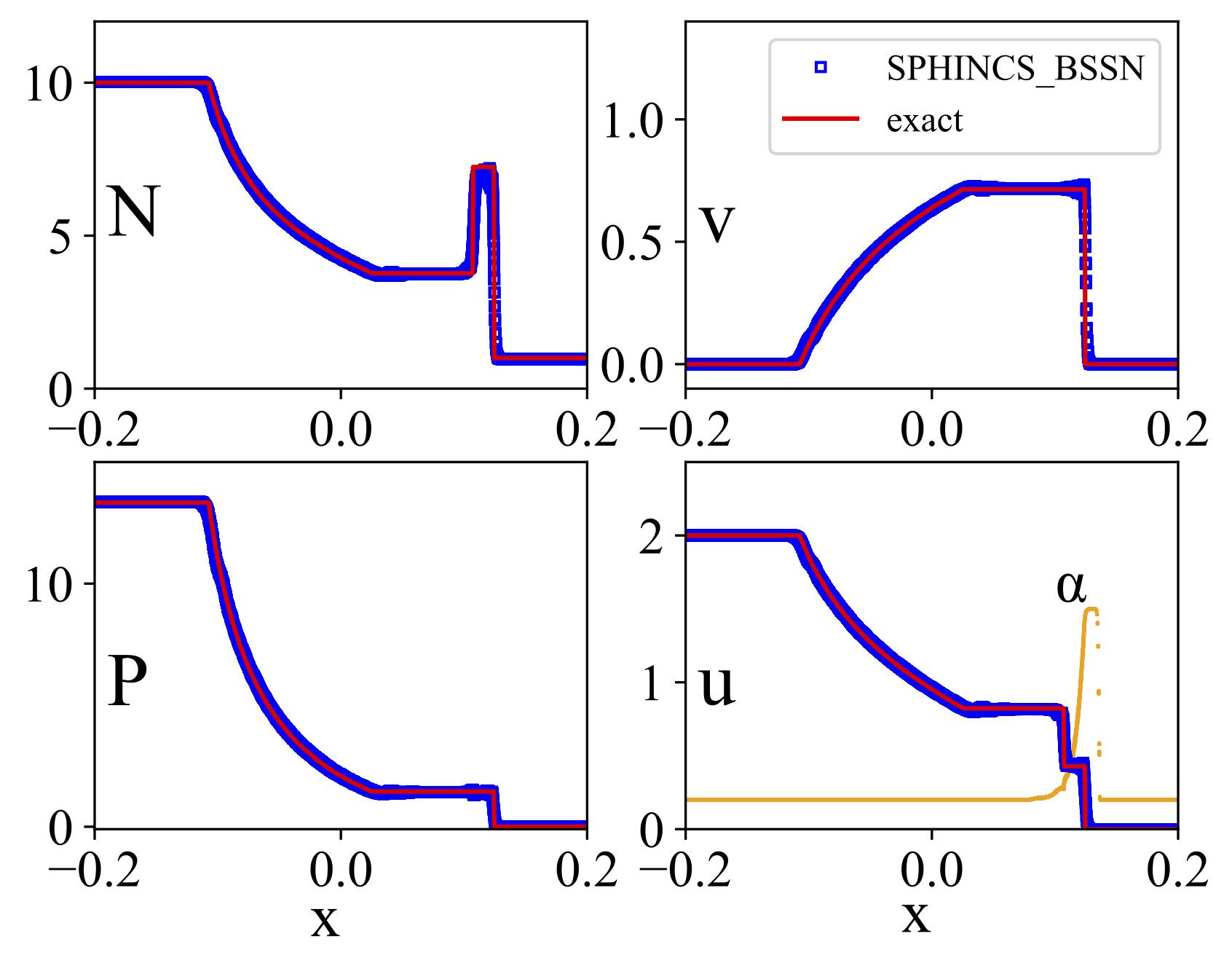}}
\caption{Results for the 3D, relativistic shock tube test. Shown are the result (density, velocity, pressure and internal energy;  $\Delta x_L=0.00075$) for our default choices, the orange dots in the "u" panel show values of the dissipation parameter $\alpha$.   Dissipation robustly switches on ahead of the shock front, remains approximately constant around the shock front and decays quickly in the post-shock region.\label{fig:shock_test}}
\end{figure}  
\subsection{Shock tube}
This test is a relativistic version of ``Sod's shocktube'' \cite{sod78} which has become
a standard benchmark for relativistic hydrodynamics codes \cite{marti96,chow97,siegler00a,delzanna02,marti03,marti15}.
Apart from demonstrating that our code solves the relativistic hydrodynamics equations correctly, we use this
test here also to show where our steering method, see Sec.~\ref{sec:AV_steering}, triggers dissipation.\\
The test uses a polytropic exponent $\Gamma=5/3$ and as initial conditions
\be
\left[ N, P \right]=   \left\{
    \begin{array}{ll}
         \left[10,\frac{40}{3}\right], & {\rm for \;}  x<0\\ \\
         \left[1,10^{-6}\right]          & {\rm for \;} x \ge 0,
   \end{array}
  \right.
\ee
with velocities initially being zero everywhere.
We place  particles with equal baryon numbers on close-packed lattices as described in \cite{rosswog15b},
so that on the left side the particle spacing  is $\Delta x_L= 0.00075$ and we have 12 particles in both $y$-
and $z$-direction. This test is performed with the full GR-code in a fixed Minkowski metric.
The result at $t= 0.15$ is shown in Fig.~\ref{fig:shock_test} with the \SpB results
marked with blue squares and the exact solution \cite{marti03} with the red line. \\
The \SpB results (blue squares) are in very good agreement with  the exact solution (red line). In the fourth sub-panel  
we show in orange the dissipation parameter $\alpha$. It  abruptly switches to $\alpha^{\rm max}$ just ahead of the shock front,
stays constant around it, and decays in the post-shock region very quickly to the floor value. Overall we have a very good agreement
with the exact solution. Only directly behind the shock front (e.g. in velocity and density) is a small amount of noise visible.
This is to some extent unavoidable, since the particles try to optimize their local distribution, see e.g. Sec. 3.2.3 in \cite{rosswog15c}, 
and need to transition from their initial arrangement into a new one.\\
We have also experimented with other slope limiters ({\tt van Leer} \cite{vanLeer77}, {\tt vanLeer Monotonized Central} \cite{vanLeer77} and {\tt superbee} \cite{sweby84}), but all of them showed an
increased velocity overshoot at the shock front and no obvious other advantage. We therefore settled on the {\tt minmod} limiter, but
we do not expect to see substantial differences when other slope limiters are used and this is also confirmed
by a number of  additional merger test simulations (not discussed further here).\\
For more special-relativistic benchmark tests with SPH the interested reader is referred to \cite{rosswog10b,rosswog11a,rosswog15b}.

\subsection{Binary mergers}
\subsubsection{Performed simulations}
Our performed simulations are summarized in Tab.~\ref{tab:runs}. We perform for each EOS several runs
with at least two different resolutions (1 and 2 million SPH particles; for the grid resolution see below) and 
for a case where the cold nuclear part corresponds to the MPA1 EOS we also vary the thermal exponent 
$\gamma_{\rm th}$. For our presumably most realistic EOSs, 
MPA1 and APR3, we also perform runs with 5 million particles, but we note that these runs are extremely 
expensive for our current simulation technology and are therefore not run for as long as the other cases. For the very compact 
stars resulting from the SLy EOS our current resolution may be at the lower end, especially for 1 million case, 
and the corresponding results should be taken with a grain of salt. This case will be re-assessed in future, 
better resolved simulations.\\
For the spacetime evolution we employ seven levels of fixed mesh refinement with the outer boundaries in each
coordinate direction at $\approx 2268$ km and  143, 193, 291 grid points in each direction for the 1, 2 and 5 million
SPH particle runs. The corresponding resolution lengths of our finest grids, $\Delta_g^{\rm min}$, are shown 
in the fourth column of Tab.~\ref{tab:runs}.
Note that due to the approach chosen in \spB, we have the freedom to choose different resolutions for the
spacetime and the hydrodynamics. For example, if the spacetime is not too strongly curved, say, for a neutron star
with a rather stiff equation of state, we may obtain reasonably accurate results with only a moderate grid resolution.
In such cases, we can instead invest the available computational resources in a higher hydrodynamic resolution, i.e. in 
larger SPH particle numbers.  Since all our simulation technology is very new, the relative
resolutions are still to some extent a matter of experiment. This is discussed in more detail in Appendix \ref{sec:resolution}.  
The minimum smoothing lengths reached in each simulation, $h_{\rm min}$,  
are also shown in Tab.~\ref{tab:runs} as a measure of the hydrodynamical resolution length. Note that today's
state-of-the-art Eulerian simulations typically have a smaller finest grid length (e.g. \cite{kashyap22} use 185 m),
but our hydrodynamic resolution length can go substantially below such length scales, see Tab.~\ref{tab:runs}.

\begin{specialtable}[H] 
\small
\caption{Simulated binary systems. All binaries are irrotational, have twice 1.3 \Msun (gravitational mass of 
each star in the binary system) and the simulations start 
from an initial separation of 45 km. Unless mentioned otherwise, the thermal exponent $\gamma_{\rm th}= 1.75$ 
is used. $\Delta_g^{\rm min}$ refers to the finest grid resolution length, $h_{\rm min}$ is the minimum resolution
length (= smoothing length) in the hydrodynamic evolution.
\label{tab:runs}}
\begin{tabular}{lccccc}
\toprule
name & EOS            & \#particles           & $\Delta_g^{\rm min}$ [m] & $h_{\rm min}$ [m]  &comment\\
\midrule
{\tt MPA1\_1mio}&MPA1	   & $1 \times 10^6$	& 499 			          & 188 &\\
{\tt MPA1\_2mio}&MPA1	   & $2 \times 10^6$	& 369 			          & 172 &\\
{\tt MPA1\_5mio}&MPA1	   & $5 \times 10^6$	& 244 			          & 117 &\\
{\tt MPA1\_2mio\_$\Gamma_{\rm th}1.5$}&MPA1	   & $2 \times 10^6$	& 369 	&	161	          & $\gamma_{\rm th}=1.5$\\
{\tt MPA1\_2mio\_$\Gamma_{\rm th}2.0$}&MPA1	   & $2 \times 10^6$	& 369 	&	149	          & $\gamma_{\rm th}=2.0$\\
{\tt APR3\_1mio}&APR3	   & $1 \times 10^6$	& 499 			          & 145 & \\
{\tt APR3\_2mio}&APR3	   & $2 \times 10^6$	& 369 			          & 136 &\\
{\tt APR3\_5mio}&APR3	   & $5 \times 10^6$	& 244 			          & 106 &\\
{\tt SLy\_1mio}&SLy   	   & $1 \times 10^6$	& 499 			          & 140 &\\
{\tt SLy\_2mio}&SLy	           & $2 \times 10^6$	& 369 			          & 106 &\\
{\tt MS1b\_1mio}&MS1b	   & $1 \times 10^6$	& 499 			          & 270 &\\
{\tt MS1b\_2mio}&MS1b	   & $2 \times 10^6$	& 369 			          & 222 &\\
\bottomrule
\end{tabular}
\end{specialtable}

\begin{figure}[H]
\centerline{\includegraphics[width=14cm]{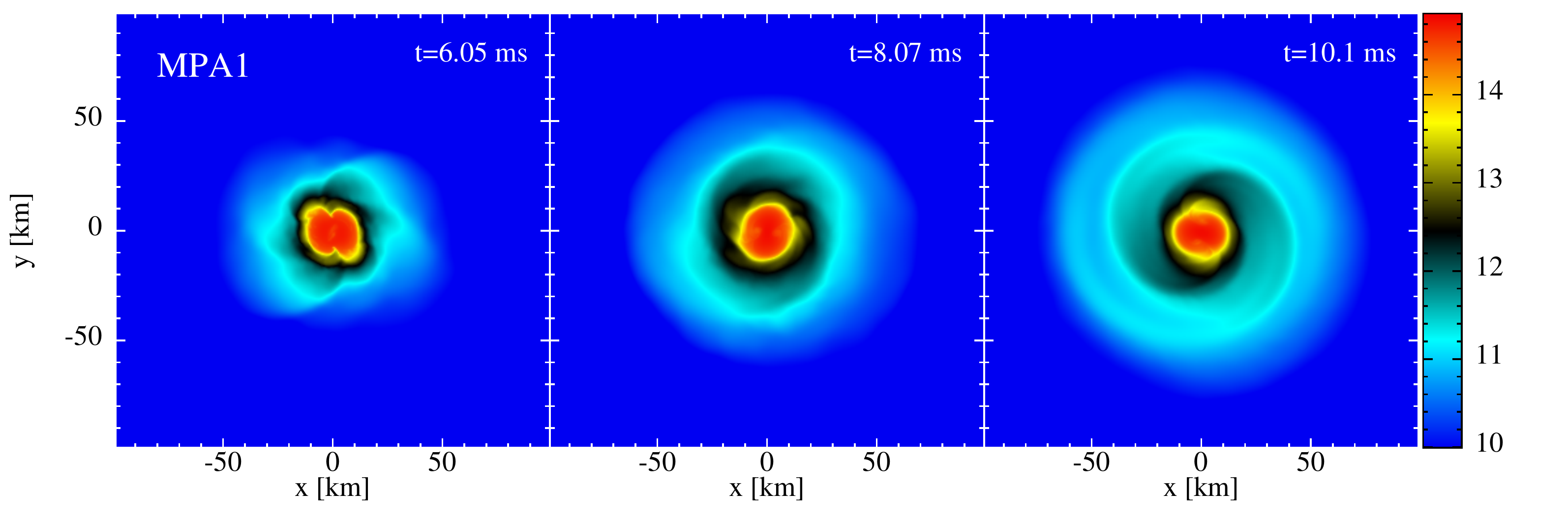}}
\centerline{\includegraphics[width=14cm]{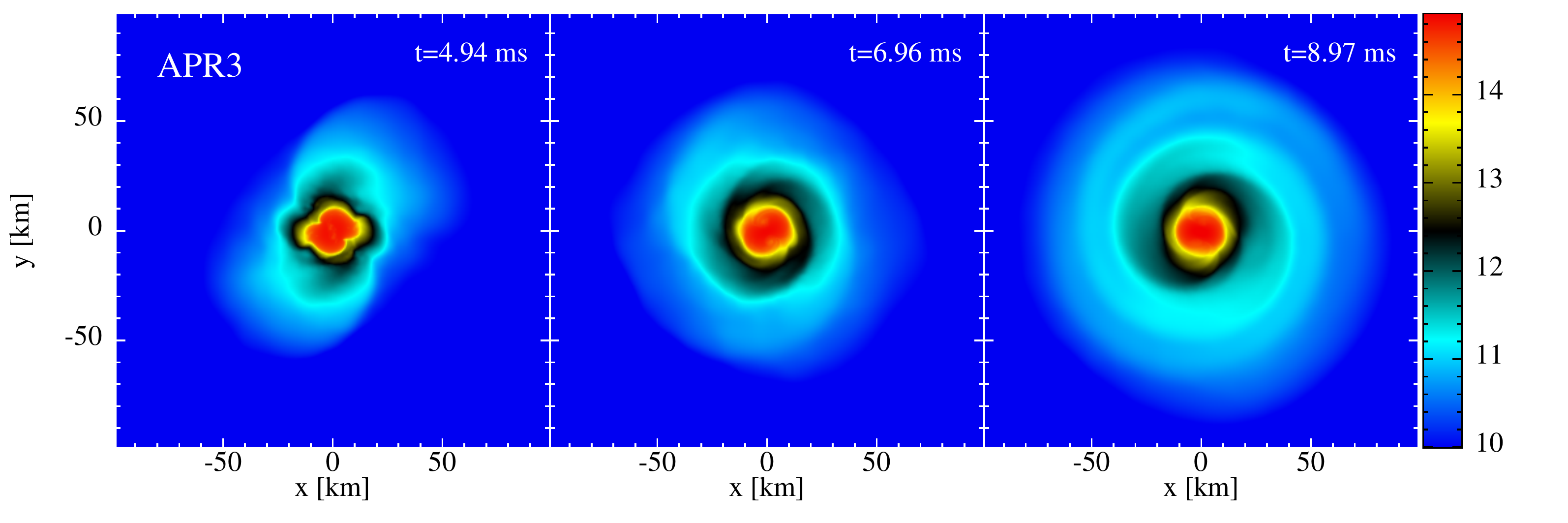}}
\centerline{\includegraphics[width=14cm]{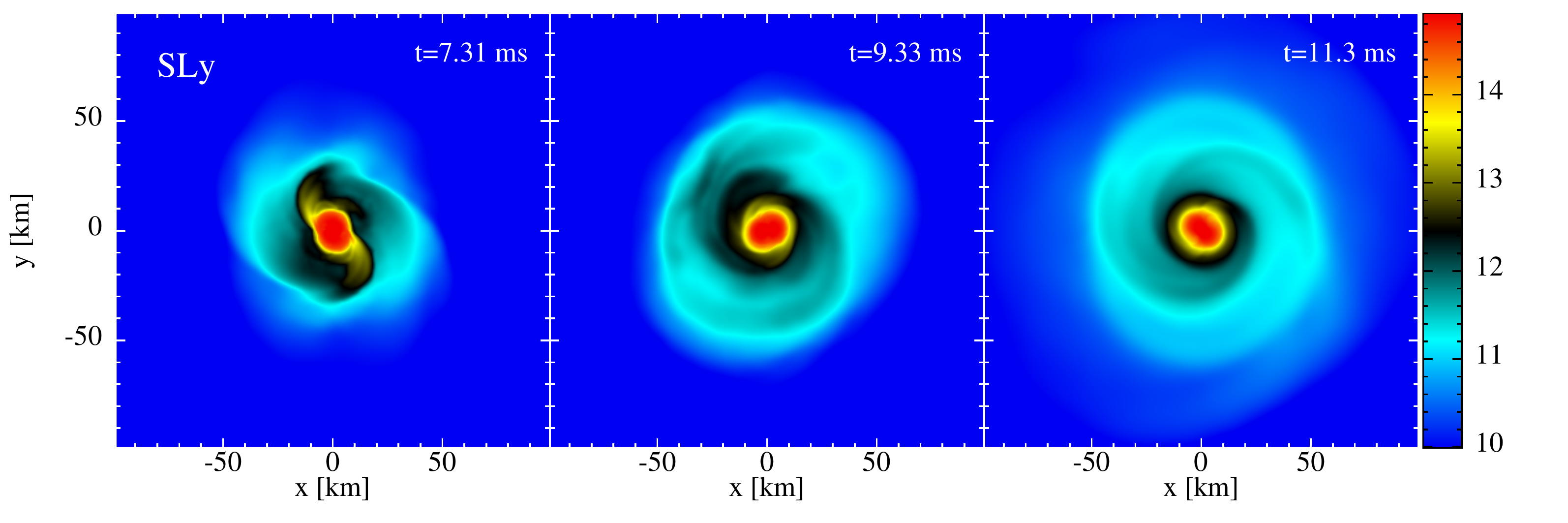}}
\centerline{\includegraphics[width=14cm]{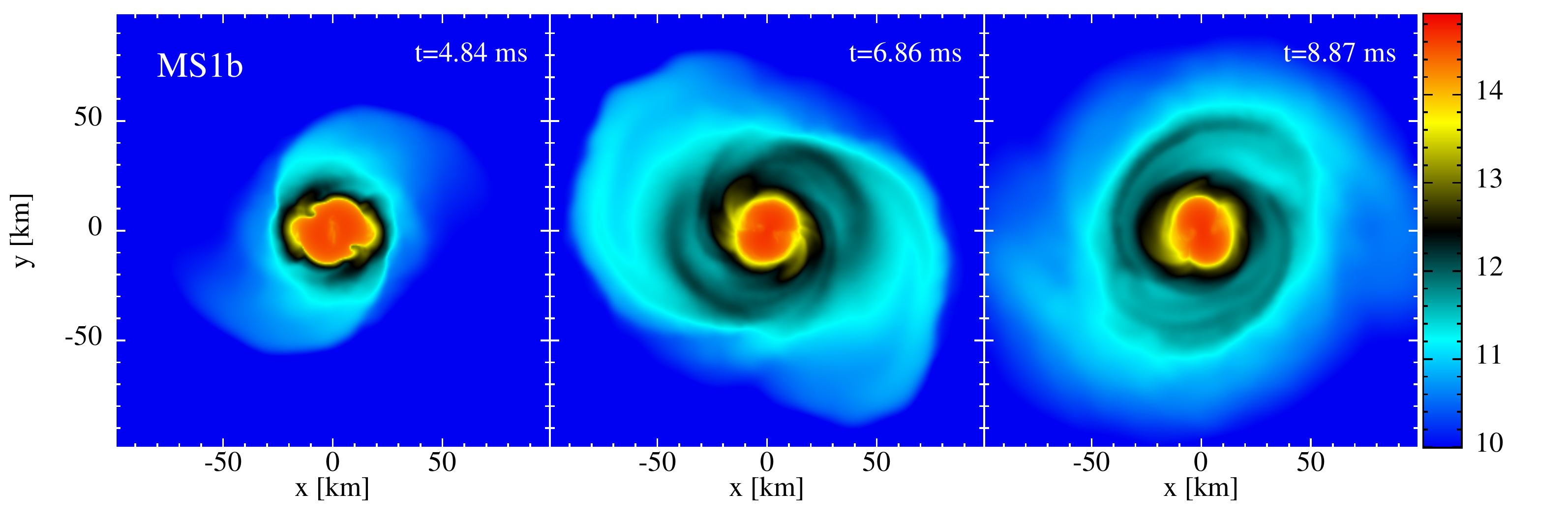}}
\caption{Density evolution (top to bottom row) for the MPA1, APR3, 
SLy and MS1b EOS. Each time  the run with 2 million particles, see 
Tab.~\ref{tab:runs}, is shown. For each EOS case, 
the times are chosen so that they correspond 
 to 1.57, 3.55 and 5.52 ms after merger (defined as the time of peak 
GW amplitude). }
\label{fig:dens_evol}
\end{figure}   
\subsubsection{Dynamical evolution}
\begin{figure}[H]
\hspace*{-0.5cm}\includegraphics[width=14cm]{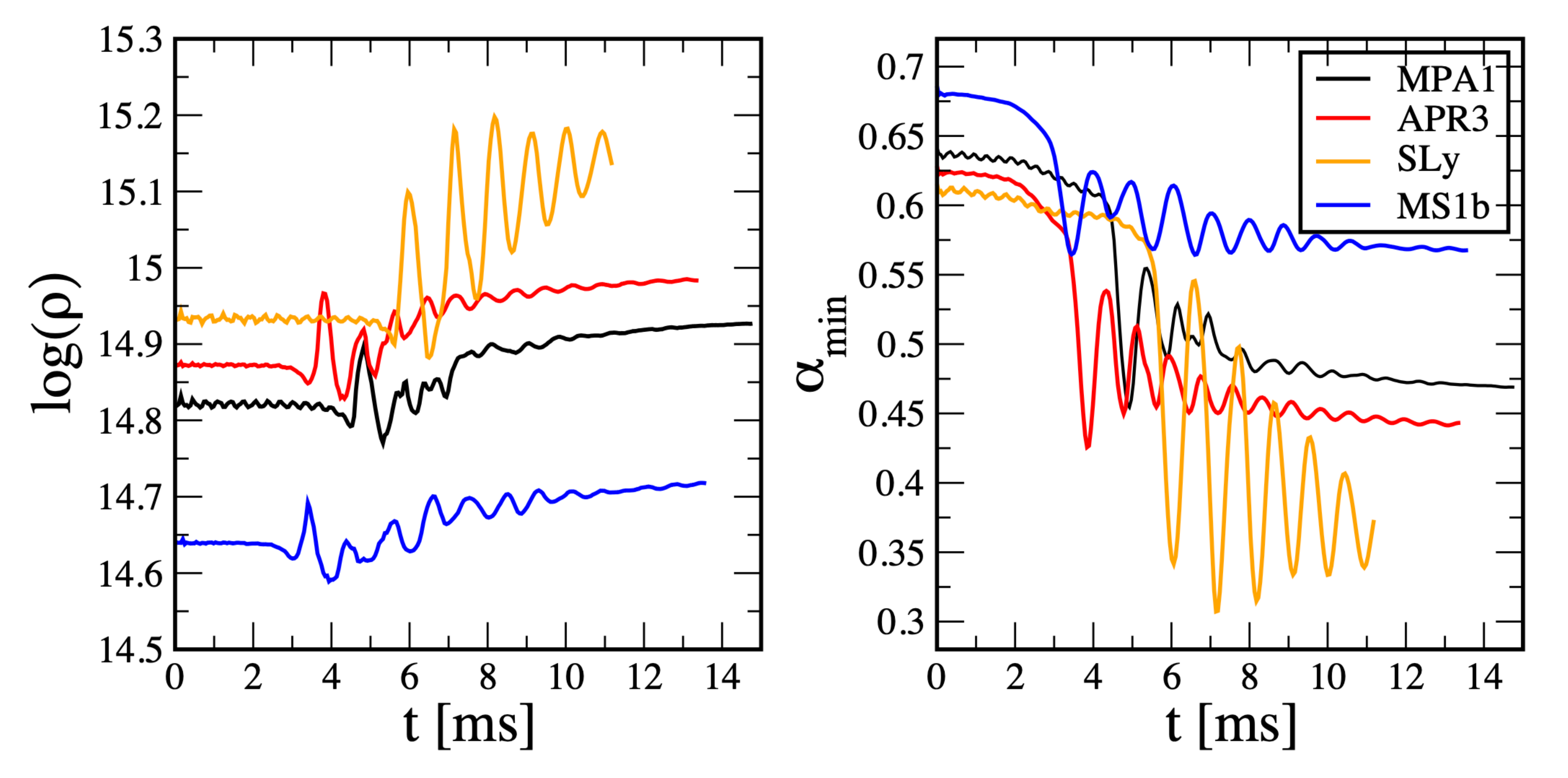}
\caption{Maximum density (in g cm$^{-3}$) and minimum lapse value for the runs with 2 million particles (i.e. runs {\tt MPA1\_2mio}, 
{\tt APR3\_2mio}, {\tt SLy\_2mio} and {\tt MS1b\_2mio}, see Tab.~\ref{tab:runs}).}
\label{fig:rho_lapse}
\end{figure}

\begin{figure}[H]
\centerline{\includegraphics[width=11cm]{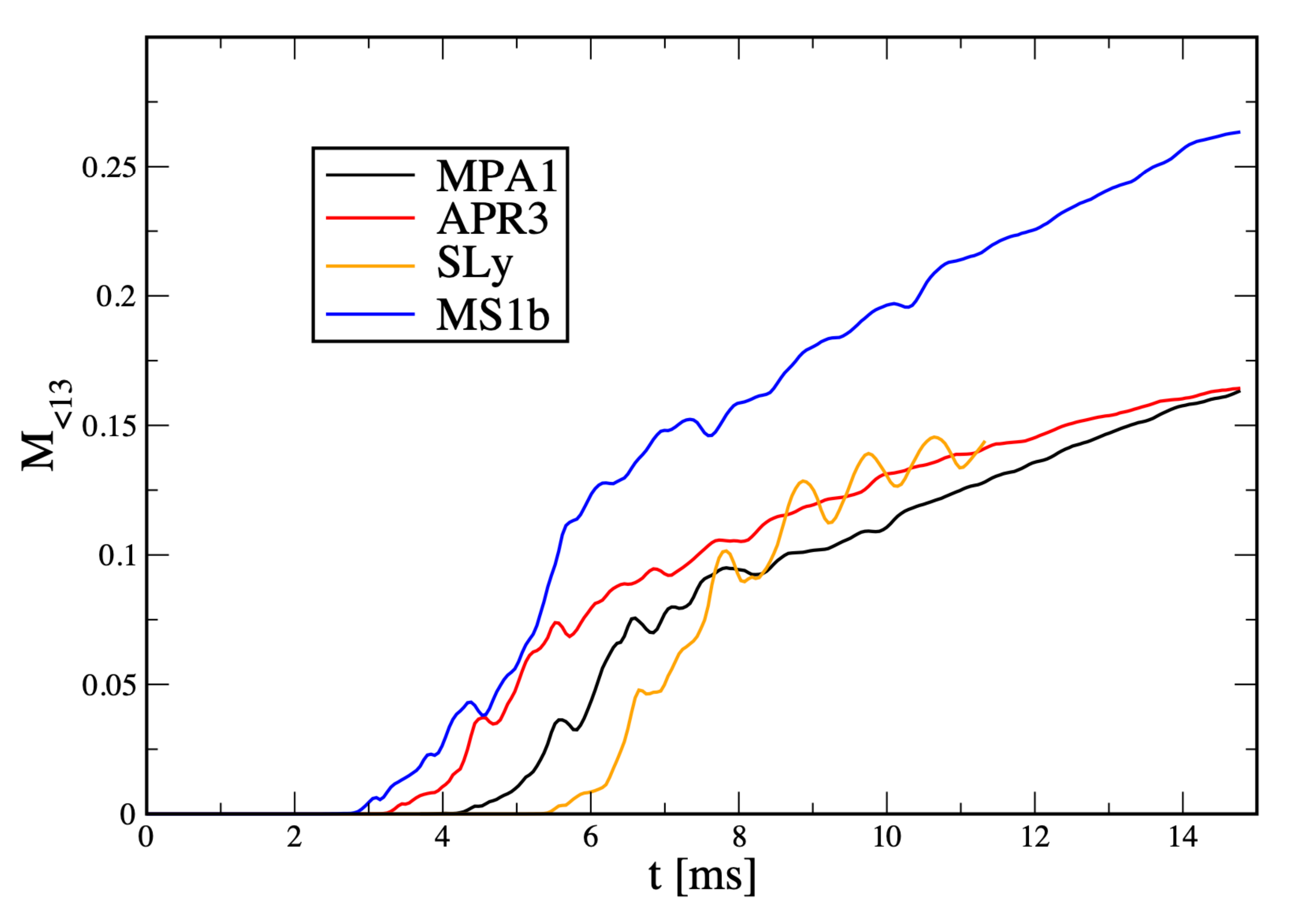}}
\caption{Disk mass evolution for the runs with 2 million particles (i.e. runs {\tt MPA1\_2mio}, 
{\tt APR3\_2mio}, {\tt SLy\_2mio} and {\tt MS1b\_2mio}, see Tab.~\ref{tab:runs}).}
\label{fig:disk_mass}
\end{figure}

In Fig.~\ref{fig:dens_evol} we show the density evolution in the orbital plane for each EOS, 
every time showing the 2 million particle runs (i.e. for {\tt MPA1\_2mio}, {\tt APR3\_2mio}, {\tt SLy\_2mio}, {\tt MS1b\_2mio}, see Tab.~\ref{tab:runs})
at 1.57, 3.55 and 5.52 ms  after merger (defined as the time of peak GW amplitude).  As expected, the EOS has a fair impact
on the last inspiral stages where tidal effects accelerate the motion and lead to an earlier merger at lower frequencies
as compared to systems without tidal effects (i.e. black hole mergers), see e.g. \cite{flanagan08,damour10,bernuzzi20a}. \\
Our softest EOS, SLy, produces the most compact remnant and undergoes very deep oscillations, see Fig.~\ref{fig:rho_lapse},
with the density in the remnant settling to more than twice the initial stellar density. Our two arguably most realistic EOSs, MPA1 and APR3,
look morphologically very similar, see the first two rows in Fig.~\ref{fig:dens_evol}, but also their peak density and minimum lapse evolution
looks alike: they show a first, dominant compression (density increase $\sim 20-25$\%) and then, within a couple of oscillation periods,
they settle onto a final central density which is approximately the same as the first compression spike. The extremely stiff MS1b EOS
shows a qualitatively similar peak density/lapse evolution, though at substantially lower densities/higher lapse values. Interestingly, none of the explored cases seem close to a collapse to a black hole, even the soft SLy EOS cases settles to a minimum lapse value of $\alpha_{\rm min}\approx 0.35$. As a rule of thumb: systems with central lapses dropping below $\approx 0.2$ are doomed to collapse to a  black hole, 
see, for example, \cite{bernuzzi20b} or Fig.16 in our first \SpB paper \cite{rosswog21a} which shows the shape of the lapse function when an 
apparent horizon is detected for the first time.
We also monitor the quantity $M_{<13}$, the (baryonic) mass of matter with a density smaller than $10^{13}$ g cm$^{-3}$ minus the ejected
mass, see below, which we consider as a proxy for the resulting torus mass. The astrophysical relevance of the torus mass 
stems from its role as an energy reservoir for powering short GRBs after the collapse to a BH \cite{nakar07,lee09,kumar15},
but also since $\sim 40$\% of this mass can potentially become unbound \cite{metzger08a,beloborodov08,siegel17a,siegel18,miller19,fernandez19}, and therefore likely contributes the lion's share
of the ejecta budget of a neutron star merger.\\
We show the temporal evolution of $M_{<13}$ in Fig.~\ref{fig:disk_mass}. None of the simulations seems to have reached a
stationary state yet, all of them keep shedding mass into the torus, but all  have already reached a torus mass exceeding 
0.15 \msun. Thus, assuming an efficiency $\epsilon$ to translate
this rest mass energy reservoir into radiation, bursts with a (true) energy of $> 10^{53} \; {\rm erg} \; (\epsilon/0.05) (M_{<13}/0.15 M_\odot) c^2$ could be reached. If a fraction of $\eta$ of the initial torus becomes unbound, one can expect neutron-rich ejecta of $> 0.06 M_\odot 
(\eta/0.4) (M_{<13}/0.15 M_\odot)$, roughly consistent  with the estimates for GW170817 \cite{kasen17,cowperthwaite17,evans17,villar17,kasliwal17,tanvir17,rosswog18a}.

\subsubsection{Impact of the thermal index $\gamma_{\rm th}$}
\begin{figure}[H]
\centerline{\includegraphics[width=14cm]{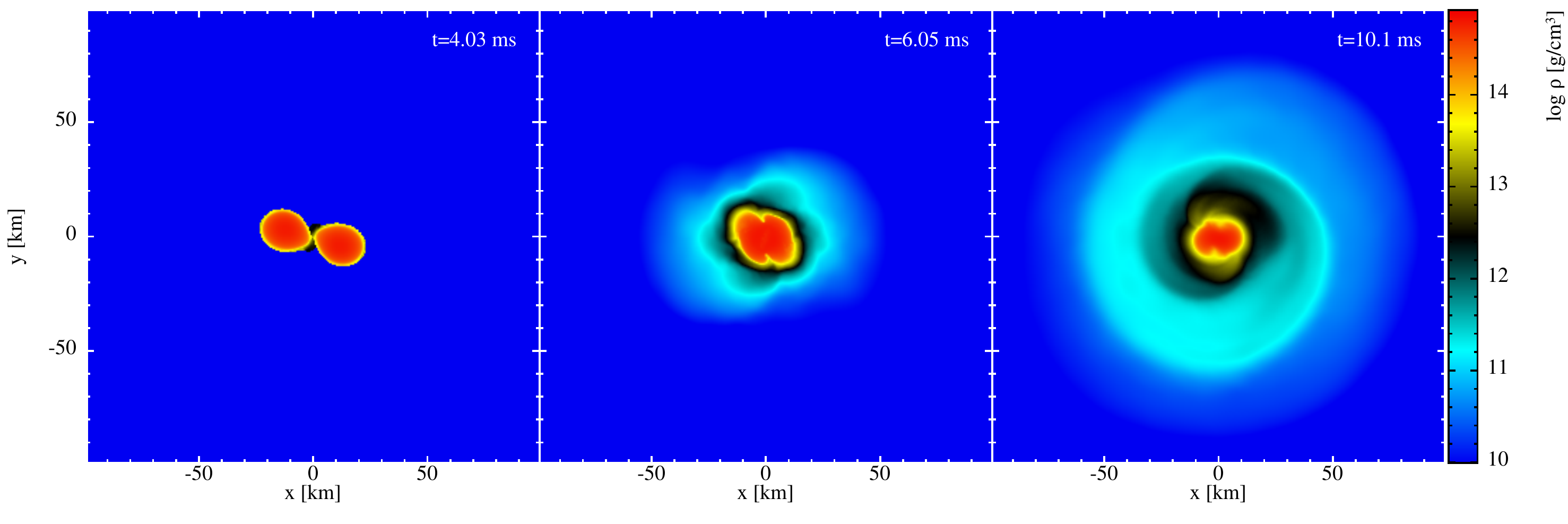}}
\centerline{\includegraphics[width=14cm]{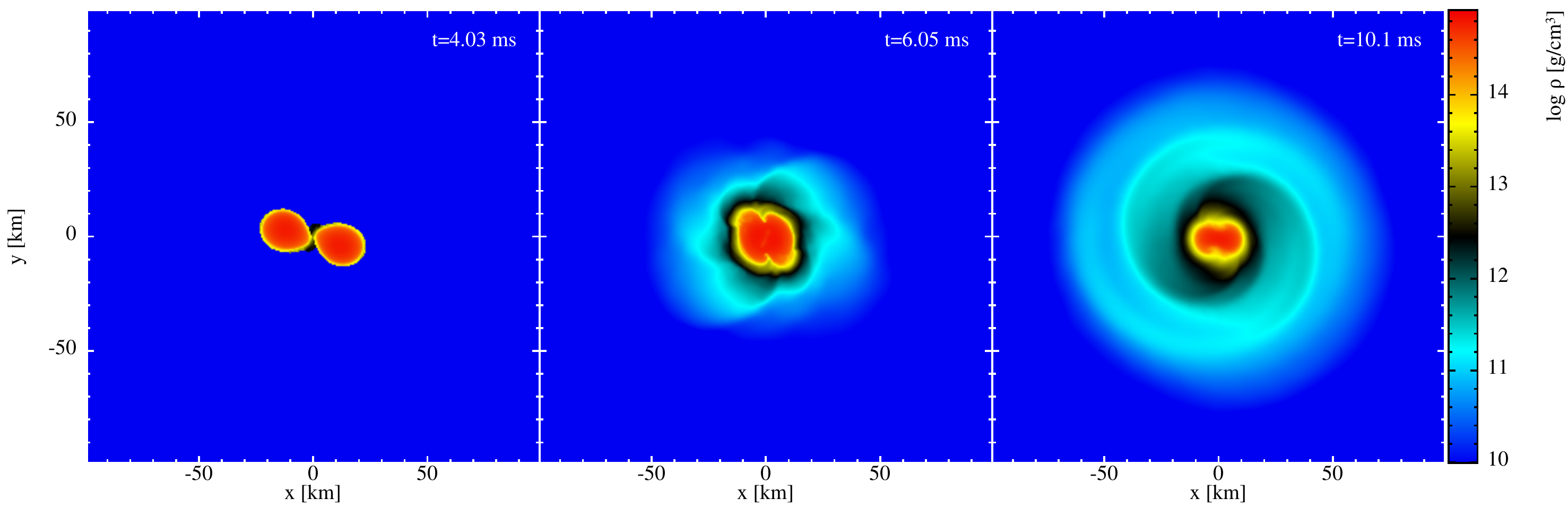}}
\centerline{\includegraphics[width=14cm]{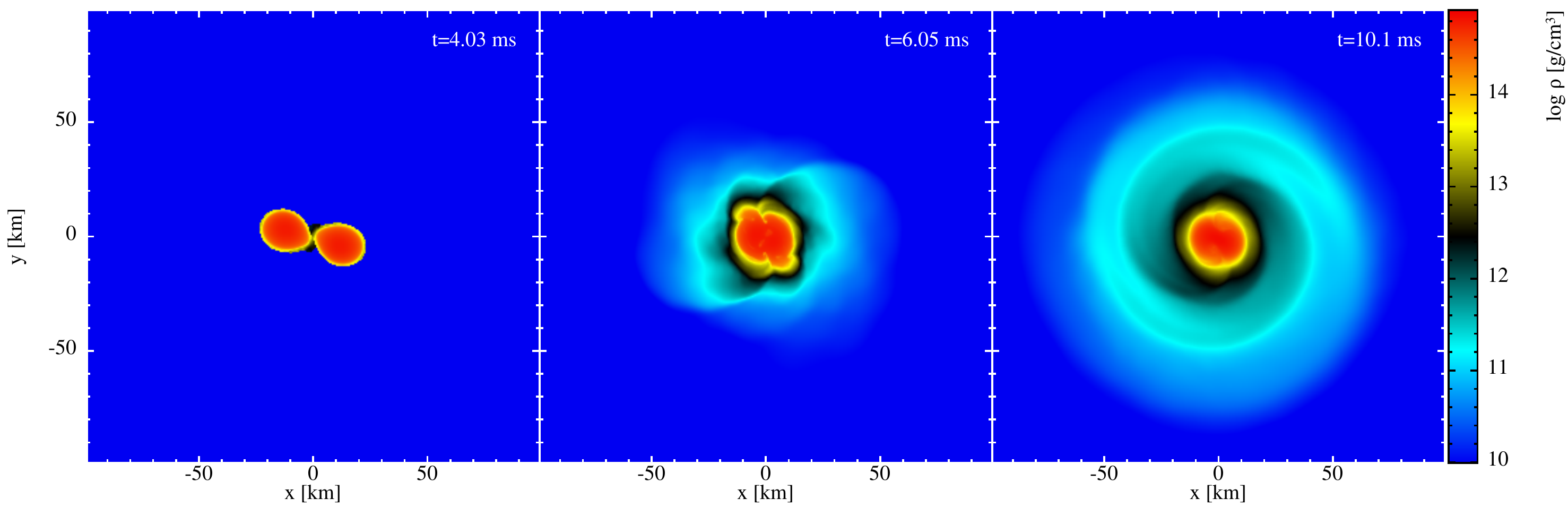}}
\caption{Impact of the thermal exponent $\gamma_{\rm th}$ on the evolution. 
Shown is a binary merger with 2$\times1.3$ \msun,
the MPA1 EOS and $\gamma_{\rm th}= 1.5$, 1.75 and 2.0 (top to bottom row).}
\label{fig:G_th}
\end{figure}

\begin{figure}[H]

\vspace*{-0.1cm}

\centerline{\includegraphics[width=13cm]{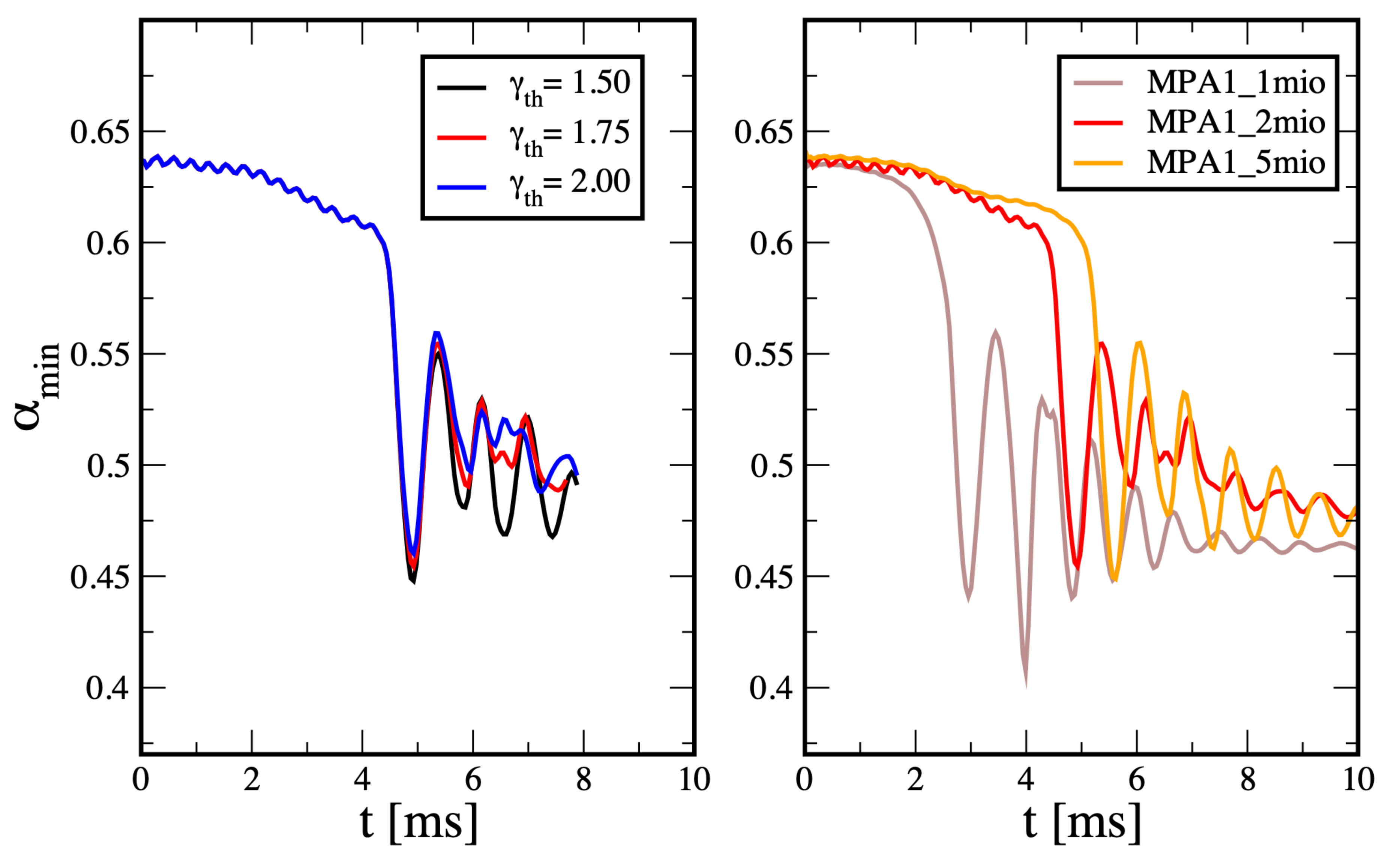}}

\vspace*{-0.3cm}

\caption{ Comparison of the evolution of the minimal lapse value (2$\times1.3$ \msun,
MPA1 EOS) for different thermal exponents $\gamma_{\rm th}$ (left) and different resolutions
(right).}
\label{fig:Gamma_th_lapse}
\end{figure}

Our treatment of thermal effects by adding an ideal gas-type pressure with a thermal index $\gamma_{\rm th}$, see Appendix A,
is clearly very simple and recently more sophisticated approaches have been developed \cite{raithel19,raithel21a}. In particular,
if thermal effects are to be described via such an ideal gas-type index $\gamma_{\rm th}$, it should vary with the local physical
conditions (e.g. density). To test for the impact of $\gamma_{\rm th}$, we perform two additional runs (2 $\times 1.3$ \Msun 
with the MPA1 EOS) where  we use,  apart from our default choice $\gamma_{\rm th}= 1.75$, also the values 1.5 and 2.0. The 
morphology of these runs is shown in Fig.~\ref{fig:G_th}. 
While the impact of $\gamma_{\rm th}$ on the mass distribution is overall moderate, it has some noticeable 
impact on the spacetime evolution as illustrated with the minimum lapse value shown in Fig.~\ref{fig:Gamma_th_lapse}, 
left panel. To get a feeling for the effects of resolution, we plot in the right panel also the MPA1 case for three different resolutions.
Smaller values of $\gamma_{\rm th}$ make the EOS overall more compressible which leads to larger amplitude 
oscillations in the minimum lapse. Since these oscillations go along with mass shedding, there is also some impact 
of $\gamma_{\rm th}$ on the amount of ejected mass, see below. 

\subsubsection{The triggering of artificial dissipation}
\begin{figure}[H]
\centerline{\includegraphics[width=15cm]{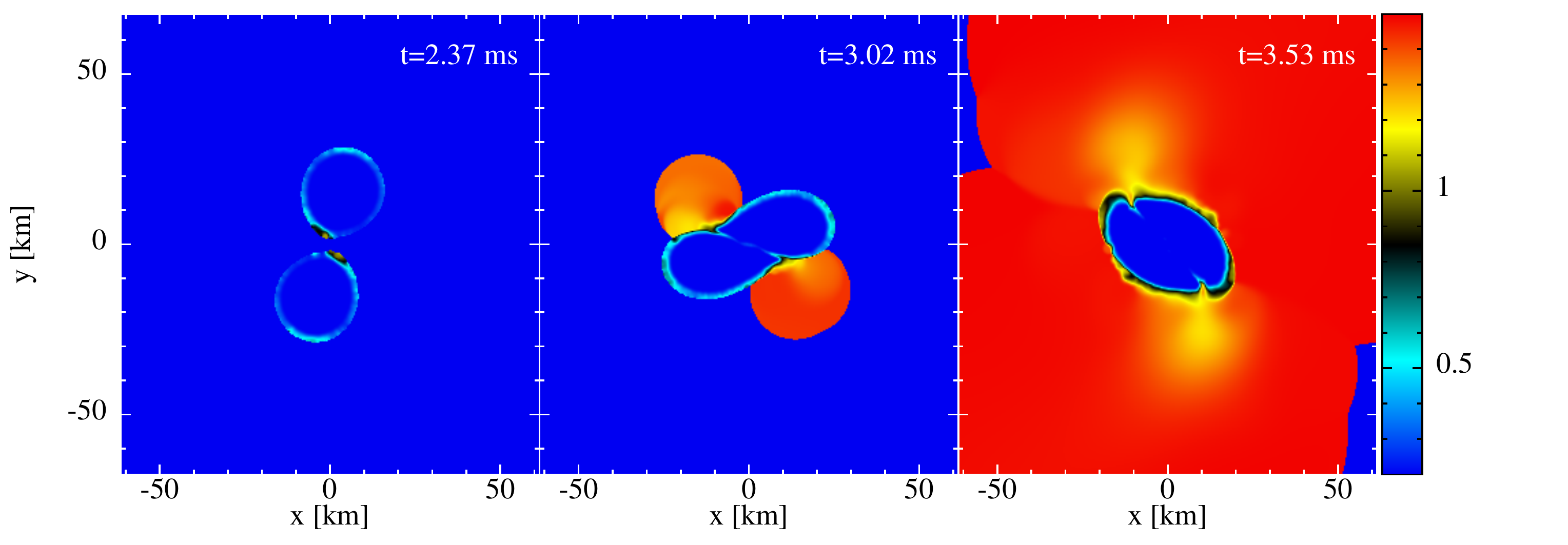}}
\caption{Evolution of the dissipation parameter $\alpha_{\rm AV}$ in
simulation {\tt MS1b\_2mio}, see Tab.~\ref{tab:runs}.}
\label{fig:AV_steering}
\end{figure}

To demonstrate where dissipation is triggered in a neutron star merger, we show in Fig.~\ref{fig:AV_steering} 
the values of the dissipation parameter $\alpha_{\rm AV}$  for simulation  {\tt MS1b\_2mio}. The left panel shows 
the dissipation parameters at the end of the  inspiral, just before merger. At this stage nearly all of the matter
has  dissipation parameters very close the floor value (here $\alpha_0= 0.2$), only
particles in a thin surface layer (e.g. at the cusps) have moderately larger values. As a side remark, we want to point
out how well-behaved the surfaces of the neutron stars are. Contrary to Eulerian hydrodynamics, in our approach 
no special treatment of the surface layers is needed, the corresponding particles are treated exactly as all the other 
particles. Inside the stars the dissipation parameter values hardly increase above the floor value, not even during the
merger, but the particles that are "squeezed out" of the shear layer between the stars have values $> 1$.
Since their sound speed drops rapidly during the decompression, their dissipation values only slowly decay towards 
lower values, see Eq.~(\ref{eq:AV_decay}). As mentioned above, we have chosen our dissipation triggers conservatively, so that
likely more dissipation is triggered than is actually needed. A possible reduction will be explored in future work. 
\begin{figure}
\centerline{\includegraphics[width=0.54\columnwidth]{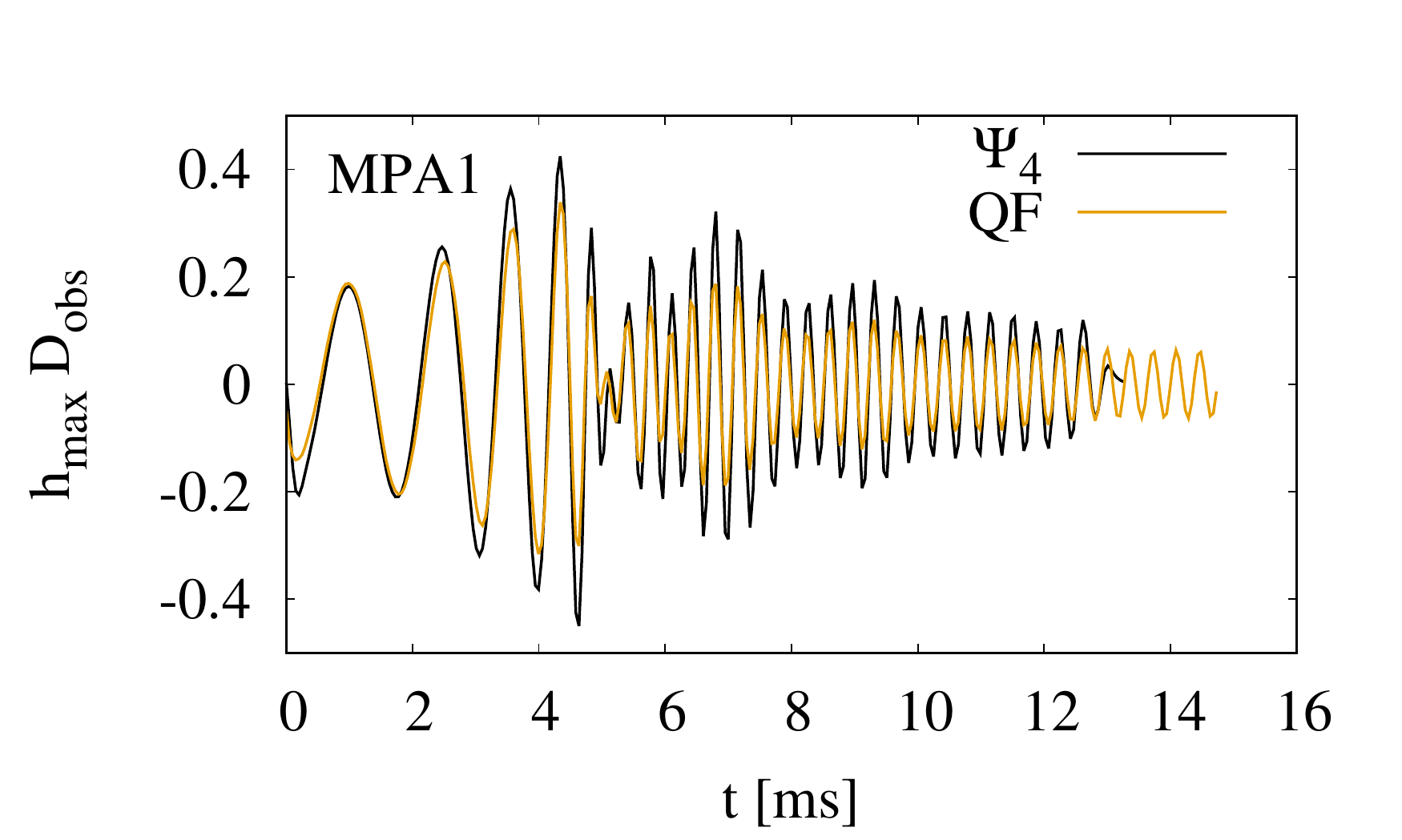}\hspace*{-0.5cm}
\includegraphics[width=0.54\columnwidth]{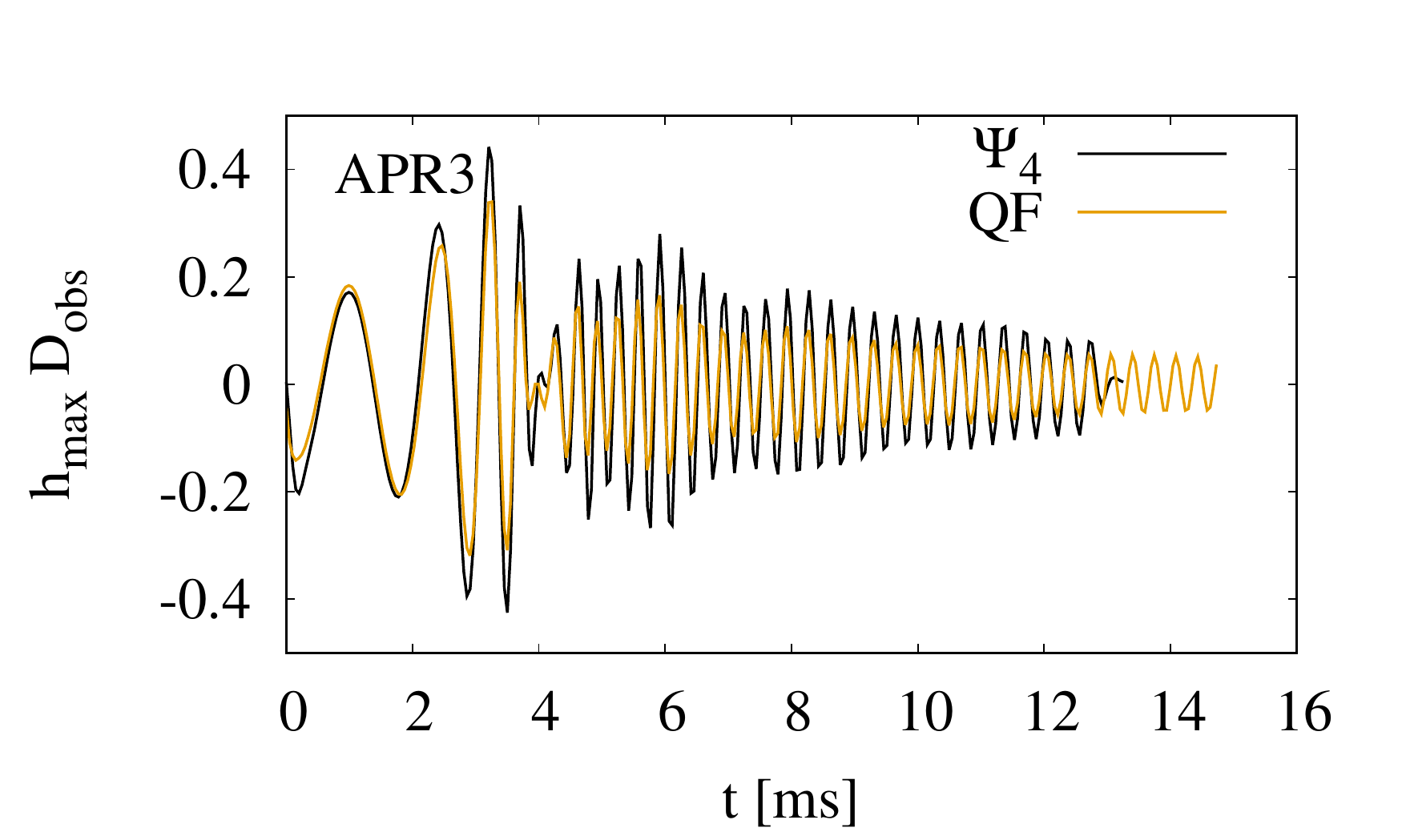}}

\vspace*{-0.5cm}

\centerline{\includegraphics[width=0.54\columnwidth]{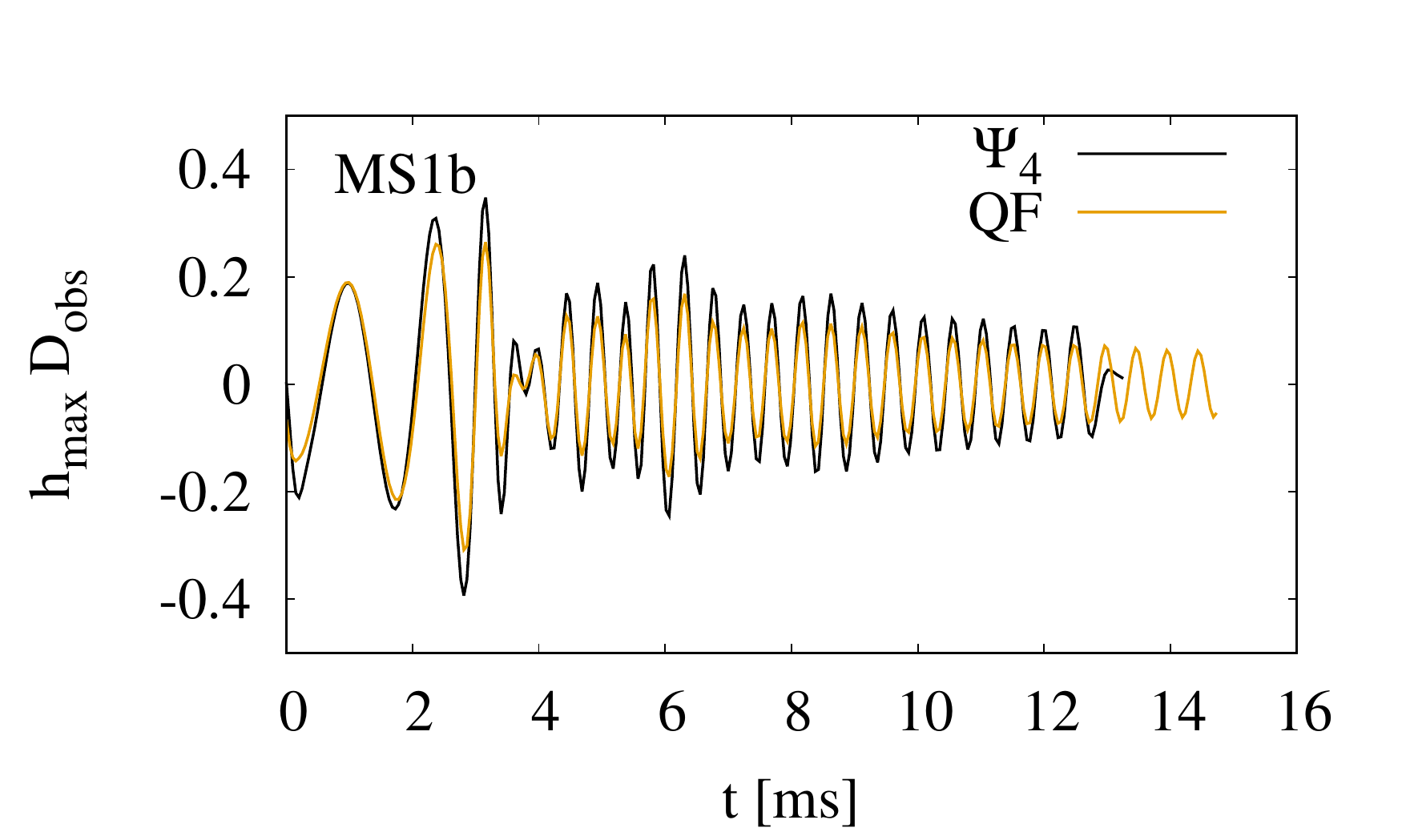}\hspace*{-0.5cm}\includegraphics[width=0.54\columnwidth]{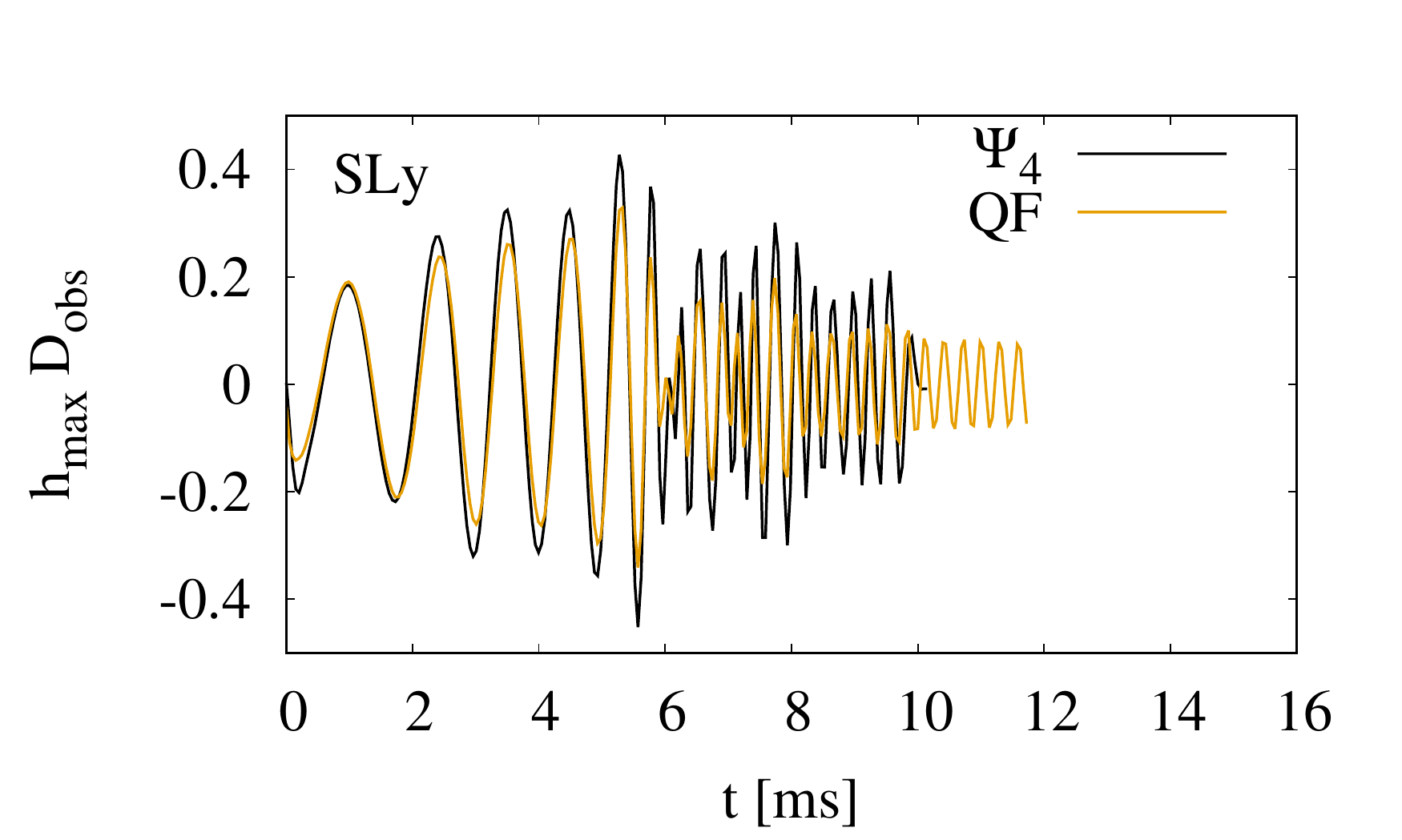}}
\caption{The extracted $\Psi_4$ gravitational wave signals for the four different
  equations of state considered here: MPA1 (top left, {\tt MPA1\_2mio}), APR3
   (top right, {\tt APR3\_2mio}), MS1b (bottom left, {\tt MS1b\_2mio}) and SLy 
  (bottom right, {\tt SLy\_2mio}). Both the 
  spacetime-extracted $\Psi_4$ (black) and quadrupole (QF; orange)
  waveforms are shown. The $\Psi_4$ waveform has been aligned with the
  quadrupole waveform by shifting it in time and hence appear to be shorter.
  The  $\Psi_4$ waveforms appear to go to zero at the end. This
  is purely an artifact of the windowing that has been applied for the time
  integration of $\Psi_4$.} \label{fig:wave_comp}
\end{figure}
\subsubsection{Gravitational wave emission}
We have extracted gravitational waves from our simulations via the quadrupole
formula (using particle information only) as well as directly from the
spacetime by calculating the Newman-Penrose Weyl scalar $\Psi_4$ (both
methods are described in Appendix A in \cite{diener22a}). After $\Psi_4$ (decomposed
into spin weight -2 spherical harmonics) is extracted at a coordinate radius
of $R=300$, we use built in functionality in {\tt kuibit}  \cite{bozzola21} to
reconstruct the strain by integrating twice in time (performed in the
frequency domain). In order to compare with the waveform extracted with the
quadrupole formula, we evaluate the sum of multipoles at the orientation that
gives the maximal signal and shift the waveform in time in order to align
the peak amplitudes to account for the difference that the $\Psi_4$ waveform
has to propagate from the source to the detector whereas the quadrupole
waveform is extracted at the source. \\
\begin{figure}
\includegraphics[width=0.95\columnwidth]{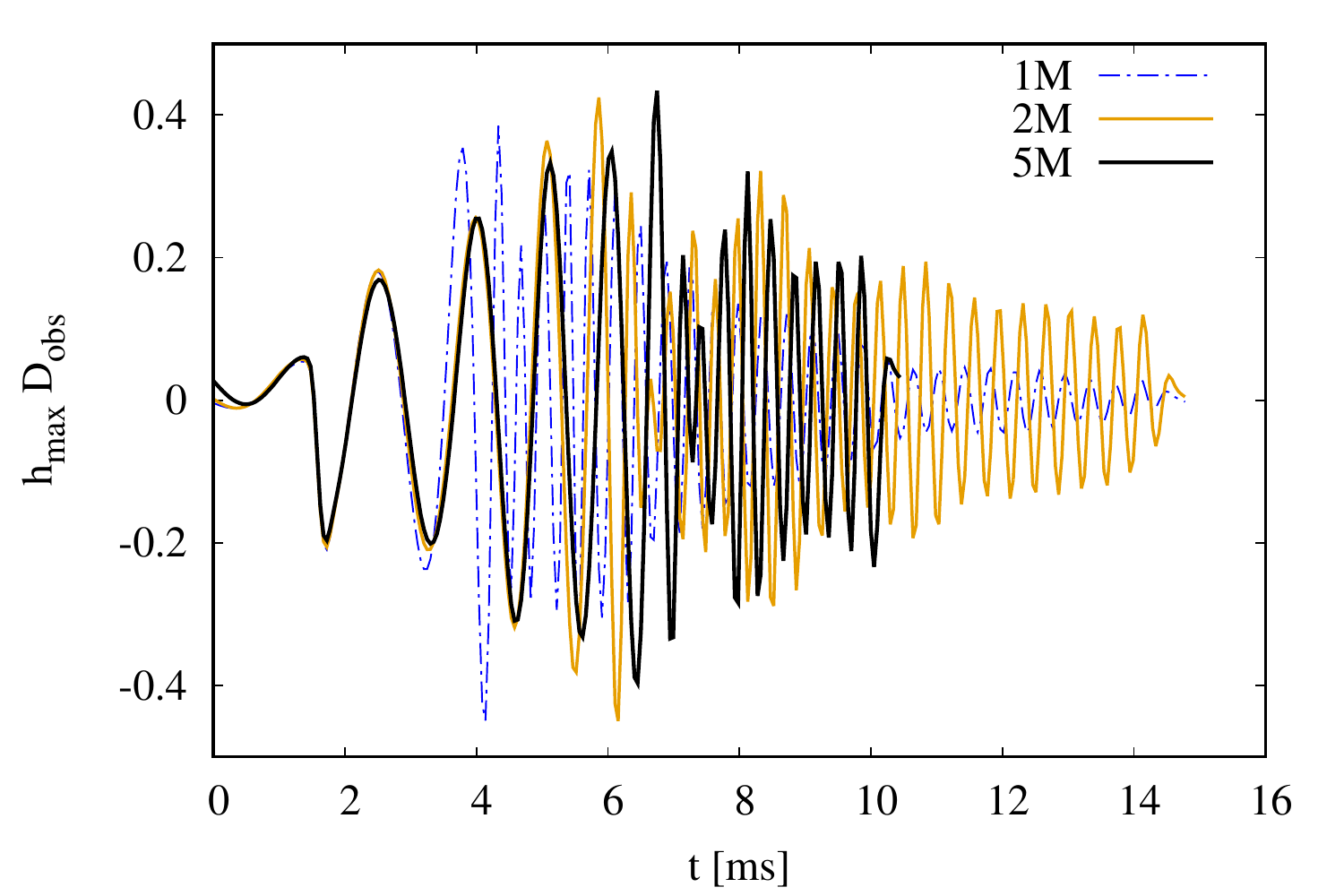}
\caption{The extracted $\Psi_4$ gravitational wave signals for MPA1 for simulations with 1 million 
  (dashed blue, {\tt MPA1\_1mio}), 2 million (solid orange, {\tt MPA1\_2mio}) and 5 million (solid black, {\tt MPA1\_5mio}) 
  particles. Notice that the 5 million particle simulation has not been run
  for quite as long as the lower resolution simulations due to its higher
  computational cost.}
\label{fig:GW_resol_dependence}
\end{figure}
\begin{figure}
\centerline{\includegraphics[width=1.05\columnwidth]{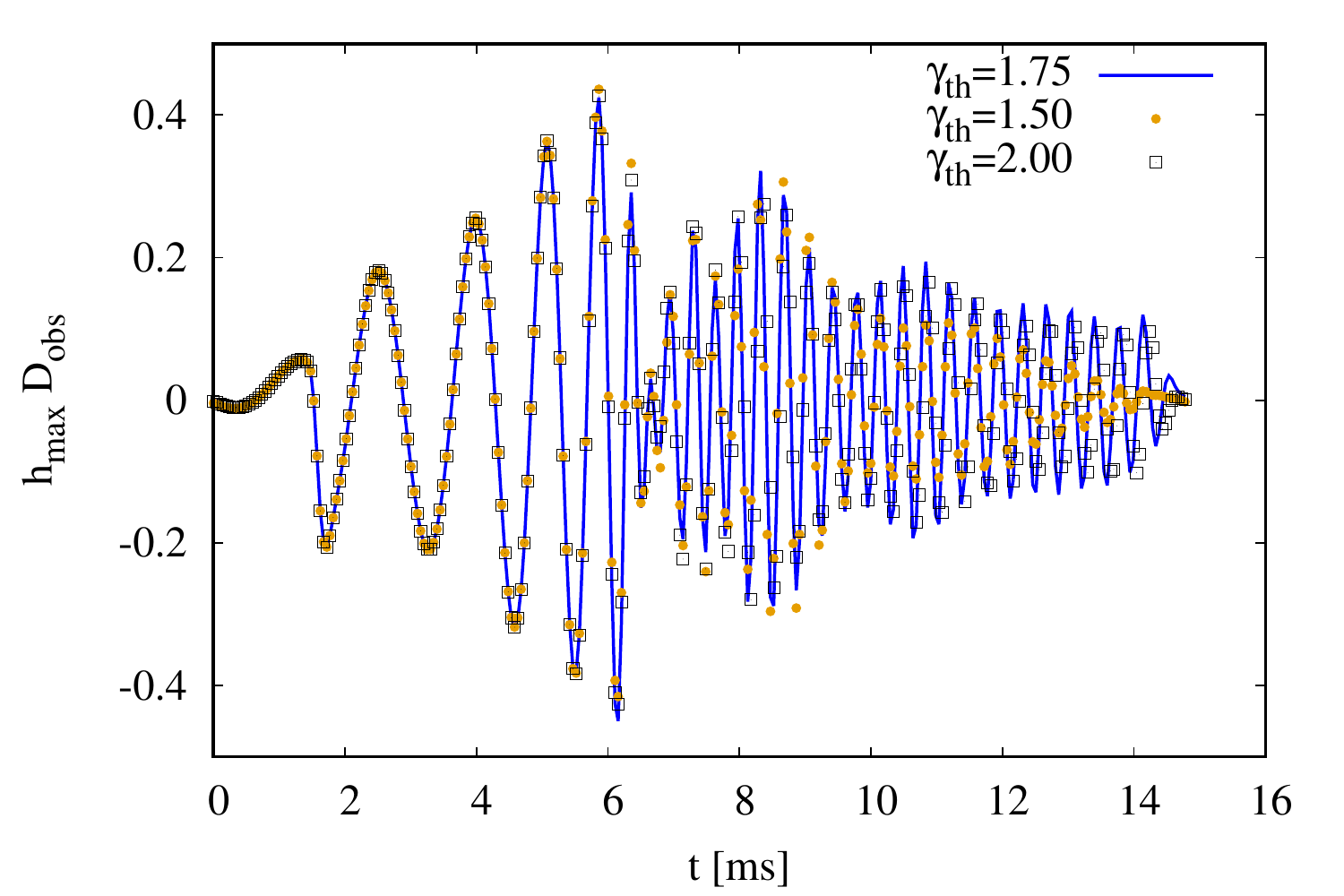}}
\caption{The extracted $\Psi_4$ gravitational wave signals for MPA1 with different
  thermal components $\gamma_{\mathrm{th}}$. Our default, $\gamma_{\mathrm{th}}=1.75$,
  is shown with a solid blue line, $\gamma_{\mathrm{th}}=1.5$ is shown with
  orange filled dots and $\gamma_{\mathrm{th}}=2.0$ is shown with black empty
  squares.
}
\label{fig:GW_Gth_dependence}
\end{figure}
In Fig.~\ref{fig:wave_comp} we show the extracted gravitational wave signals
for the four different equations of state considered here: MPA1 (top left) 
with thermal component $\gamma_{\mathrm{th}}=1.75$, APR3 (top right), 
MS1b (bottom left) and SLy (bottom right) extracted from the simulations with
2 million particles. As can be seen the quadrupole formula does a good job
in tracking the phase of the waves, but typically underestimates (by up to
60\%) the amplitude of the waveform. As mentioned earlier, the larger tidal
effects with harder equations of state lead to a faster inspiral and an
earlier merger. This is clearly also an effect that is visible in the 
waveforms, where the SLy (the softest EOS) waveform has a significantly 
longer inspiral part and the MS1b (the hardest EOS) waveform has the
shortest inspiral part. In the post-merger part of the waveform it is also
clear that the amplitude of the wave is larger for softer equations of 
state. \\
In Fig.~\ref{fig:GW_resol_dependence} we show the extracted gravitational wave
signal for MPA1 (with $\gamma_{\mathrm{th}}=1.75$) at 3 different resolutions
(1, 2 and 5 million particles). As can be seen, the effect of low resolution
(and higher dissipation) is to drive the neutron stars to faster mergers. In
addition, there is also a rapid decay in gravitational wave amplitude in the
post-merger phase at low resolution. It is clear that we have not quite reached 
convergence in these simulations even at the highest resolution, but it is 
encouraging to see that the differences between the two lower resolution
simulations are much larger than the differences between the two higher
resolutions.\\
We have checked that the recent changes to the code do not alter the 
convergence properties in any significant way compared to our recent study \cite{diener22a}. 
We find that the constraint violation behaviour is virtually identical to what we found there, 
and we therefore refer the interested reader to Fig.~12 of that study. From
that plot the conclusion is that the convergence of the Hamiltonian constraint 
is consistent with 2nd order.\\
In Fig.~\ref{fig:GW_Gth_dependence} we explore the effect of the different thermal
components on the extracted gravitational waveform for the 2 million particle
simulations with the MPA1 EOS and $\gamma_{\mathrm{th}}=1.75$, $1.50$ and 
$2.00$. As expected, the thermal component does not make a difference during
the inspiral as the waveforms are practically indistinguishable before the
merger. Interestingly, after the merger the softer $\gamma_{\mathrm{th}}=1.5$
simulation shows a significantly faster decay of the gravitational wave signal
than the harder ones.\\
In Fig.~\ref{fig:spectra} we plot the amplitude of the Fourier transform
of the dominant $\ell=2, m=2$ mode of the strain as computed from $\Psi_4$
for the four different EOS considered here in the left panel, whereas the
right panel shows the spectra for the MPA1 EOS with different thermal
components. In both plots, the solid lines show the Fourier transforms
of the full waveforms, while the dashed lines show the Fourier transforms of
the post-merger part of the waveforms only. Consistent with findings reported
in the literature \cite{bauswein15a,bernuzzi15a,dietrich15b,bauswein15b,clark15,ciolfi17,maione17,sarin20,sun22}, 
the spectra at low ($<1$ kHz) frequencies
are dominated by the inspiral with increasing amplitude up to a maximum at
the frequency of the binary at the merger. This feature is absent in the
post-merger spectra. For all the different EOS, the spectra show a dominant
peak at higher frequency and the location of that peak is strongly dependent
on the EOS. The softer the EOS, the higher the frequency with values ranging
from about $2.1$ kHZ for MS1b to $3.4$ kHz for SLy. For the softest EOS (SLy)
there is clear evidence of sub-dominant peaks (also reported in the 
literature) at both lower and higher frequency separated from the dominant
peak by about $\pm \Delta f\approx 1$ kHz.\\
In Table III in \cite{takami14} several peak frequencies (in the authors' convention
$f_{\mathrm{max}}$, the instantaneous frequency at maximal GW amplitude, $f_1$, 
the first sub-dominant peak after merger, and $f_2$, the dominant peak after
merger) are reported for a number of different equations of state and masses of
the binary systems. In particular their SLy-q10-M1300 case is very similar to one
of our systems (but they use a $\gamma_{\rm th}=2.00$ rather our value of 1.75). 
Unfortunately, this is our softest equation of state that
has the highest resolution requirements and hence is our least trusted 
case. Nevertheless we find frequencies that are in reasonable agreement.
For $f_1$ and $f_2$ we do see only small differences between
the values extracted from our low and medium resolution runs. We find $f_1=2.55$
kHz and $f_2=3.36$ kHz for 1 million particles and $f_1=2.52$ kHZ and 
$f_2=3.39$ kHz for 2 million particles. These are slightly larger than the
$f_1=2.13$ kHz and $f_2=3.23$ kHz values reported in \cite{takami14}. On the
other hand, we do find substantial differences in the value for $f_{\mathrm{max}}$
at the two different values with $f_{\mathrm{max}}= 2.14$ kHZ for 1 million
particles and $f_{\mathrm{max}}= 1.76$ kHz for 2 million particles. This is
to be compared with a value of $f_{\mathrm{max}}= 1.95$ kHZ in \cite{takami14}.
We suspect that these values are quite sensitive to the details of the 
inspiral and at current resolutions we still see significant differences for
the SLy equation of state. Note that we have not used the fitting procedure
described in \cite{takami14} to extract the $f_1$ and $f_2$ frequency peaks, but
have simply found the peaks numerically from the raw power spectral density.\\
Turning now to a comparison of the spectra for MPA1 with different thermal
components (right plot), it is clear that there is very little dependence
of the location of the dominant peak with $\gamma_{\mathrm{th}}$. However,
consistent with the rapid decay of the post-merger waveform for
$\gamma_{\mathrm{th}}=1.5$, in Fig.~\ref{fig:GW_Gth_dependence} we do see a smaller
amplitude of the peak for that case.
\begin{figure}
\centerline{\includegraphics[width=0.55\columnwidth]{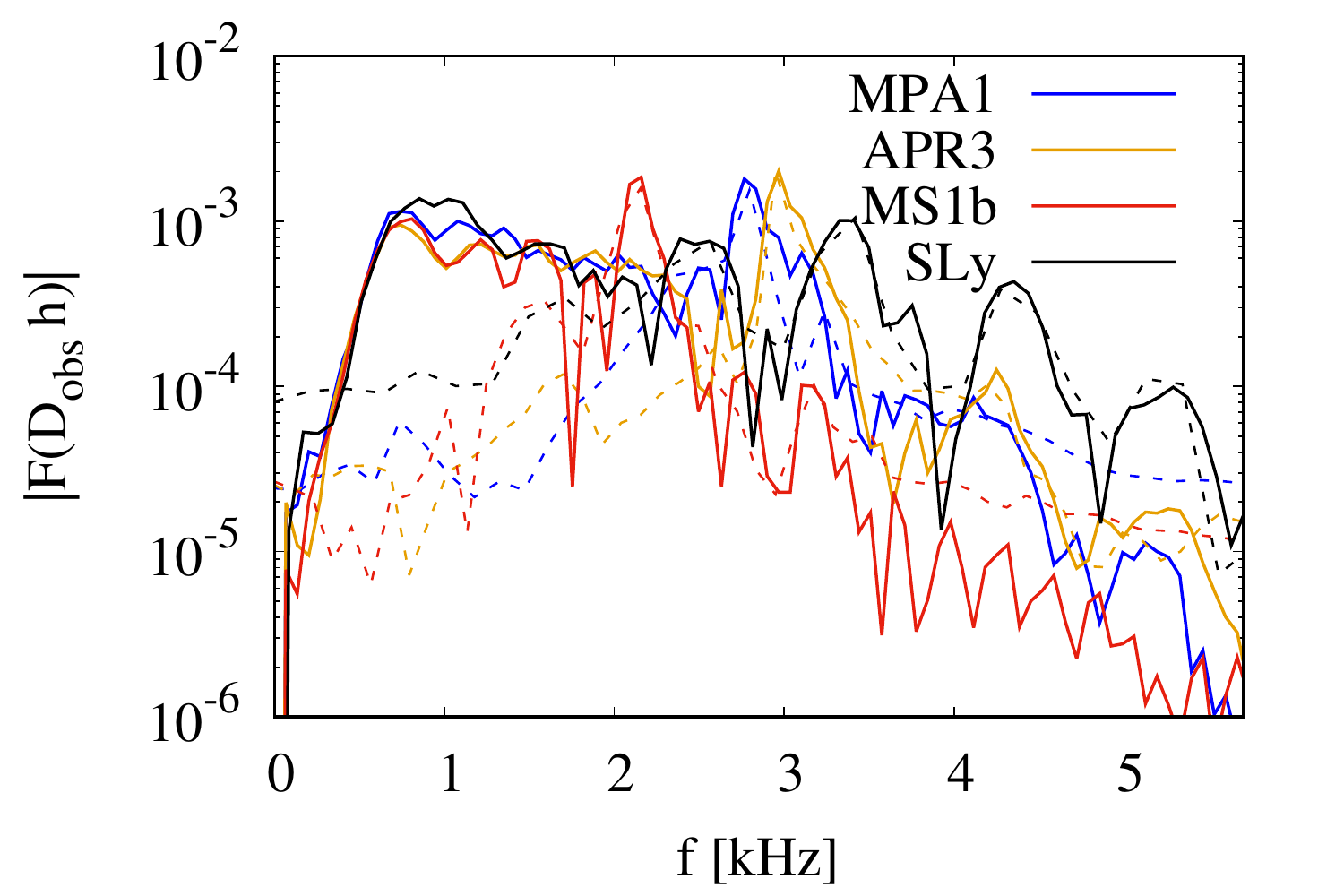} \hspace*{-0.5cm}
\includegraphics[width=0.55\columnwidth]{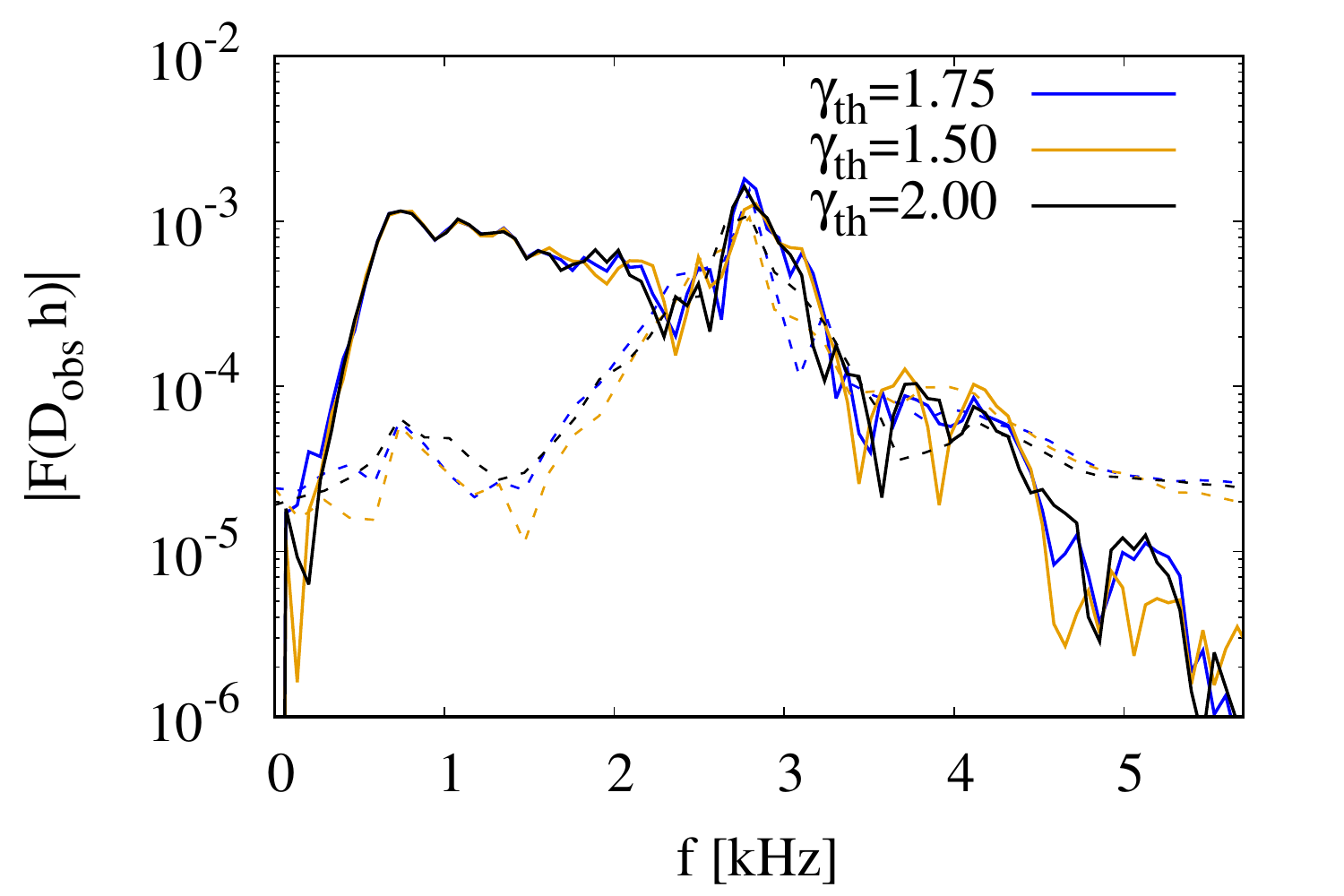}}
\caption{Fourier spectra of the dominant $\ell=2, m=2$ mode of the strain for
MPA1 ($\gamma_{\mathrm{th}}=1.75$), APR3, MS1b and SLy (left panel) and for
MPA1 ($\gamma_{\mathrm{th}}=1.5, 1.75$ and $2.0$) (right panel). In both cases
the solid lines shows the spectra of the full waveforms, while the dashed lines
show the spectra of the post-merger waveforms only.}
\label{fig:spectra}
\end{figure}
\subsubsection{Ejecta}
\begin{figure}[H]
\centerline{\includegraphics[width=10cm]{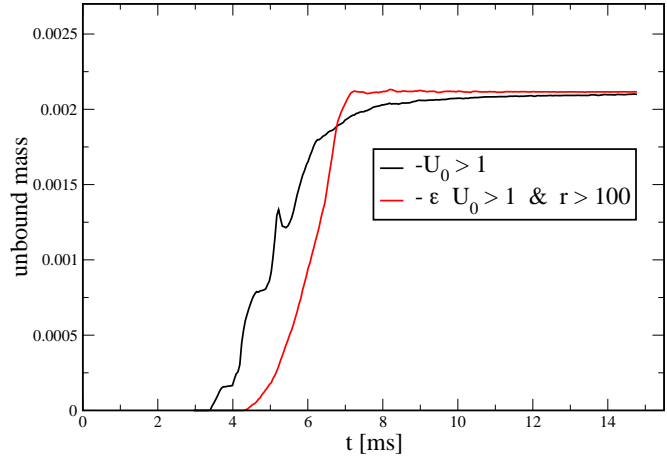}}
\caption{Comparison of the "geodesic" and the (slightly modified) "Bernoulli" criterion to identify unbound mass.}
\label{fig:geodesic_vs_Bernoulli}
\end{figure}

\begin{figure}[H]
\centerline{\includegraphics[width=13cm]{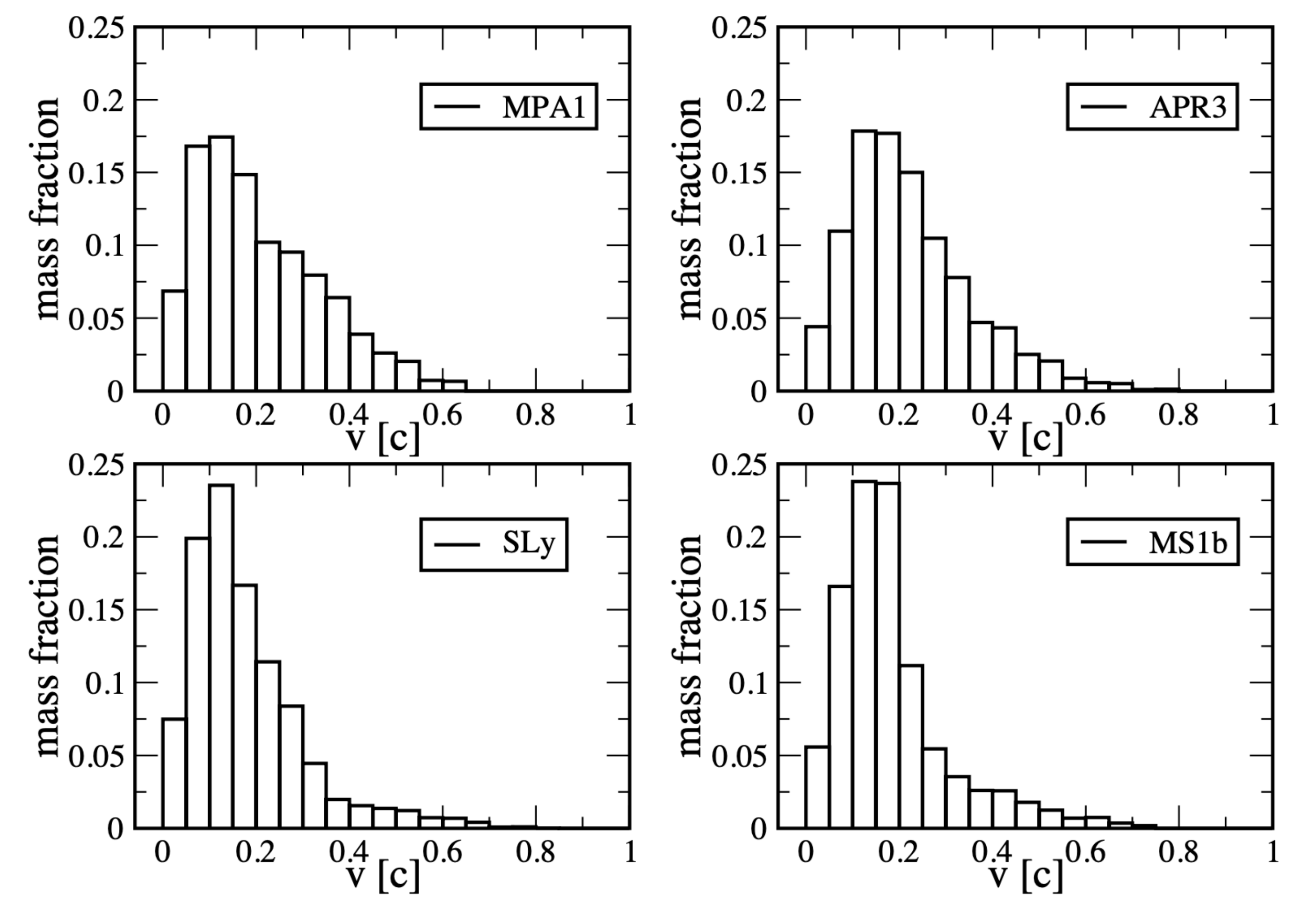}}
\caption{Fraction of the ejected mass binned according to their velocities at infinity, for each 
equation of state the simulation results with 2 million particles is shown.}
\label{fig:ejecta_velocities}
\end{figure}

\begin{figure}[H]
\centerline{\includegraphics[width=8cm]{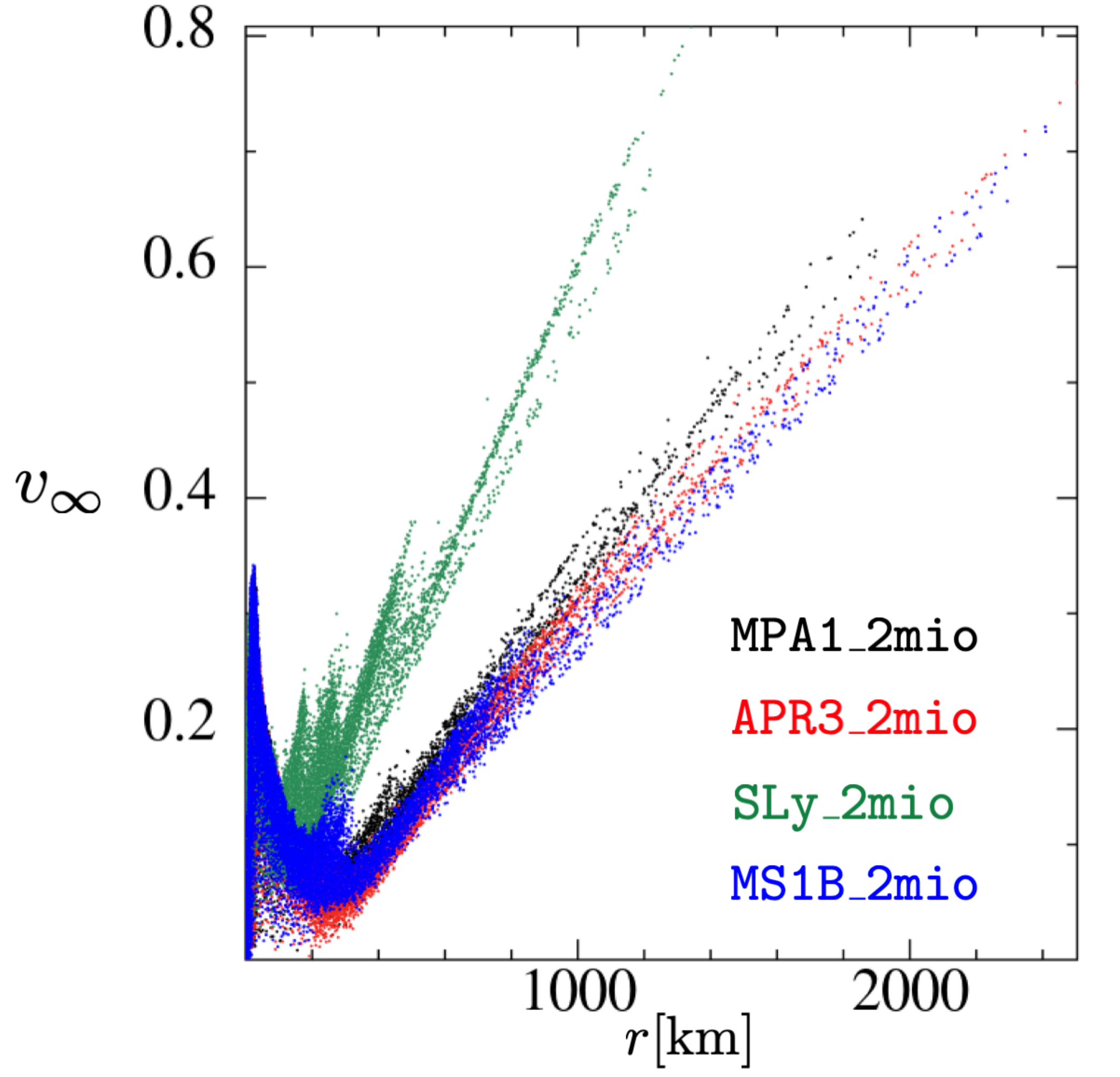}}
\caption{Asymptotic velocities as a function of radius of the 2 million particle simulations.}
\label{fig:ejecta_vel_r}
\end{figure}

\begin{figure}[H]
\centerline{
\includegraphics[width=6cm]{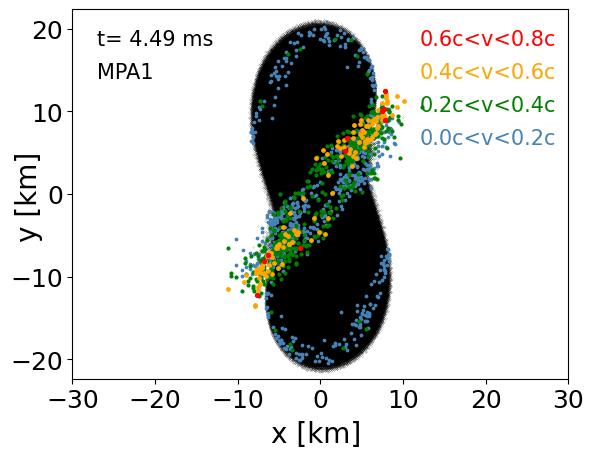}
\includegraphics[width=6cm]{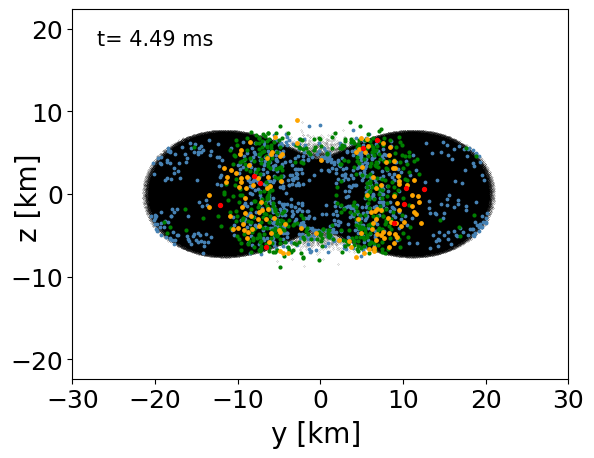}

}
\centerline{\hspace*{-0.4cm}
\includegraphics[width=6cm]{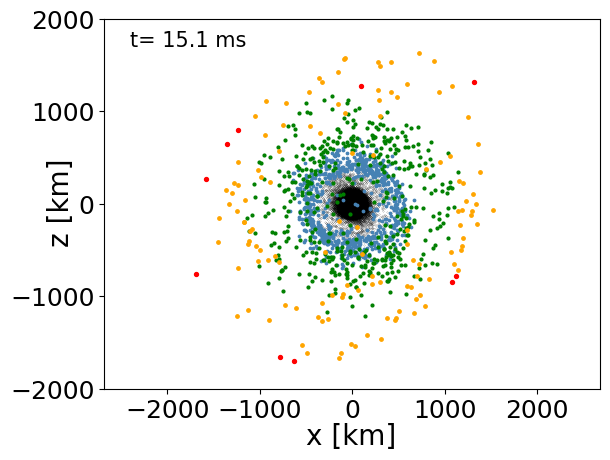}
\includegraphics[width=6cm]{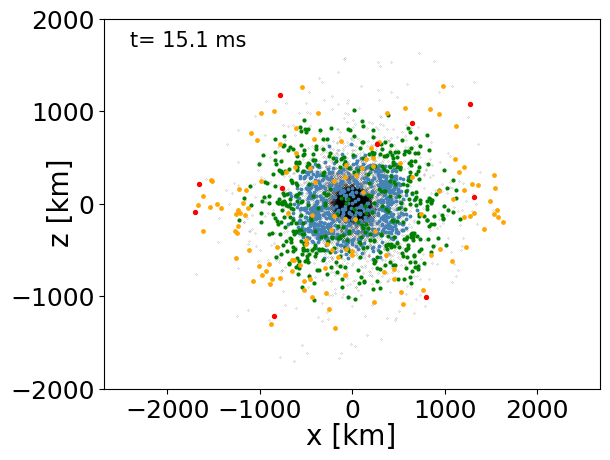}
}
\caption{ Projections of the particle positions for the MPA1 EOS case (run {\tt MPA1\_2mio}). All particles are shown in black, ejecta with $v_\infty < 0.2 c$ in blue,  ejecta with $0.2 c \le v_\infty < 0.4 c$ in green, ejecta with $0.4 c \le v_\infty < 0.6 c$ in orange and ejecta with $0.6 c \le v_\infty < 0.8 c$ in red.}
\label{fig:ejecta_pos_vel_MPA1}
\end{figure}

\begin{figure}[H]
\centerline{
\includegraphics[width=6.5cm]{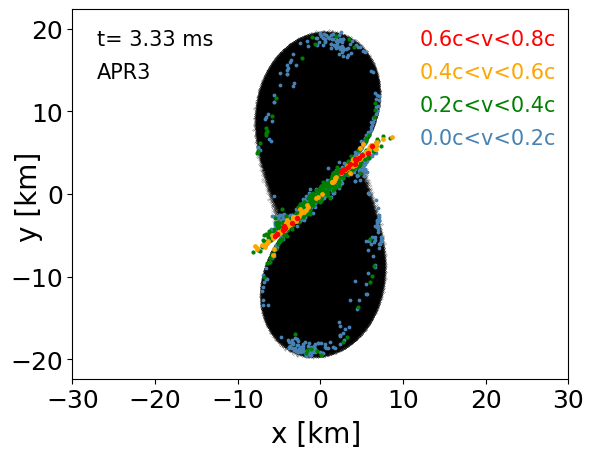}
\includegraphics[width=6.5cm]{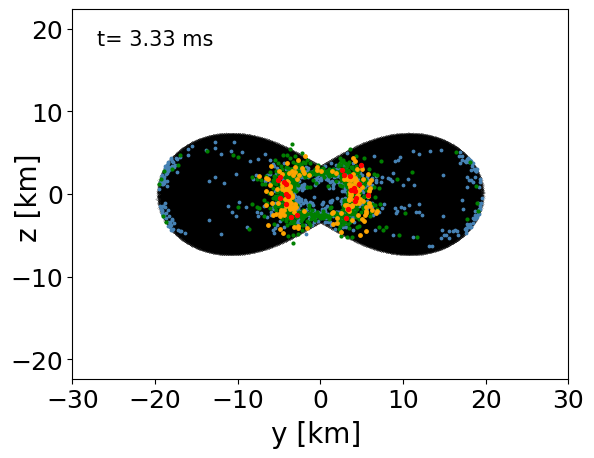}

}
\centerline{\hspace*{-0.5cm}
\includegraphics[width=6.5cm]{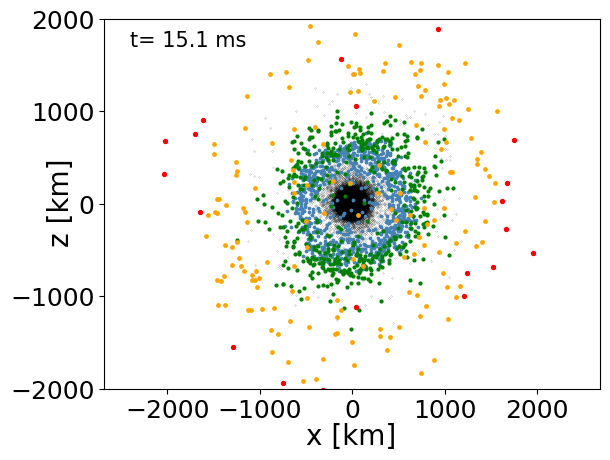}
\includegraphics[width=6.5cm]{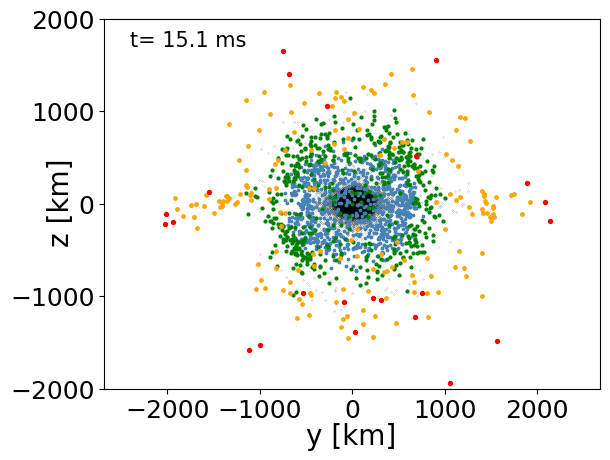}
}
\caption{Projections of the particle positions for the APR3 EOS case (run {\tt APR3\_2mio}). All particles are shown in black, ejecta with $v_\infty < 0.2 c$ in blue,  ejecta with $0.2 c \le v_\infty < 0.4 c$ in green, ejecta with $0.4 c \le v_\infty < 0.6 c$ in orange and ejecta with $0.6 c \le v_\infty < 0.8 c$ in red.}
\label{fig:pos_vel_APR3}
\end{figure}

\begin{figure}[H]
\centerline{
\includegraphics[width=6.5cm]{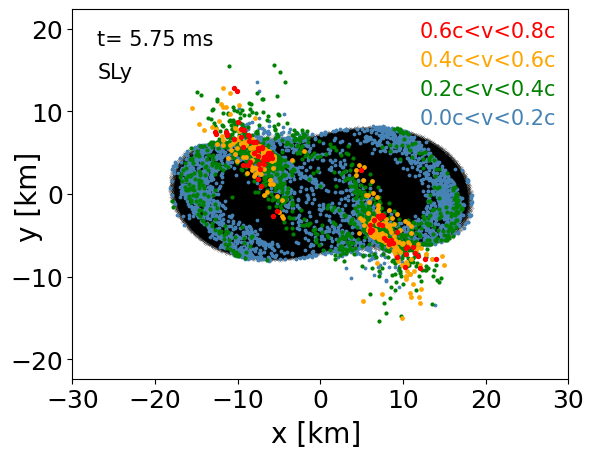}
\includegraphics[width=6.5cm]{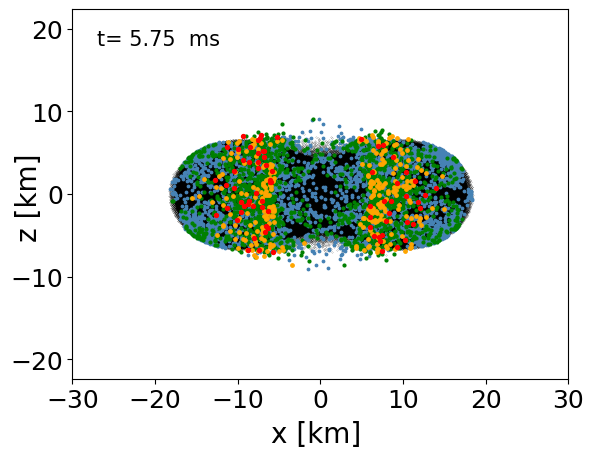}

}
\centerline{\hspace*{-0.5cm}
\includegraphics[width=6.5cm]{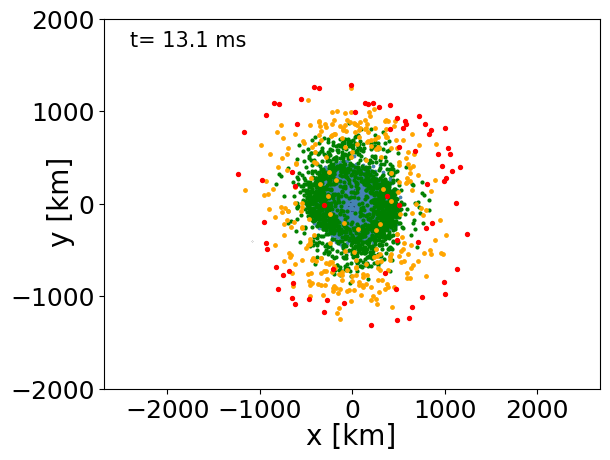}
\includegraphics[width=6.5cm]{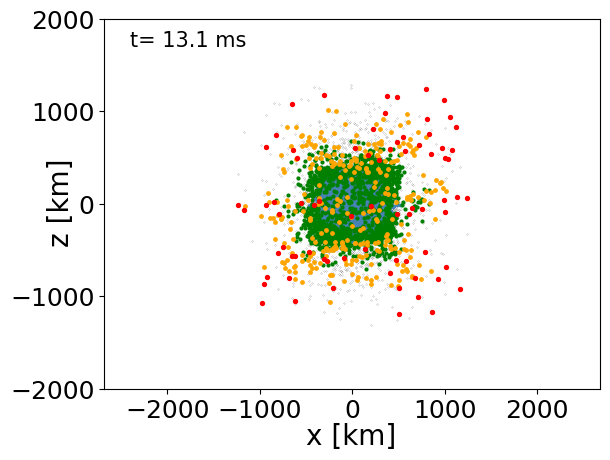}
}
\caption{ Projections of the particle positions for the SLy EOS case (run {\tt SLy\_2mio}). All particles are shown in black, ejecta with $v_\infty < 0.2 c$ in blue,  ejecta with $0.2 c \le v_\infty < 0.4 c$ in green, ejecta with $0.4 c \le v_\infty < 0.6 c$ in orange and ejecta with $0.6 c \le v_\infty < 0.8 c$ in red.}
\label{fig:pos_vel_SLy}
\end{figure}

\begin{figure}[H]
\centerline{
\includegraphics[width=6.5cm]{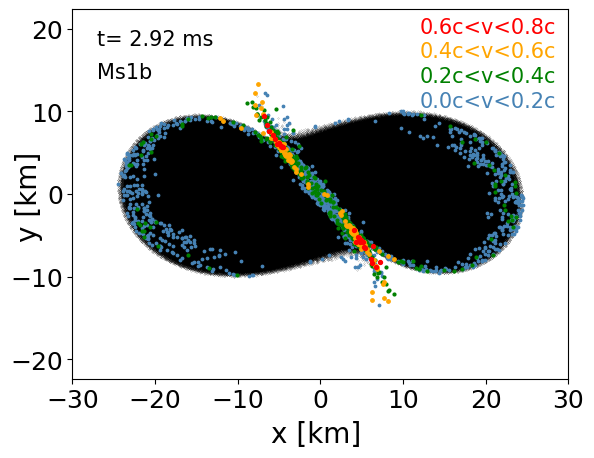}
\includegraphics[width=6.5cm]{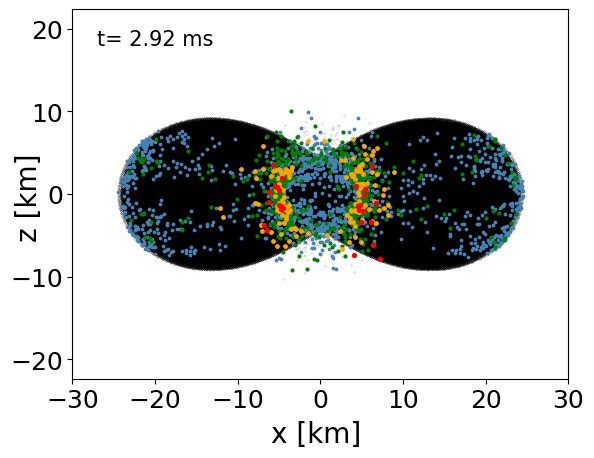}

}
\centerline{\hspace*{-0.5cm}
\includegraphics[width=6.5cm]{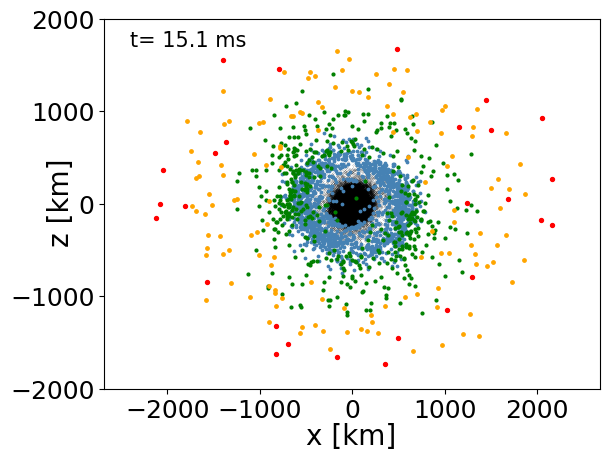}
\includegraphics[width=6.5cm]{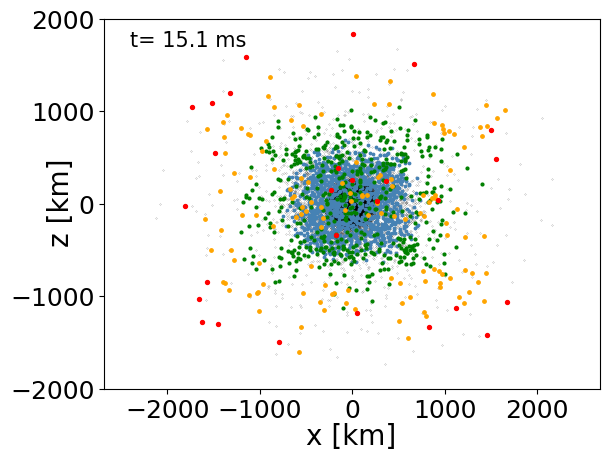}
}
\caption{ Projections of the particle positions for the Ms1b EOS case (run {\tt Ms1b\_2mio}). All particles are shown in black, ejecta with $v_\infty < 0.2 c$ in blue,  ejecta with $0.2 c \le v_\infty < 0.4 c$ in green, ejecta with $0.4 c \le v_\infty < 0.6 c$ in orange and ejecta with $0.6 c \le v_\infty < 0.8 c$ in red.}
\label{fig:pos_vel_Ms1b}
\end{figure}

The ejection of  neutron-rich matter  is arguably one of
most important aspects of a neutron star merger \cite{rosswog99,freiburghaus99b,bauswein13a,hotokezaka13a,radice18a}, 
it is responsible for enriching the
cosmos with heavy elements and for all of the electromagnetic emission. To identify
ejecta, we apply two basic criteria, the "geodesic criterion" and the "Bernoulli criterion",
both of which can be augmented by additional conditions (e.g. an outward pointing radial
velocity). The geodesic criterion assumes that a fluid element is moving along a geodesic
in a time-independent, asymptotically flat spacetime, e.g. \cite{hotokezaka13a,foucart21b}.
Under these assumptions, $-U_0$ corresponds to the Lorentz factor of the fluid element
at spatial infinity and therefore a fluid element with
\be
-U_0 > 1
\ee
is considered as unbound since it still has a finite velocity. In practice however, this criterion ignores potential further acceleration
due to internal energy degrees of freedom and we therefore consider the results from the geodesic 
criterion as a lower limit.\\
The internal degrees of freedom are included in the Bernoulli criterion, see e.g. \cite{rezzolla13a}, which multiplies 
the left hand side of the above criterion with the specific enthalpy
\be
- \mathcal{E} U_0 > 1.
\ee
To avoid falsely identifying hot matter near the centre as unbound,
we apply the Bernoulli criterion only to matter outside of a coordinate radius
of 100 ($\approx 150$ km). We find  very good agreement between both these criteria, with the Bernoulli criterion
identifying only a slightly larger amount unbound mass. This is illustrated in Fig.~\ref{fig:geodesic_vs_Bernoulli}  
for run  {\tt APR3\_2mio}. In the following, we only use the above described version of the Bernoulli 
criterion\footnote{The ejecta will undergo r-process nucleosynthesis
and, at later times, radioactive decay. This additional energy input can unbind otherwise nearly unbound matter
and may therefore further enhance the amount of unbound mass.}.
The ejecta mass and average velocities for our runs are summarized in Tab.~\ref{tab:ejecta}. Consistent with other studies (e.g.
\cite{hotokezaka13a,bauswein13a,radice18a}) we find that the studied equal mass systems eject only a few times $10^{-3}$ \Msun
dynamically, i.e. the dynamical ejecta channel falls short by about an order of magnitude in reproducing the ejecta amounts that
have been inferred from GW170817 \cite{kasen17,cowperthwaite17,evans17,villar17,kasliwal17,tanvir17,rosswog18a}. 
While part of these large inferred masses may be explained by the fact that the interpretations of the observations largely 
neglect the 3D ejecta geometry and assume sphericity instead \cite{korobkin21}, it seems obvious that complementary 
ejecta channels are needed.\\
Ordering our equations of state in terms of stiffness from soft to hard (either based on $M_{\rm TOV}^{\rm max}$ or $\Lambda_{1.4}$,
see Sec.~\ref{sec:EOS}),
SLy, APR3, MPA1, MS1b, we see that they eject more mass the softer they are, consistent with shocks (that emerge easier
in soft EOSs with lower sound speed) being the major ejection mechanism and confirming earlier results \cite{bauswein13a,hotokezaka13a}. The thermal exponent $\gamma_{\rm th}$ has a 
noticeable impact on the ejecta masses with the $\gamma_{\rm th}=1.5$ case ejecting nearly twice as much as our standard case.
Reference \cite{bauswein13a} actually finds that ejecta from tabulated EOSs are best approximated by $\gamma_{\rm th}=1.5$, therefore
the masses in Tab.~\ref{tab:ejecta} may be considered as lower limits.
Given the simplicity of how the thermal contribution is modelled, and its impact on both the GW signal, see Fig.~\ref{fig:GW_Gth_dependence}, and the ejecta this should also be a warning sign that a more sophisticated modelling
of the thermal EOS is needed.\\
The ejecta velocities in GW170817 have provided additional constraints on the physical origin of the ejecta. Keeping in mind
that, within the assumptions entering the Bernoulli criterion, the physical interpretation of $- \mathcal{E} U_0$ is that of the
Lorentz factor at infinity, we bin the asymptotic velocities
\be
v_{\infty}= \sqrt{1 - \frac{1}{(\mathcal{E} U_0)^2}}
\ee
 in Fig.~\ref{fig:ejecta_velocities}. The baryon number weighted average velocities at infinity 
\be
\langle v_{\infty} \rangle = \frac{\sum_b \nu_b \; v_{\infty,b}}{\sum_b \nu_b},
\ee
where the index $b$ runs over all unbound particles and $\nu$ is, as before, the baryon number carried by an SPH particle,
is typically around $\sim 0.2$ c, but in each of the cases $\sim 10^{-4}$ \Msun escapes with velocities above 0.5 c, extending
up to $\sim0.7$ c, see Fig.~\ref{fig:ejecta_velocities} and Tab.~\ref{tab:ejecta}. Such high-velocity ejecta have been reported
also by other studies \cite{hotokezaka13a,just15,metzger15a,radice18a}.
While we cannot claim that these small amounts 
of mass are fully converged, Fig.~\ref{fig:ejecta_vel_r} shows that the velocity distribution in all cases smoothly extends to such 
large velocity values. We are therefore confident that this high-velocity ejecta component is not a numerical artefact.
\begin{specialtable}[H] 
\small
\caption{Dynamical ejecta masses and velocities of the simulated binary systems. $m_{>Xc}$ refers to the amount of mass
that has a velocity in excess of $Xc$.
\label{tab:ejecta}}
\begin{tabular}{lccccccc}
\toprule
name & $m_{\rm ej}$ [$10^{-3}$ \msun]    & $\langle v_\infty \rangle [c]$      & $m_{>0.5c}$ [\msun]  & $m_{>0.6c}$ [\msun] & $m_{>0.7c}$ [\msun]\\
\midrule
{\tt MPA1\_1mio}	   & 3.6	 & 0.23  & $1.7 \times10^{-4}$ & $5.2 \times 10^{-5}$ & $6.3 \times 10^{-6}$\\
{\tt MPA1\_2mio}          & 1.6      & 0.21 & $5.6 \times10^{-5}$ & $1.1 \times 10^{-5}$ & 0\\
{\tt MPA1\_5mio}          & 1.2      & 0.24 & $1.0 \times10^{-4}$ & $3.5 \times 10^{-5}$ & $3.4 \times 10^{-6}$\\
{\tt MPA1\_2mio\_$\gamma_{\rm th}1.5$}	   & 2.8	& 0.16 & $4.7 \times10^{-5}$ & $3.6 \times 10^{-6}$ & 0 \\
{\tt MPA1\_2mio\_$\gamma_{\rm th}2.0$}	   & 1.8     & 0.22 & $6.9 \times10^{-5}$ & $1.7 \times 10^{-5}$ & $1.0\times 10^{-6}$ \\
{\tt APR3\_1mio}	   & 	9.7 & 		0.27         &  $7.1 \times10^{-4}$ & $2.7 \times 10^{-4}$ & $8.5 \times 10^{-5}$\\
{\tt APR3\_2mio}	   & 	2.1 & 		0.22	        &  $9.0 \times10^{-5}$ & $2.8 \times 10^{-5}$ & $4.8 \times 10^{-6}$ \\
{\tt APR3\_5mio}	   & 	1.9 & 		0.21	        &  $8.3\times10^{-5}$ & $2.0 \times 10^{-5}$ & $8.4 \times 10^{-7}$   \\
{\tt SLy\_1mio}  	   & $5.4$	&  		0.21	        &  $2.7\times10^{-4}$ & $8.4 \times 10^{-5}$ & $2.1 \times 10^{-6}$  \\
{\tt SLy\_2mio}             & $>6.7$& 0.18                      &  $2.2\times10^{-4}$ & $8.8 \times 10^{-5}$ & $1.4 \times 10^{-5}$\\
{\tt MS1b\_1mio}	   & 	2.9 & 0.16                         &  $1.1\times10^{-4}$ & $2.8 \times 10^{-5}$ & 0\\
{\tt MS1b\_2mio}	   & 	2.7 & 0.18                         &  $8.8\times10^{-5}$ & $3.5 \times 10^{-5}$ & $5.0 \times 10^{-6}$\\
\bottomrule
\end{tabular}
\end{specialtable}
To identify how the high-velocity ejecta are launched, we sort the ejecta 
in the last data output in four groups: $\mathcal{G}_1$: $ v_{\infty} < 0.2 c$, 
$\mathcal{G}_2$: $ 0.2 c \le v_{\infty} < 0.4 c$, $\mathcal{G}_3$: $ 0.4 c \le v_{\infty} < 0.6 c$
and $\mathcal{G}_4$: $ 0.6 c \le v_{\infty} < 0.8 c$. In Figs.~\ref{fig:ejecta_pos_vel_MPA1} 
to \ref{fig:pos_vel_Ms1b} we plot these groups
 of particles at the approximate times of the merger and for the final data 
dumps of our 2 million particle runs. The highest velocity particles, $\mathcal{G}_3$ (orange) and 
$\mathcal{G}_4$ (red) emerge  from the
shock-heated interface between the two neutron stars. While the high-velocity ejecta are still not 
well-resolved, we note that our simulations here have an order of magnitude more 
particles than the approximate GR simulations in which these fast ejecta were originally
identified \cite{metzger15a}. This high-velocity component could have important observational 
consequences: it may produce an early blue/UV transient on
a time scale of several minutes to an hour preceeding the main kilonova event \cite{metzger15a}
and, at late times, it may be responsible for synchrotron emission 
\cite{mooley17,hotokezaka18a,hajela22}.\\
Another interesting result in a multi-messenger context
is that the ejecta distribution only shows moderate deviations from spherical 
symmetry, so that the resulting electromagnetic emission could be reasonably modelled
with simple approaches. This result, however, may be specific for our equal mass binaries and
for the currently implemented physics and it is possible that different equations of
state, neutrinos and/or magnetic fields could modify this.


\section{Summary}
\label{sec:summary}
We have presented here a further methodological refinement of  our Lagrangian Numerical Relativity
code \SpB \cite{rosswog21a,diener22a}. The new methodological elements in \SpB include a new 
way to steer where artificial dissipation is applied, see Sec.~\ref{sec:AV}.
We use both a compression that increases in time (as suggested in \cite{cullen10}) and a numerical noise indicator
following \cite{rosswog15b} to determine how much dissipation is used. We have also further refined our MOOD 
algorithm that we use in our "particle-to-mesh" step, see Sec.~\ref{sec:PM_coupling}. As a step towards more realistic 
neutron star merger simulations, we have implemented piecewise polytropic equations of state that approximate the 
cold nuclear matter equations of state MPA1, APR3, SLy and MS1b and we have augmented these with a  thermal 
contribution, see Sec.~\ref{sec:EOS} and Appendix~\ref{sec:recovery} for details.\\
In this first \SpB study using piecewise polytropic EOSs, we have restricted ourselves to neutron star binary systems 
with 2 $\times$ 1.3 \Msun and we have explored how the results depend on both resolution and the choice of the thermal
polytropic exponent $\gamma_{\rm th}$. None of the explored cases seems to be prone to a BH collapse, at least not 
on the simulated time scale of $\sim15$ ms. But the SLy EOS cases undergo particularly deep pulsations during which
they shed mass into the surrounding torus and  they also eject more mass than the stiffer equations of state (MPA1,APR3, 
MS1b). When our simulations end, the remnant has not yet settled into a stationary state and the torus mass
is still increasing.  All of the torus masses are large enough to power short GRBs and, if indeed tori unbind several 10\% 
of their mass on secular time scales  \cite{metzger08a,beloborodov08,siegel17a,siegel18,miller19,fernandez19}, then 
in all cases their ejecta amount to a few percent of a solar mass.\\
For all our runs we extract the gravitational waves, both directly from the particles via the quadrupole approximation
and from the spacetime by means of the Newman-Penrose Weyl scalar $\Psi_4$. Overall, we find rather good agreement,
the waves phases are practically perfectly tracked in the quadrupole approximation, but in the post-merger phase the amplitudes
can be underestimated by several 10\%. As expected, the softest EOS leads to the longest inspiral wave train and
also to larger post-merger amplitudes. We further explore the impact of the thermal adiabatic exponent $\gamma_{\rm th}$
on the gravitational wave and spectrum.\\
Consistent with earlier studies, we find that these equal mass systems eject only a few $10^{-3}$ \Msun dynamically 
and the ejection is driven by shocks.  Based on quasi-Newtonian \cite{rosswog99,rosswog00,korobkin12a,rosswog13b}, conformal
flatness approximation \cite{bauswein13a} and full-GR simulations \cite{hotokezaka13a,radice18a}, however, we expect that
asymmetric systems with mass ratio $q \ne 1$ eject substantially more matter and in particular have a larger contribution
from tidal ejecta. Overall, the small amount of ejecta underlines the need for additional ejection channels such 
as torus unbinding or neutrino-driven winds in order to  reach the ejecta masses estimated for GW170817. The matter being
predominantly ejected via shocks probably means that its electron fraction is increased with respect to the cold, $\beta$-equilibrium
values inside the original neutron stars ($\sim 0.05$, see e.g. Fig. 21 in \cite{farouqi22}). Again consistent with earlier studies, we
find that the softer equation of state cases eject more mass and also reducing the exponent $\gamma_{\rm th}$
seems to enhance mass ejection. \\
Interestingly, we find in all cases that $\sim 10^{-4}$ \Msun are escaping at velocities exceeding 0.5c and this high-velocity 
part of the ejecta originates from the interface between the two neutron stars during merger. While we cannot claim
that the properties of this matter are well converged, we see this fast component in all the simulations and their velocity
distribution, see Fig.~\ref{fig:ejecta_vel_r}, extends smoothly to large velocities, so that we have  confidence in the physical presence of
these high-velocity ejecta. This neutron-rich matter expands sufficiently fast for most neutrons to avoid capture and the $\beta$-decay
of these free neutrons has been discussed as a source of early, blue "precursor" emission before the main kilonova \cite{metzger15a}.
To "bracket" the kilonova emission, such a high-velocity ejecta component has also been suggested to be responsible for a X-ray emission
excess observed three years after GW170817 \cite{hajela22}.\\
While the piecewise polytropic equations of state  are an important improvement over our previous merger simulations 
with \SpB \cite{diener22a}, they are still only a far cry from realistic microphysics. This topic is a major target for our future work.



\vspace{6pt} 




\funding{
SR has been supported by the Swedish Research Council (VR) under grant number 2020-05044, by 
the Swedish National Space Board under grant number  Dnr. 107/16,  by the 
research environment grant ``Gravitational Radiation and Electromagnetic Astrophysical
Transients (GREAT)'' funded by the Swedish Research Council (VR) under Dnr 2016-06012, 
by which also FT is supported, and by the Knut and Alice Wallenberg Foundation under grant Dnr. KAW 2019.0112. }

\acknowledgments{We thank E. Gourgoulhon, R. Haas and J. Novak for useful clarifications concerning \Lo and S.V. Chaurasia 
for sharing his insights into the BAM code. 
We gratefully  acknowledge inspiring interactions via the COST Action CA16104 
``Gravitational waves, black holes and fundamental physics'' (GWverse) and  COST Action CA16214
``The multi-messenger physics and astrophysics of neutron stars'' (PHAROS).
PD would like to thank the Astronomy Department at SU and the Oscar Klein Centre for their hospitality during
numerous visits in the course of the development of \spB.
The simulations for this paper were performed on the facilities of the North-German Supercomputing Alliance (HLRN),
on the resources provided by the Swedish National Infrastructure for Computing (SNIC) 
in Link\"oping  partially funded by the Swedish Research Council through grant agreement no. 2016-07213
and on the {\em SUNRISE} HPC facility supported by the Technical Division at the Department of Physics, 
Stockholm University. Special thanks go to Holger Motzkau and Mikica Kocic for their excellent support in
upgrading and maintaining {\em SUNRISE}. Some of the plots in this paper were produced with the software 
SPLASH \cite{price07d}. }

\abbreviations{Abbreviations}{
The following abbreviations are used in this manuscript:\\

\noindent 
\begin{tabular}{@{}ll}
ADM   & Arnowitt, Deser, Misner\\
BH     & black hole\\
BSSN & formulation according to Baumgarte, Shapiro, Shibata, Nakamura\\
EOS    & equation of state\\
GR      & General Relativity\\
GW    & gravitational waves\\
SPH & Smooth Partice Hydrodynamics\\
SPHINCS & Smooth Partice Hydrodynamics in Curved Spacetime\\
\end{tabular}}

\appendixtitles{yes} 
\appendixstart
\appendix
\section{Recovery procedure for piecewise polytropic equations of state}
\label{sec:recovery}
In this work we use piecewise polytropic equations of state.
The part resulting from the cold, nuclear matter pressure $P_{\rm cold}$ is described by several
polytropic pieces as discussed in \cite{read09}, and a thermal part, $P_{\rm th}$, is
added under the assumption that it also follows a polytropic equation of state with some thermal
exponent $\gamma_{\rm th}$, for which we choose a default value of 1.75. The total pressure is then given by
\be
P= P_{\rm cold} + P_{\rm th},
\ee
where the cold part is given by pieces
\be
P_{\rm cold}= K_i \rho^{\gamma_i},
\ee
where the values $\{K_i, \gamma_i\}$ are chosen according to the density (for $\rho_i \le \rho < \rho_{i+1}$).
The thermal pressure is calculated only from the "thermal" (i.e. non-degenerate) part of the internal energy
and has a separate (smaller) polytropic exponent
\be
P_{\rm th}= (\gamma_{\rm th}-1) \rho u_{\rm th}.
\ee
The thermal part of $u$ is found by subtracting the "cold/degenerate" value of the internal energy. This cold value
is
\be
u_{\rm cold}= a_i + \frac{K_i}{\gamma_i - 1} \rho^{\gamma_i-1}
\ee
and the integration constants $a_i$ make the internal energy continuous between different pieces, see Eq.(7) of \cite{read09}.\\
From now onwards, we will again use our conventions and measure energies in units of $m_0 c^2$, $m_0$ being the baryon mass,
so that our pressure is given as $P= (\gamma-1) \, n \, u$.
The general strategy is similar to the purely polytropic case: we express both $n$ and $u$ in terms of $S_i$, $e$, $N$ and the pressure, 
substitute everything into the (here analytically known) equation of state
\bea
f(P) &\equiv& P - (P_{\rm cold} + P_{\rm th})\nonumber\\
       &=& P - \left\{K_i n^{\gamma_i}  + (\gamma_{\rm th}-1) n \left[ u - a_i - \frac{K_i}{\gamma_i - 1} n^{\gamma_i-1} \right] \right\}
       = 0. \label{eq:solve_pwp}
\eea
and solve numerically for the new pressure that is consistent with the current values of $S_i$, $e$ and $N$.
We need
\be
n= n(S_i,e,N,P) \quad {\rm and} \quad u= u(S_i,e,N,P).
\ee
We find that the generalized Lorentz factor $\Theta$ can be expressed as 
\be
\Theta(S_i,e,N,P)= \sqrt{\frac{-g^{00}}{1 + \frac{A}{B^2}}},
\label{eq:theta_pwp}
\ee
where
\be
A \equiv g^{00} g^{jk}S_j S_k - (g^{0j} S_j)^2
\quad {\rm and} \quad
B \equiv g^{0j} S_j - g^{00} \left(\frac{\sqrt{-g}}{N} P + e \right).
\ee
This provides us with the internal energy
\be
u(S_i,e,N,P)= \frac{g^{0j} S_j}{\Theta} - \frac{g^{00}e}{\Theta} - \frac{\sqrt{-g}P}{\Theta N} \left( g^{00} + \Theta^2\right) - 1.
\ee
and, from our earlier definition, we have
\be
n(S_i,e,N,P)= \frac{N}{\sqrt{-g} \Theta}.
\label{eq:n_N}
\ee

\noindent The solution procedure is then the following. 
We first find the new pressure  that fulfills Eq.~(\ref{eq:solve_pwp})
for the new values of $S_i,e,N$. For this root finding we use Ridders' method
\cite{ridders82,press92}. This method is a robust variant of the {\em regula falsi} method 
and does not require any derivates. With the consistent pressure at hand, 
we obtain the new $\Theta$ from Eq.~(\ref{eq:theta_pwp}). It can be shown that
\be
    \Theta \ent= g^{0j} S_j - g^{00} \left( \frac{\sqrt{-g} P}{N} + e \right),
     \label{eq:theta_enth}
\ee
and this provides us with the enthalpy $\enth$ and 
the covariant spatial velocity components $v_i= S_i/(\Theta \ent).$  
The time component is found from the equation for the generalized Lorentz factor, Eq.(\ref{eq:theta_def}),
\be
v_0= \frac{1-g^{0i}v_i}{g^{00}}.
\ee
The contravariant velocity is then straight-forwardly calculated via  $v^i= g^{i\lambda} v_\lambda$,  $n$ from Eq.~(\ref{eq:n_N})  
and the internal energy as $u= \enth - \frac{P}{n} - 1$.

\section{Which resolution?}
\label{sec:resolution}
In our new simulation methodology, where we evolve the spacetime on a mesh and the matter with particles,  we have 
 two different resolution lengths and it its not a priori clear how they should be related. We therefore present 
here some numerical experiments to shed some light on the effects of the grid- and particle resolution. To keep
the parameter space under control, we restrict ourselves here to one of our "most realistic" equations of state, MPA1.
We perform the following test simulations:
\bi
\i {\tt TS1}: the outer boundary in each coordinate direction is located at 375 ($\approx 554$ km), five refinement levels and $175^3$ grid points
 which corresponds to the finest grid resolution length of $\Delta_{\rm g}^{\rm min}\approx 400$ m. We use here our default, i.e. 6th order, 
 Finite Differencing ("FD6").
\i  {\tt TS2}: same as  {\tt TS1}, but FD4
\i  {\tt TS3}: same as  {\tt TS1}, but FD8
\i  {\tt TS4}: same as  {\tt TS1}, but $233^3$ grid points, i.e. $\Delta_{\rm g}^{\rm min}\approx 300$ m
\i  {\tt TS5}: same as  {\tt TS1}, but $351^3$ grid points, i.e. $\Delta_{\rm g}^{\rm min}\approx 200$ m.
\i  {\tt TS6}: same as  {\tt TS1}, but $401^3$ grid points, i.e. $\Delta_{\rm g}^{\rm min}\approx 175$ m.
\ei
All the variations related to the spacetime evolution accuracy indicate that the changes in the inspiral are only minor
and our default of 175$^3$ grid points together with 6th order finite differencing is a good choice.\\
\begin{figure}[H]
\centerline{\includegraphics[width=13cm]{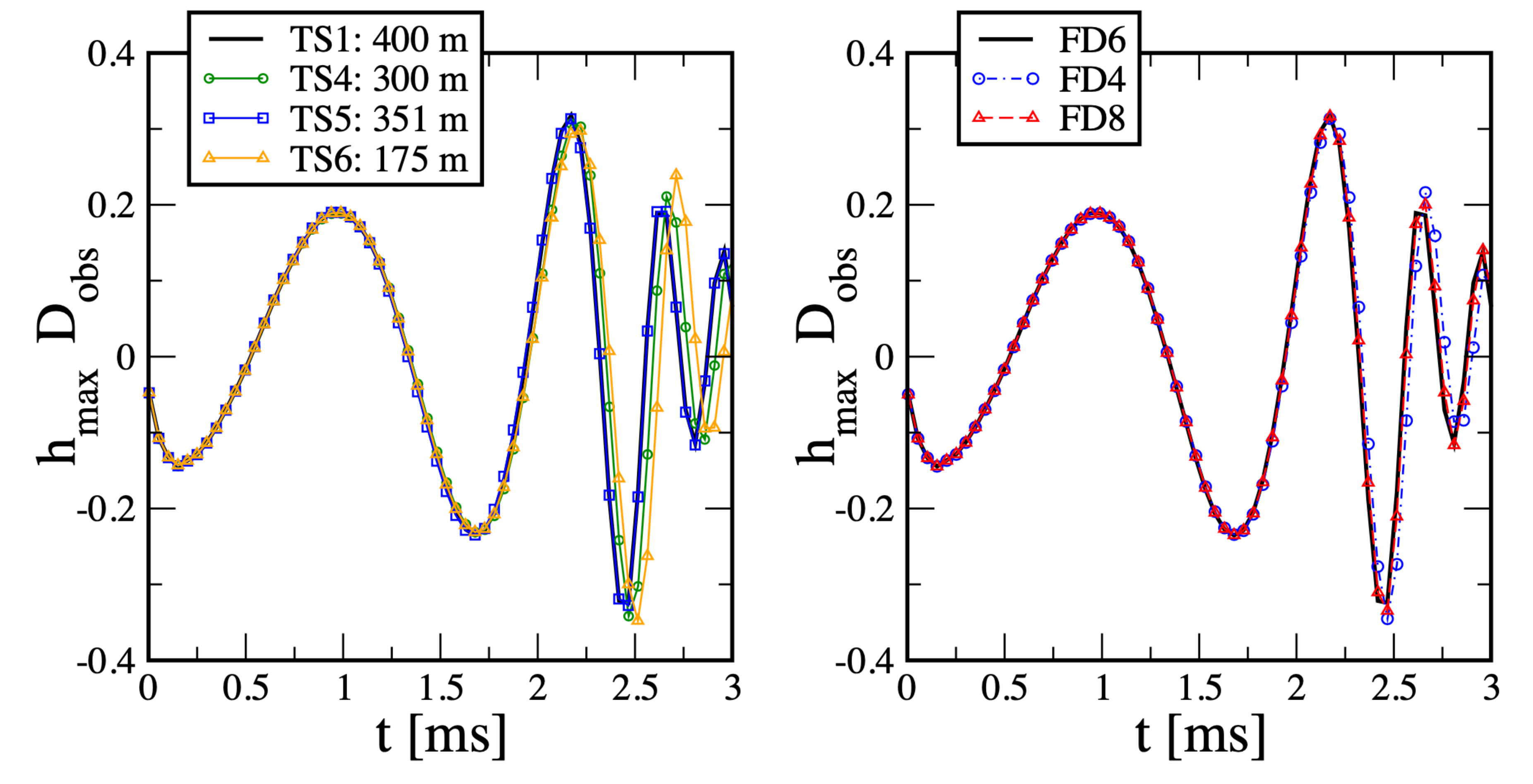}}
\caption{Dependence of the inspiral on the grid resolution (left) and Finite Differencing order (right). The length labels in the left panel refer to the finest grid resolution $\Delta_{\rm g}^{\rm min}$.}
\label{fig:GW_dependence_ngrid_FD}
\end{figure}   
Given that spacetime resolution has only a very minor impact on the inspiral, but we still see substantial 
difference between the 2 million and the 5 million particle run in Fig.~ \ref{fig:GW_resol_dependence},
this suggests that it is the particle number that, at the available resolutions, has the largest impact.  While we 
currently have no accurate estimate for the number of particles that is required for a fully
converged inspiral, Fig.~ \ref{fig:GW_resol_dependence} seems to indicate that we need at least $\sim$ 5 million
particles, but possibly more.  This issue will be explored in more detail in future work.



\end{paracol}
\reftitle{References}


\externalbibliography{yes}
\bibliography{astro_SKR.bib}

\begin{thebibliography}{999}

\bibitem[{Abbott} \em{et~al.}(2021){Abbott}, {Abbott}, {Abraham}, {Acernese},
  {Ackley}, {Adams}, {Adams}, {Adhikari}, {Adya}, {Affeldt}, {Agathos},
  {Agatsuma}, {Aggarwal}, {Aguiar}, {Aiello}, {Ain}, {Ajith}, {Akcay}, {Allen},
  {Allocca}, {Altin}, {Amato}, {Anand}, {Ananyeva}, {Anderson}, {Anderson},
  {Angelova}, {Ansoldi}, {Antelis}, {Antier}, {Appert}, {Arai}, {Araya},
  {Areeda}, {Ar{\`e}ne}, {Arnaud}, {Aronson}, {Arun}, {Asali}, {Ascenzi},
  {Ashton}, {Aston}, {Astone}, {Aubin}, {Aufmuth}, {AultONeal}, {Austin},
  {Avendano}, {Babak}, {Badaracco}, {Bader}, {Bae}, {Baer}, {Bagnasco},
  {Baird}, {Ball}, {Ballardin}, {Ballmer}, {Bals}, {Balsamo}, {Baltus},
  {Banagiri}, {Bankar}, {Bankar}, {Barayoga}, {Barbieri}, {Barish}, {Barker},
  {Barneo}, {Barnum}, {Barone}, {Barr}, {Barsotti}, {Barsuglia}, {Barta},
  {Bartlett}, {Bartos}, {Bassiri}, {Basti}, {Bawaj}, {Bayley}, {Bazzan},
  {Becher}, {B{\'e}csy}, {Bedakihale}, {Bejger}, {Belahcene}, {Beniwal},
  {Benjamin}, {Bennett}, {Bentley}, {Bergamin}, {Berger}, {Bergmann},
  {Bernuzzi}, {Berry}, {Bersanetti}, {Bertolini}, {Betzwieser}, {Bhandare}, ,
  {LIGO Scientific Collaboration}, and {Virgo Collaboration}]{abbott21b}
{Abbott}, R.; {Abbott}, T.D.; {Abraham}, S.; {Acernese}, F.; {Ackley}, K.;
  {Adams}, A.; {Adams}, C.; {Adhikari}, R.X.; {Adya}, V.B.; {Affeldt}, C.;
  {Agathos}, M.; {Agatsuma}, K.; {Aggarwal}, N.; {Aguiar}, O.D.; {Aiello}, L.;
  {Ain}, A.; {Ajith}, P.; {Akcay}, S.; {Allen}, G.; {Allocca}, A.; {Altin},
  P.A.; {Amato}, A.; {Anand}, S.; {Ananyeva}, A.; {Anderson}, S.B.; {Anderson},
  W.G.; {Angelova}, S.V.; {Ansoldi}, S.; {Antelis}, J.M.; {Antier}, S.;
  {Appert}, S.; {Arai}, K.; {Araya}, M.C.; {Areeda}, J.S.; {Ar{\`e}ne}, M.;
  {Arnaud}, N.; {Aronson}, S.M.; {Arun}, K.G.; {Asali}, Y.; {Ascenzi}, S.;
  {Ashton}, G.; {Aston}, S.M.; {Astone}, P.; {Aubin}, F.; {Aufmuth}, P.;
  {AultONeal}, K.; {Austin}, C.; {Avendano}, V.; {Babak}, S.; {Badaracco}, F.;
  {Bader}, M.K.M.; {Bae}, S.; {Baer}, A.M.; {Bagnasco}, S.; {Baird}, J.;
  {Ball}, M.; {Ballardin}, G.; {Ballmer}, S.W.; {Bals}, A.; {Balsamo}, A.;
  {Baltus}, G.; {Banagiri}, S.; {Bankar}, D.; {Bankar}, R.S.; {Barayoga}, J.C.;
  {Barbieri}, C.; {Barish}, B.C.; {Barker}, D.; {Barneo}, P.; {Barnum}, S.;
  {Barone}, F.; {Barr}, B.; {Barsotti}, L.; {Barsuglia}, M.; {Barta}, D.;
  {Bartlett}, J.; {Bartos}, I.; {Bassiri}, R.; {Basti}, A.; {Bawaj}, M.;
  {Bayley}, J.C.; {Bazzan}, M.; {Becher}, B.R.; {B{\'e}csy}, B.; {Bedakihale},
  V.M.; {Bejger}, M.; {Belahcene}, I.; {Beniwal}, D.; {Benjamin}, M.G.;
  {Bennett}, T.F.; {Bentley}, J.D.; {Bergamin}, F.; {Berger}, B.K.; {Bergmann},
  G.; {Bernuzzi}, S.; {Berry}, C.P.L.; {Bersanetti}, D.; {Bertolini}, A.;
  {Betzwieser}, J.; {Bhandare}, R.; .; {LIGO Scientific Collaboration}.; {Virgo
  Collaboration}.
\newblock {GWTC-2: Compact Binary Coalescences Observed by LIGO and Virgo
  during the First Half of the Third Observing Run}.
\newblock {\em Physical Review X} {\bf 2021}, {\em 11},~021053,
  \href{http://xxx.lanl.gov/abs/2010.14527}{{\normalfont
  [arXiv:gr-qc/2010.14527]}}.
\newblock
  doi:{\changeurlcolor{black}\href{https://doi.org/10.1103/PhysRevX.11.021053}{\detokenize{10.1103/PhysRevX.11.021053}}}.

\bibitem[{Baiotti}(2019)]{baiotti19}
{Baiotti}, L.
\newblock {Gravitational waves from neutron star mergers and their relation to
  the nuclear equation of state}.
\newblock {\em Progress in Particle and Nuclear Physics} {\bf 2019}, {\em
  109},~103714,  \href{http://xxx.lanl.gov/abs/1907.08534}{{\normalfont
  [arXiv:astro-ph.HE/1907.08534]}}.
\newblock
  doi:{\changeurlcolor{black}\href{https://doi.org/10.1016/j.ppnp.2019.103714}{\detokenize{10.1016/j.ppnp.2019.103714}}}.

\bibitem[{Ruffert} \em{et~al.}(1997){Ruffert}, {Janka}, {Takahashi}, and
  {Schaefer}]{ruffert97a}
{Ruffert}, M.; {Janka}, H.; {Takahashi}, K.; {Schaefer}, G.
\newblock {Coalescing neutron stars - a step towards physical models. II.
  Neutrino emission, neutron tori, and gamma-ray bursts.}
\newblock {\em A \& A} {\bf 1997}, {\em 319},~122--153.

\bibitem[{Rosswog} and {Liebend{\"o}rfer}(2003)]{rosswog03a}
{Rosswog}, S.; {Liebend{\"o}rfer}, M.
\newblock {High-resolution calculations of merging neutron stars - II. Neutrino
  emission}.
\newblock {\em MNRAS} {\bf 2003}, {\em 342},~673--689.
\newblock
  doi:{\changeurlcolor{black}\href{https://doi.org/10.1046/j.1365-8711.2003.06579.x}{\detokenize{10.1046/j.1365-8711.2003.06579.x}}}.

\bibitem[{Sekiguchi} \em{et~al.}(2011){Sekiguchi}, {Kiuchi}, {Kyutoku}, and
  {Shibata}]{sekiguchi11}
{Sekiguchi}, Y.; {Kiuchi}, K.; {Kyutoku}, K.; {Shibata}, M.
\newblock {Gravitational Waves and Neutrino Emission from the Merger of Binary
  Neutron Stars}.
\newblock {\em Physical Review Letters} {\bf 2011}, {\em 107},~051102,
  \href{http://xxx.lanl.gov/abs/1105.2125}{{\normalfont
  [arXiv:gr-qc/1105.2125]}}.
\newblock
  doi:{\changeurlcolor{black}\href{https://doi.org/10.1103/PhysRevLett.107.051102}{\detokenize{10.1103/PhysRevLett.107.051102}}}.

\bibitem[{Perego} \em{et~al.}(2014){Perego}, {Rosswog}, {Cabez{\'o}n},
  {Korobkin}, {K{\"a}ppeli}, {Arcones}, and {Liebend{\"o}rfer}]{perego14b}
{Perego}, A.; {Rosswog}, S.; {Cabez{\'o}n}, R.M.; {Korobkin}, O.;
  {K{\"a}ppeli}, R.; {Arcones}, A.; {Liebend{\"o}rfer}, M.
\newblock {Neutrino-driven winds from neutron star merger remnants}.
\newblock {\em MNRAS} {\bf 2014}, {\em 443},~3134--3156.
\newblock
  doi:{\changeurlcolor{black}\href{https://doi.org/10.1093/mnras/stu1352}{\detokenize{10.1093/mnras/stu1352}}}.

\bibitem[{Just} \em{et~al.}(2015){Just}, {Bauswein}, {Pulpillo}, {Goriely}, and
  {Janka}]{just15}
{Just}, O.; {Bauswein}, A.; {Pulpillo}, R.A.; {Goriely}, S.; {Janka}, H.T.
\newblock {Comprehensive nucleosynthesis analysis for ejecta of compact binary
  mergers}.
\newblock {\em MNRAS} {\bf 2015}, {\em 448},~541--567,
  \href{http://xxx.lanl.gov/abs/1406.2687}{{\normalfont
  [arXiv:astro-ph.SR/1406.2687]}}.
\newblock
  doi:{\changeurlcolor{black}\href{https://doi.org/10.1093/mnras/stv009}{\detokenize{10.1093/mnras/stv009}}}.

\bibitem[{Fujibayashi} \em{et~al.}(2020){Fujibayashi}, {Shibata}, {Wanajo},
  {Kiuchi}, {Kyutoku}, and {Sekiguchi}]{fujibayashi20}
{Fujibayashi}, S.; {Shibata}, M.; {Wanajo}, S.; {Kiuchi}, K.; {Kyutoku}, K.;
  {Sekiguchi}, Y.
\newblock {Mass ejection from disks surrounding a low-mass black hole: Viscous
  neutrino-radiation hydrodynamics simulation in full general relativity}.
\newblock {\em Phys. Rev. D} {\bf 2020}, {\em 101},~083029,
  \href{http://xxx.lanl.gov/abs/2001.04467}{{\normalfont
  [arXiv:astro-ph.HE/2001.04467]}}.
\newblock
  doi:{\changeurlcolor{black}\href{https://doi.org/10.1103/PhysRevD.101.083029}{\detokenize{10.1103/PhysRevD.101.083029}}}.

\bibitem[{Foucart} \em{et~al.}(2021){Foucart}, {Duez}, {H{\'e}bert}, {Kidder},
  {Kovarik}, {Pfeiffer}, and {Scheel}]{foucart21a}
{Foucart}, F.; {Duez}, M.D.; {H{\'e}bert}, F.; {Kidder}, L.E.; {Kovarik}, P.;
  {Pfeiffer}, H.P.; {Scheel}, M.A.
\newblock {Implementation of Monte Carlo Transport in the General Relativistic
  SpEC Code}.
\newblock {\em \apj} {\bf 2021}, {\em 920},~82,
  \href{http://xxx.lanl.gov/abs/2103.16588}{{\normalfont
  [arXiv:astro-ph.HE/2103.16588]}}.
\newblock
  doi:{\changeurlcolor{black}\href{https://doi.org/10.3847/1538-4357/ac1737}{\detokenize{10.3847/1538-4357/ac1737}}}.

\bibitem[{Just} \em{et~al.}(2022){Just}, {Goriely}, {Janka}, {Nagataki}, and
  {Bauswein}]{just22}
{Just}, O.; {Goriely}, S.; {Janka}, H.T.; {Nagataki}, S.; {Bauswein}, A.
\newblock {Neutrino absorption and other physics dependencies in
  neutrino-cooled black hole accretion discs}.
\newblock {\em MNRAS} {\bf 2022}, {\em 509},~1377--1412,
  \href{http://xxx.lanl.gov/abs/2102.08387}{{\normalfont
  [arXiv:astro-ph.HE/2102.08387]}}.
\newblock
  doi:{\changeurlcolor{black}\href{https://doi.org/10.1093/mnras/stab2861}{\detokenize{10.1093/mnras/stab2861}}}.

\bibitem[{Radice} \em{et~al.}(2022){Radice}, {Bernuzzi}, {Perego}, and
  {Haas}]{radice22}
{Radice}, D.; {Bernuzzi}, S.; {Perego}, A.; {Haas}, R.
\newblock {A New Moment-Based General-Relativistic Neutrino-Radiation Transport
  Code: Methods and First Applications to Neutron Star Mergers}.
\newblock {\em MNRAS} {\bf 2022},
  \href{http://xxx.lanl.gov/abs/2111.14858}{{\normalfont
  [arXiv:astro-ph.HE/2111.14858]}}.
\newblock
  doi:{\changeurlcolor{black}\href{https://doi.org/10.1093/mnras/stac589}{\detokenize{10.1093/mnras/stac589}}}.

\bibitem[Price and Rosswog(2006)]{price06}
Price, D.; Rosswog, S.
\newblock Producing ultra-strong magnetic fields in neutron star mergers.
\newblock {\em Science} {\bf 2006}, {\em 312},~719.

\bibitem[{Kiuchi} \em{et~al.}(2015){Kiuchi}, {Cerd{\'a}-Dur{\'a}n}, {Kyutoku},
  {Sekiguchi}, and {Shibata}]{kiuchi15}
{Kiuchi}, K.; {Cerd{\'a}-Dur{\'a}n}, P.; {Kyutoku}, K.; {Sekiguchi}, Y.;
  {Shibata}, M.
\newblock {Efficient magnetic-field amplification due to the Kelvin-Helmholtz
  instability in binary neutron star mergers}.
\newblock {\em Phys. Rev. D} {\bf 2015}, {\em 92},~124034,
  \href{http://xxx.lanl.gov/abs/1509.09205}{{\normalfont
  [arXiv:astro-ph.HE/1509.09205]}}.
\newblock
  doi:{\changeurlcolor{black}\href{https://doi.org/10.1103/PhysRevD.92.124034}{\detokenize{10.1103/PhysRevD.92.124034}}}.

\bibitem[{Palenzuela} \em{et~al.}(2015){Palenzuela}, {Liebling}, {Neilsen},
  {Lehner}, {Caballero}, {O'Connor}, and {Anderson}]{palenzuela15}
{Palenzuela}, C.; {Liebling}, S.L.; {Neilsen}, D.; {Lehner}, L.; {Caballero},
  O.L.; {O'Connor}, E.; {Anderson}, M.
\newblock {Effects of the microphysical equation of state in the mergers of
  magnetized neutron stars with neutrino cooling}.
\newblock {\em Phys. Rev. D} {\bf 2015}, {\em 92},~044045,
  \href{http://xxx.lanl.gov/abs/1505.01607}{{\normalfont
  [arXiv:gr-qc/1505.01607]}}.
\newblock
  doi:{\changeurlcolor{black}\href{https://doi.org/10.1103/PhysRevD.92.044045}{\detokenize{10.1103/PhysRevD.92.044045}}}.

\bibitem[{Rosswog} and {Diener}(2021)]{rosswog21a}
{Rosswog}, S.; {Diener}, P.
\newblock {SPHINCS\_BSSN: a general relativistic smooth particle hydrodynamics
  code for dynamical spacetimes}.
\newblock {\em Classical and Quantum Gravity} {\bf 2021}, {\em 38},~115002,
  \href{http://xxx.lanl.gov/abs/2012.13954}{{\normalfont
  [arXiv:gr-qc/2012.13954]}}.
\newblock
  doi:{\changeurlcolor{black}\href{https://doi.org/10.1088/1361-6382/abee65}{\detokenize{10.1088/1361-6382/abee65}}}.

\bibitem[{Diener} \em{et~al.}(2022){Diener}, {Rosswog}, and
  {Torsello}]{diener22a}
{Diener}, P.; {Rosswog}, S.; {Torsello}, F.
\newblock {Simulating neutron star mergers with the Lagrangian Numerical
  Relativity code SPHINCS\_BSSN}.
\newblock {\em European Physical Journal A} {\bf 2022}, {\em 58},~74,
  \href{http://xxx.lanl.gov/abs/2203.06478}{{\normalfont
  [arXiv:astro-ph.HE/2203.06478]}}.
\newblock
  doi:{\changeurlcolor{black}\href{https://doi.org/10.1140/epja/s10050-022-00725-7}{\detokenize{10.1140/epja/s10050-022-00725-7}}}.

\bibitem[{Shibata} and {Nakamura}(1995)]{shibata95}
{Shibata}, M.; {Nakamura}, T.
\newblock {Evolution of three-dimensional gravitational waves: Harmonic slicing
  case}.
\newblock {\em Phys. Rev. D} {\bf 1995}, {\em 52},~5428--5444.
\newblock
  doi:{\changeurlcolor{black}\href{https://doi.org/10.1103/PhysRevD.52.5428}{\detokenize{10.1103/PhysRevD.52.5428}}}.

\bibitem[{Baumgarte} and {Shapiro}(1999)]{baumgarte99}
{Baumgarte}, T.W.; {Shapiro}, S.L.
\newblock {Numerical integration of Einstein's field equations}.
\newblock {\em Phys. Rev. D} {\bf 1999}, {\em 59},~024007,
  \href{http://xxx.lanl.gov/abs/arXiv:gr-qc/9810065}{{\normalfont
  [arXiv:gr-qc/9810065]}}.
\newblock
  doi:{\changeurlcolor{black}\href{https://doi.org/10.1103/PhysRevD.59.024007}{\detokenize{10.1103/PhysRevD.59.024007}}}.

\bibitem[Rosswog \em{et~al.}(1999)Rosswog, Liebend\"orfer, Thielemann, Davies,
  Benz, and Piran]{rosswog99}
Rosswog, S.; Liebend\"orfer, M.; Thielemann, F.K.; Davies, M.; Benz, W.; Piran,
  T.
\newblock Mass ejection in neutron star mergers.
\newblock {\em A \&\ A} {\bf 1999}, {\em 341},~499--526.

\bibitem[{Oechslin} and {Janka}(2007)]{oechslin07b}
{Oechslin}, R.; {Janka}, H.
\newblock {Gravitational Waves from Relativistic Neutron-Star Mergers with
  Microphysical Equations of State}.
\newblock {\em Physical Review Letters} {\bf 2007}, {\em 99},~121102,
  \href{http://xxx.lanl.gov/abs/arXiv:astro-ph/0702228}{{\normalfont
  [arXiv:astro-ph/0702228]}}.
\newblock
  doi:{\changeurlcolor{black}\href{https://doi.org/10.1103/PhysRevLett.99.121102}{\detokenize{10.1103/PhysRevLett.99.121102}}}.

\bibitem[{Bauswein} \em{et~al.}(2013){Bauswein}, {Goriely}, and
  {Janka}]{bauswein13a}
{Bauswein}, A.; {Goriely}, S.; {Janka}, H.T.
\newblock {Systematics of Dynamical Mass Ejection, Nucleosynthesis, and
  Radioactively Powered Electromagnetic Signals from Neutron-star Mergers}.
\newblock {\em ApJ} {\bf 2013}, {\em 773},~78.
\newblock
  doi:{\changeurlcolor{black}\href{https://doi.org/10.1088/0004-637X/773/1/78}{\detokenize{10.1088/0004-637X/773/1/78}}}.

\bibitem[{Hotokezaka} \em{et~al.}(2013){Hotokezaka}, {Kiuchi}, {Kyutoku},
  {Okawa}, {Sekiguchi}, {Shibata}, and {Taniguchi}]{hotokezaka13a}
{Hotokezaka}, K.; {Kiuchi}, K.; {Kyutoku}, K.; {Okawa}, H.; {Sekiguchi}, Y.i.;
  {Shibata}, M.; {Taniguchi}, K.
\newblock {Mass ejection from the merger of binary neutron stars}.
\newblock {\em Phys. Rev. D} {\bf 2013}, {\em 87},~024001.
\newblock
  doi:{\changeurlcolor{black}\href{https://doi.org/10.1103/PhysRevD.87.024001}{\detokenize{10.1103/PhysRevD.87.024001}}}.

\bibitem[{Radice} \em{et~al.}(2018){Radice}, {Perego}, {Hotokezaka}, {Fromm},
  {Bernuzzi}, and {Roberts}]{radice18a}
{Radice}, D.; {Perego}, A.; {Hotokezaka}, K.; {Fromm}, S.A.; {Bernuzzi}, S.;
  {Roberts}, L.F.
\newblock {Binary Neutron Star Mergers: Mass Ejection, Electromagnetic
  Counterparts, and Nucleosynthesis}.
\newblock {\em ApJ} {\bf 2018}, {\em 869},~130,
  \href{http://xxx.lanl.gov/abs/1809.11161}{{\normalfont
  [arXiv:astro-ph.HE/1809.11161]}}.
\newblock
  doi:{\changeurlcolor{black}\href{https://doi.org/10.3847/1538-4357/aaf054}{\detokenize{10.3847/1538-4357/aaf054}}}.

\bibitem[{Schoepe} \em{et~al.}(2018){Schoepe}, {Hilditch}, and
  {Bugner}]{schoepe18}
{Schoepe}, A.; {Hilditch}, D.; {Bugner}, M.
\newblock {Revisiting hyperbolicity of relativistic fluids}.
\newblock {\em Phys. Rev. D} {\bf 2018}, {\em 97},~123009,
  \href{http://xxx.lanl.gov/abs/1712.09837}{{\normalfont
  [arXiv:gr-qc/1712.09837]}}.
\newblock
  doi:{\changeurlcolor{black}\href{https://doi.org/10.1103/PhysRevD.97.123009}{\detokenize{10.1103/PhysRevD.97.123009}}}.

\bibitem[{Rosswog}(2020)]{rosswog20a}
{Rosswog}, S.
\newblock {The Lagrangian hydrodynamics code MAGMA2}.
\newblock {\em MNRAS} {\bf 2020}, {\em 498},~4230--4255,
  \href{http://xxx.lanl.gov/abs/1911.13093}{{\normalfont
  [arXiv:astro-ph.IM/1911.13093]}}.
\newblock
  doi:{\changeurlcolor{black}\href{https://doi.org/10.1093/mnras/staa2591}{\detokenize{10.1093/mnras/staa2591}}}.

\bibitem[lor()]{lor}
LORENE library.
\newblock https://lorene.obspm.fr.

\bibitem[{Read} \em{et~al.}(2009){Read}, {Lackey}, {Owen}, and
  {Friedman}]{read09}
{Read}, J.S.; {Lackey}, B.D.; {Owen}, B.J.; {Friedman}, J.L.
\newblock {Constraints on a phenomenologically parametrized neutron-star
  equation of state}.
\newblock {\em Phys. Rev. D} {\bf 2009}, {\em 79},~124032,
  \href{http://xxx.lanl.gov/abs/0812.2163}{{\normalfont
  [arXiv:astro-ph/0812.2163]}}.
\newblock
  doi:{\changeurlcolor{black}\href{https://doi.org/10.1103/PhysRevD.79.124032}{\detokenize{10.1103/PhysRevD.79.124032}}}.

\bibitem[{Cullen} and {Dehnen}(2010)]{cullen10}
{Cullen}, L.; {Dehnen}, W.
\newblock {Inviscid smoothed particle hydrodynamics}.
\newblock {\em MNRAS} {\bf 2010}, {\em 408},~669--683,
  \href{http://xxx.lanl.gov/abs/1006.1524}{{\normalfont
  [arXiv:astro-ph.IM/1006.1524]}}.
\newblock
  doi:{\changeurlcolor{black}\href{https://doi.org/10.1111/j.1365-2966.2010.17158.x}{\detokenize{10.1111/j.1365-2966.2010.17158.x}}}.

\bibitem[{Rosswog}(2015)]{rosswog15b}
{Rosswog}, S.
\newblock {Boosting the accuracy of SPH techniques: Newtonian and
  special-relativistic tests}.
\newblock {\em MNRAS} {\bf 2015}, {\em 448},~3628--3664,
  \href{http://xxx.lanl.gov/abs/1405.6034}{{\normalfont
  [arXiv:astro-ph.IM/1405.6034]}}.
\newblock
  doi:{\changeurlcolor{black}\href{https://doi.org/10.1093/mnras/stv225}{\detokenize{10.1093/mnras/stv225}}}.

\bibitem[Rosswog(2009)]{rosswog09b}
Rosswog, S.
\newblock Astrophysical Smooth Particle Hydrodynamics.
\newblock {\em New Astronomy Reviews} {\bf 2009}, {\em 53},~78--104.

\bibitem[{Rosswog}(2020)]{rosswog20b}
{Rosswog}, S.
\newblock {A Simple, Entropy-based Dissipation Trigger for SPH}.
\newblock {\em ApJ} {\bf 2020}, {\em 898},~60,
  \href{http://xxx.lanl.gov/abs/1912.01095}{{\normalfont
  [arXiv:astro-ph.IM/1912.01095]}}.
\newblock
  doi:{\changeurlcolor{black}\href{https://doi.org/10.3847/1538-4357/ab9a2e}{\detokenize{10.3847/1538-4357/ab9a2e}}}.

\bibitem[{Alcubierre}(2008)]{alcubierre08}
{Alcubierre}, M.
\newblock {\em {Introduction to 3+1 Numerical Relativity}}; Oxford University
  Press,  2008.

\bibitem[{Baumgarte} and {Shapiro}(2010)]{baumgarte10}
{Baumgarte}, T.W.; {Shapiro}, S.L.
\newblock {\em {Numerical Relativity: Solving Einstein's Equations on the
  Computer}};  2010.

\bibitem[{Rezzolla} and {Zanotti}(2013)]{rezzolla13a}
{Rezzolla}, L.; {Zanotti}, O.
\newblock {\em {Relativistic Hydrodynamics}};  2013.

\bibitem[{Shibata}(2016)]{shibata16}
{Shibata}, M.
\newblock {\em {Numerical Relativity}};  2016.
\newblock
  doi:{\changeurlcolor{black}\href{https://doi.org/10.1142/9692}{\detokenize{10.1142/9692}}}.

\bibitem[{W}endland(1995)]{wendland95}
{W}endland, H.
\newblock Piecewise polynomial, positive definite and compactly supported
  radial functions of minimal degree.
\newblock {\em Advances in Computational Mathematics} {\bf 1995}, {\em
  4},~389--296.

\bibitem[{Gafton} and {Rosswog}(2011)]{gafton11}
{Gafton}, E.; {Rosswog}, S.
\newblock {A fast recursive coordinate bisection tree for neighbour search and
  gravity}.
\newblock {\em MNRAS} {\bf 2011}, {\em 418},~770--781,
  \href{http://xxx.lanl.gov/abs/1108.0028}{{\normalfont
  [arXiv:astro-ph.IM/1108.0028]}}.
\newblock
  doi:{\changeurlcolor{black}\href{https://doi.org/10.1111/j.1365-2966.2011.19528.x}{\detokenize{10.1111/j.1365-2966.2011.19528.x}}}.

\bibitem[{Laguna} \em{et~al.}(1993){Laguna}, {Miller}, and {Zurek}]{laguna93a}
{Laguna}, P.; {Miller}, W.A.; {Zurek}, W.H.
\newblock {Smoothed particle hydrodynamics near a black hole}.
\newblock {\em ApJ} {\bf 1993}, {\em 404},~678--685.
\newblock
  doi:{\changeurlcolor{black}\href{https://doi.org/10.1086/172321}{\detokenize{10.1086/172321}}}.

\bibitem[{von Neumann} and {Richtmyer}(1950)]{vonNeumann50}
{von Neumann}, J.; {Richtmyer}, R.D.
\newblock {A Method for the Numerical Calculation of Hydrodynamic Shocks}.
\newblock {\em Journal of Applied Physics} {\bf 1950}, {\em 21},~232--237.

\bibitem[{Liptai} and {Price}(2019)]{liptai19}
{Liptai}, D.; {Price}, D.J.
\newblock {General relativistic smoothed particle hydrodynamics}.
\newblock {\em MNRAS} {\bf 2019}, {\em 485},~819--842,
  \href{http://xxx.lanl.gov/abs/1901.08064}{{\normalfont
  [arXiv:astro-ph.IM/1901.08064]}}.
\newblock
  doi:{\changeurlcolor{black}\href{https://doi.org/10.1093/mnras/stz111}{\detokenize{10.1093/mnras/stz111}}}.

\bibitem[Christensen(1990)]{christensen90}
Christensen, R.B.
\newblock Godunov methods on a staggered mesh–an improved artificial
  viscosity.
\newblock {\em Nuclear Explosives Code Developers Conference, volume
  UCRL-JC-105269. Lawrence Livermore National Lab, Lawrence Livermore Technical
  Report.} {\bf 1990}, {\em UCRL-JC-105269}.

\bibitem[{Frontiere} \em{et~al.}(2017){Frontiere}, {Raskin}, and
  {Owen}]{frontiere17}
{Frontiere}, N.; {Raskin}, C.D.; {Owen}, J.M.
\newblock {CRKSPH - A Conservative Reproducing Kernel Smoothed Particle
  Hydrodynamics Scheme}.
\newblock {\em Journal of Computational Physics} {\bf 2017}, {\em
  332},~160--209,  \href{http://xxx.lanl.gov/abs/1605.00725}{{\normalfont
  [arXiv:physics.comp-ph/1605.00725]}}.
\newblock
  doi:{\changeurlcolor{black}\href{https://doi.org/10.1016/j.jcp.2016.12.004}{\detokenize{10.1016/j.jcp.2016.12.004}}}.

\bibitem[Brown \em{et~al.}(2009)Brown, Diener, Sarbach, Schnetter, and
  Tiglio]{Brown:2008sb}
Brown, J.D.; Diener, P.; Sarbach, O.; Schnetter, E.; Tiglio, M.
\newblock {Turduckening black holes: an analytical and computational study}.
\newblock {\em Phys. Rev. D} {\bf 2009}, {\em 79},~044023,
  \href{http://xxx.lanl.gov/abs/arXiv:0809.3533 [gr-qc]}{{\normalfont
  [arXiv:0809.3533 [gr-qc]]}}.

\bibitem[{Einstein Toolkit web page}(2020)]{ETK:web}
{Einstein Toolkit web page}.
\newblock https://einsteintoolkit.org/,  2020.
\newblock [Online; accessed 9-December-2020].

\bibitem[L{\"{o}}ffler \em{et~al.}(2012)L{\"{o}}ffler, Faber, Bentivegna, Bode,
  Diener, Haas, Hinder, Mundim, Ott, Schnetter, Allen, Campanelli, and
  Laguna]{Loffler:2011ay}
L{\"{o}}ffler, F.; Faber, J.; Bentivegna, E.; Bode, T.; Diener, P.; Haas, R.;
  Hinder, I.; Mundim, B.C.; Ott, C.D.; Schnetter, E.; Allen, G.; Campanelli,
  M.; Laguna, P.
\newblock {{T}he {E}instein {T}oolkit: {A} {C}ommunity {C}omputational
  {I}nfrastructure for {R}elativistic {A}strophysics}.
\newblock {\em Class. Quantum Grav.} {\bf 2012}, {\em 29},~115001,
  \href{http://xxx.lanl.gov/abs/arXiv:1111.3344 [gr-qc]}{{\normalfont
  [arXiv:1111.3344 [gr-qc]]}}.
\newblock
  doi:{\changeurlcolor{black}\href{https://doi.org/doi:10.1088/0264-9381/29/11/115001}{\detokenize{doi:10.1088/0264-9381/29/11/115001}}}.

\bibitem[{Cottet} and {Koumoutsakos}(2000)]{cottet00}
{Cottet}, G.H.; {Koumoutsakos}, P.D.
\newblock {\em {Vortex Methods}};  2000.

\bibitem[{Cottet} \em{et~al.}(2014){Cottet}, {Etancelin}, {Perignon}, and
  {Picard}]{cottet14}
{Cottet}, G.; {Etancelin}, J.; {Perignon}, F.; {Picard}, C.
\newblock {High order Semi-Lagrangian particle methods for tranport equations}.
\newblock {\em ESAIM: Mathematical Modelling and Numerical Analysis, EDP
  Sciences} {\bf 2014}, {\em 48},~1029 -- 1060.
\newblock
  doi:{\changeurlcolor{black}\href{https://doi.org/10.1086/185939}{\detokenize{10.1086/185939}}}.

\bibitem[{Douchin} and {Haensel}(2001)]{SLY_eos}
{Douchin}, F.; {Haensel}, P.
\newblock {A unified equation of state of dense matter and neutron star
  structure}.
\newblock {\em A \& A} {\bf 2001}, {\em 380},~151--167,
  \href{http://xxx.lanl.gov/abs/astro-ph/0111092}{{\normalfont
  [arXiv:astro-ph/astro-ph/0111092]}}.
\newblock
  doi:{\changeurlcolor{black}\href{https://doi.org/10.1051/0004-6361:20011402}{\detokenize{10.1051/0004-6361:20011402}}}.

\bibitem[{Akmal} \em{et~al.}(1998){Akmal}, {Pandharipande}, and
  {Ravenhall}]{akmal98}
{Akmal}, A.; {Pandharipande}, V.R.; {Ravenhall}, D.G.
\newblock {Equation of state of nucleon matter and neutron star structure}.
\newblock {\em Phys. Rev. C} {\bf 1998}, {\em 58},~1804--1828,
  \href{http://xxx.lanl.gov/abs/hep-ph/9804388}{{\normalfont
  [hep-ph/9804388]}}.
\newblock
  doi:{\changeurlcolor{black}\href{https://doi.org/10.1103/PhysRevC.58.1804}{\detokenize{10.1103/PhysRevC.58.1804}}}.

\bibitem[{M{\"u}ther} \em{et~al.}(1987){M{\"u}ther}, {Prakash}, and
  {Ainsworth}]{MPA1_eos}
{M{\"u}ther}, H.; {Prakash}, M.; {Ainsworth}, T.L.
\newblock {The nuclear symmetry energy in relativistic Brueckner-Hartree-Fock
  calculations}.
\newblock {\em Physics Letters B} {\bf 1987}, {\em 199},~469--474.
\newblock
  doi:{\changeurlcolor{black}\href{https://doi.org/10.1016/0370-2693(87)91611-X}{\detokenize{10.1016/0370-2693(87)91611-X}}}.

\bibitem[{M{\"u}ller} and {Serot}(1996)]{MS1_EOS}
{M{\"u}ller}, H.; {Serot}, B.D.
\newblock {Relativistic mean-field theory and the high-density nuclear equation
  of state}.
\newblock {\em Nuc. Phys. A} {\bf 1996}, {\em 606},~508--537,
  \href{http://xxx.lanl.gov/abs/nucl-th/9603037}{{\normalfont
  [arXiv:nucl-th/nucl-th/9603037]}}.
\newblock
  doi:{\changeurlcolor{black}\href{https://doi.org/10.1016/0375-9474(96)00187-X}{\detokenize{10.1016/0375-9474(96)00187-X}}}.

\bibitem[{Pacilio} \em{et~al.}(2022){Pacilio}, {Maselli}, {Fasano}, and
  {Pani}]{pacilio22}
{Pacilio}, C.; {Maselli}, A.; {Fasano}, M.; {Pani}, P.
\newblock {Ranking Love Numbers for the Neutron Star Equation of State: The
  Need for Third-Generation Detectors}.
\newblock {\em Phys. Rev. Lett.} {\bf 2022}, {\em 128},~101101,
  \href{http://xxx.lanl.gov/abs/2104.10035}{{\normalfont
  [arXiv:gr-qc/2104.10035]}}.
\newblock
  doi:{\changeurlcolor{black}\href{https://doi.org/10.1103/PhysRevLett.128.101101}{\detokenize{10.1103/PhysRevLett.128.101101}}}.

\bibitem[{Cromartie} \em{et~al.}(2020){Cromartie}, {Fonseca}, {Ransom},
  {Demorest}, {Arzoumanian}, {Blumer}, {Brook}, {DeCesar}, {Dolch}, {Ellis},
  {Ferdman}, {Ferrara}, {Garver-Daniels}, {Gentile}, {Jones}, {Lam}, {Lorimer},
  {Lynch}, {McLaughlin}, {Ng}, {Nice}, {Pennucci}, {Spiewak}, {Stairs},
  {Stovall}, {Swiggum}, and {Zhu}]{cromartie20}
{Cromartie}, H.T.; {Fonseca}, E.; {Ransom}, S.M.; {Demorest}, P.B.;
  {Arzoumanian}, Z.; {Blumer}, H.; {Brook}, P.R.; {DeCesar}, M.E.; {Dolch}, T.;
  {Ellis}, J.A.; {Ferdman}, R.D.; {Ferrara}, E.C.; {Garver-Daniels}, N.;
  {Gentile}, P.A.; {Jones}, M.L.; {Lam}, M.T.; {Lorimer}, D.R.; {Lynch}, R.S.;
  {McLaughlin}, M.A.; {Ng}, C.; {Nice}, D.J.; {Pennucci}, T.T.; {Spiewak}, R.;
  {Stairs}, I.H.; {Stovall}, K.; {Swiggum}, J.K.; {Zhu}, W.W.
\newblock {Relativistic Shapiro delay measurements of an extremely massive
  millisecond pulsar}.
\newblock {\em Nature Astronomy} {\bf 2020}, {\em 4},~72--76,
  \href{http://xxx.lanl.gov/abs/1904.06759}{{\normalfont
  [arXiv:astro-ph.HE/1904.06759]}}.
\newblock
  doi:{\changeurlcolor{black}\href{https://doi.org/10.1038/s41550-019-0880-2}{\detokenize{10.1038/s41550-019-0880-2}}}.

\bibitem[{Fryer} \em{et~al.}(2015){Fryer}, {Belczynski}, {Ramirez-Ruiz},
  {Rosswog}, {Shen}, and {Steiner}]{fryer15}
{Fryer}, C.L.; {Belczynski}, K.; {Ramirez-Ruiz}, E.; {Rosswog}, S.; {Shen}, G.;
  {Steiner}, A.W.
\newblock {The Fate of the Compact Remnant in Neutron Star Mergers}.
\newblock {\em ApJ} {\bf 2015}, {\em 812},~24,
  \href{http://xxx.lanl.gov/abs/1504.07605}{{\normalfont
  [arXiv:astro-ph.HE/1504.07605]}}.
\newblock
  doi:{\changeurlcolor{black}\href{https://doi.org/10.1088/0004-637X/812/1/24}{\detokenize{10.1088/0004-637X/812/1/24}}}.

\bibitem[{Margalit} and {Metzger}(2017)]{margalit17}
{Margalit}, B.; {Metzger}, B.D.
\newblock {Constraining the Maximum Mass of Neutron Stars from Multi-messenger
  Observations of GW170817}.
\newblock {\em \apjl} {\bf 2017}, {\em 850},~L19,
  \href{http://xxx.lanl.gov/abs/1710.05938}{{\normalfont
  [arXiv:astro-ph.HE/1710.05938]}}.
\newblock
  doi:{\changeurlcolor{black}\href{https://doi.org/10.3847/2041-8213/aa991c}{\detokenize{10.3847/2041-8213/aa991c}}}.

\bibitem[{Bauswein} \em{et~al.}(2017){Bauswein}, {Just}, {Janka}, and
  {Stergioulas}]{bauswein17}
{Bauswein}, A.; {Just}, O.; {Janka}, H.T.; {Stergioulas}, N.
\newblock {Neutron-star Radius Constraints from GW170817 and Future
  Detections}.
\newblock {\em ApJL} {\bf 2017}, {\em 850},~L34,
  \href{http://xxx.lanl.gov/abs/1710.06843}{{\normalfont
  [arXiv:astro-ph.HE/1710.06843]}}.
\newblock
  doi:{\changeurlcolor{black}\href{https://doi.org/10.3847/2041-8213/aa9994}{\detokenize{10.3847/2041-8213/aa9994}}}.

\bibitem[{Shibata} \em{et~al.}(2017){Shibata}, {Fujibayashi}, {Hotokezaka},
  {Kiuchi}, {Kyutoku}, {Sekiguchi}, and {Tanaka}]{shibata17c}
{Shibata}, M.; {Fujibayashi}, S.; {Hotokezaka}, K.; {Kiuchi}, K.; {Kyutoku},
  K.; {Sekiguchi}, Y.; {Tanaka}, M.
\newblock {Modeling GW170817 based on numerical relativity and its
  implications}.
\newblock {\em Phys. Rev. D} {\bf 2017}, {\em 96},~123012,
  \href{http://xxx.lanl.gov/abs/1710.07579}{{\normalfont
  [arXiv:astro-ph.HE/1710.07579]}}.
\newblock
  doi:{\changeurlcolor{black}\href{https://doi.org/10.1103/PhysRevD.96.123012}{\detokenize{10.1103/PhysRevD.96.123012}}}.

\bibitem[{Rezzolla} \em{et~al.}(2018){Rezzolla}, {Most}, and
  {Weih}]{rezzolla18}
{Rezzolla}, L.; {Most}, E.R.; {Weih}, L.R.
\newblock {Using Gravitational-wave Observations and Quasi-universal Relations
  to Constrain the Maximum Mass of Neutron Stars}.
\newblock {\em \apjl} {\bf 2018}, {\em 852},~L25,
  \href{http://xxx.lanl.gov/abs/1711.00314}{{\normalfont
  [arXiv:astro-ph.HE/1711.00314]}}.
\newblock
  doi:{\changeurlcolor{black}\href{https://doi.org/10.3847/2041-8213/aaa401}{\detokenize{10.3847/2041-8213/aaa401}}}.

\bibitem[{Biswas} and {Datta}(2021)]{biswas22}
{Biswas}, B.; {Datta}, S.
\newblock {Constraining neutron star properties with a new equation of state
  insensitive approach}.
\newblock {\em arXiv e-prints} {\bf 2021}, p. arXiv:2112.10824,
  \href{http://xxx.lanl.gov/abs/2112.10824}{{\normalfont
  [arXiv:astro-ph.HE/2112.10824]}}.

\bibitem[{Rhoades} and {Ruffini}(1974)]{rhoades74}
{Rhoades}, C.E.; {Ruffini}, R.
\newblock {Maximum Mass of a Neutron Star}.
\newblock {\em Physical Review Letters} {\bf 1974}, {\em 32},~324--327.
\newblock
  doi:{\changeurlcolor{black}\href{https://doi.org/10.1103/PhysRevLett.32.324}{\detokenize{10.1103/PhysRevLett.32.324}}}.

\bibitem[{Kalogera} and {Baym}(1996)]{kalogera96}
{Kalogera}, V.; {Baym}, G.
\newblock {The Maximum Mass of a Neutron Star}.
\newblock {\em ApJL} {\bf 1996}, {\em 470},~L61,
  \href{http://xxx.lanl.gov/abs/astro-ph/9608059}{{\normalfont
  [arXiv:astro-ph/astro-ph/9608059]}}.
\newblock
  doi:{\changeurlcolor{black}\href{https://doi.org/10.1086/310296}{\detokenize{10.1086/310296}}}.

\bibitem[Schaffner-Bielich(2020)]{schaffner_bielich20}
Schaffner-Bielich, J.
\newblock {\em Compact Stars}, 1. ed.; Cambridge University Press: Cambridge,
  UK,  2020.

\bibitem[{Abbott} \em{et~al.}(2017){Abbott}, {Abbott}, {Abbott}, {Acernese},
  {Ackley}, {Adams}, {Adams}, {Addesso}, {Adhikari}, {Adya}, and
  et~al.]{abbott17b}
{Abbott}, B.P.; {Abbott}, R.; {Abbott}, T.D.; {Acernese}, F.; {Ackley}, K.;
  {Adams}, C.; {Adams}, T.; {Addesso}, P.; {Adhikari}, R.X.; {Adya}, V.B.;
  et~al..
\newblock {GW170817: Observation of Gravitational Waves from a Binary Neutron
  Star Inspiral}.
\newblock {\em Physical Review Letters} {\bf 2017}, {\em 119},~161101,
  \href{http://xxx.lanl.gov/abs/1710.05832}{{\normalfont
  [arXiv:gr-qc/1710.05832]}}.
\newblock
  doi:{\changeurlcolor{black}\href{https://doi.org/10.1103/PhysRevLett.119.161101}{\detokenize{10.1103/PhysRevLett.119.161101}}}.

\bibitem[Sod(1978)]{sod78}
Sod, G.
\newblock A survey of several finite difference methods for systems of
  nonlinear hyperbolic conservation laws.
\newblock {\em J. Comput. Phys.} {\bf 1978}, {\em 43},~1--31.

\bibitem[Marti and M\"uller(1996)]{marti96}
Marti, J.; M\"uller, E.
\newblock {\em J. Comp. Phys.} {\bf 1996}, {\em 123},~1.

\bibitem[Chow and Monaghan(1997)]{chow97}
Chow, J.E.; Monaghan, J.
\newblock Ultrarelativistic SPH.
\newblock {\em J. Computat. Phys.} {\bf 1997}, {\em 134},~296.

\bibitem[{Siegler} and {Riffert}(2000)]{siegler00a}
{Siegler}, S.; {Riffert}, H.
\newblock {Smoothed Particle Hydrodynamics Simulations of Ultrarelativistic
  Shocks with Artificial Viscosity}.
\newblock {\em ApJ} {\bf 2000}, {\em 531},~1053--1066,
  \href{http://xxx.lanl.gov/abs/arXiv:astro-ph/9904070}{{\normalfont
  [arXiv:astro-ph/9904070]}}.
\newblock
  doi:{\changeurlcolor{black}\href{https://doi.org/10.1086/308482}{\detokenize{10.1086/308482}}}.

\bibitem[{Del Zanna} and {Bucciantini}(2002)]{delzanna02}
{Del Zanna}, L.; {Bucciantini}, N.
\newblock {An efficient shock-capturing central-type scheme for
  multidimensional relativistic flows. I. Hydrodynamics}.
\newblock {\em A\&A} {\bf 2002}, {\em 390},~1177--1186,
  \href{http://xxx.lanl.gov/abs/arXiv:astro-ph/0205290}{{\normalfont
  [arXiv:astro-ph/0205290]}}.
\newblock
  doi:{\changeurlcolor{black}\href{https://doi.org/10.1051/0004-6361:20020776}{\detokenize{10.1051/0004-6361:20020776}}}.

\bibitem[{Marti} and {M{\"u}ller}(2003)]{marti03}
{Marti}, J.M.; {M{\"u}ller}, E.
\newblock {Numerical Hydrodynamics in Special Relativity}.
\newblock {\em Living Reviews in Relativity} {\bf 2003}, {\em 6},~7.

\bibitem[{Marti} and {M{\"u}ller}(2015)]{marti15}
{Marti}, J.M.; {M{\"u}ller}, E.
\newblock {Grid-based Methods in Relativistic Hydrodynamics and
  Magnetohydrodynamics}.
\newblock {\em Living Reviews in Computational Astrophysics} {\bf 2015}, {\em
  1},~3.
\newblock
  doi:{\changeurlcolor{black}\href{https://doi.org/10.1007/lrca-2015-3}{\detokenize{10.1007/lrca-2015-3}}}.

\bibitem[{Rosswog}(2015)]{rosswog15c}
{Rosswog}, S.
\newblock {SPH Methods in the Modelling of Compact Objects}.
\newblock {\em Living Reviews of Computational Astrophysics (2015)} {\bf 2015},
  {\em 1},  \href{http://xxx.lanl.gov/abs/1406.4224}{{\normalfont
  [arXiv:astro-ph.IM/1406.4224]}}.
\newblock
  doi:{\changeurlcolor{black}\href{https://doi.org/10}{\detokenize{10}}}.

\bibitem[{van Leer}(1977)]{vanLeer77}
{van Leer}, B.
\newblock {Towards the Ultimate Conservative Difference Scheme. IV. A New
  Approach to Numerical Convection}.
\newblock {\em Journal of Computational Physics} {\bf 1977}, {\em 23},~276.
\newblock
  doi:{\changeurlcolor{black}\href{https://doi.org/10.1016/0021-9991(77)90095-X}{\detokenize{10.1016/0021-9991(77)90095-X}}}.

\bibitem[{Sweby}(1984)]{sweby84}
{Sweby}, P.K.
\newblock {High Resolution Schemes Using Flux Limiters for Hyperbolic
  Conservation Laws}.
\newblock {\em SIAM Journal on Numerical Analysis} {\bf 1984}, {\em
  21},~995--1011.
\newblock
  doi:{\changeurlcolor{black}\href{https://doi.org/10.1137/0721062}{\detokenize{10.1137/0721062}}}.

\bibitem[{Rosswog}(2010)]{rosswog10b}
{Rosswog}, S.
\newblock {Conservative, special-relativistic smooth particle hydrodynamics}.
\newblock {\em J. Comp. Phys.} {\bf 2010}, {\em 229},~8591--8612,
  \href{http://xxx.lanl.gov/abs/0907.4890}{{\normalfont [0907.4890]}}.

\bibitem[{Rosswog}(2011)]{rosswog11a}
{Rosswog}, S.
\newblock {Special-relativistic Smoothed Particle Hydrodynamics: a benchmark
  suite}.
\newblock {\em Springer Lecture Notes in Computational Science and Engineering,
  "Meshfree Methods for Partial Differential Equations V", Eds. M. Griebel,
  M.A. Schweitzer, Heidelberg, p. 89-103} {\bf 2011}.

\bibitem[{Kashyap} \em{et~al.}(2021){Kashyap}, {Das}, {Radice}, {Padamata},
  {Prakash}, {Logoteta}, {Perego}, {Godzieba}, {Bernuzzi}, {Bombaci},
  {Fattoyev}, {Reed}, and {da Silva Schneider}]{kashyap22}
{Kashyap}, R.; {Das}, A.; {Radice}, D.; {Padamata}, S.; {Prakash}, A.;
  {Logoteta}, D.; {Perego}, A.; {Godzieba}, D.A.; {Bernuzzi}, S.; {Bombaci},
  I.; {Fattoyev}, F.J.; {Reed}, B.T.; {da Silva Schneider}, A.
\newblock {Numerical relativity simulations of prompt collapse mergers:
  threshold mass and phenomenological constraints on neutron star properties
  after GW170817}.
\newblock {\em arXiv e-prints} {\bf 2021}, p. arXiv:2111.05183,
  \href{http://xxx.lanl.gov/abs/2111.05183}{{\normalfont
  [arXiv:astro-ph.HE/2111.05183]}}.

\bibitem[{Flanagan} and {Hinderer}(2008)]{flanagan08}
{Flanagan}, E.; {Hinderer}, T.
\newblock {Constraining neutron-star tidal Love numbers with gravitational-wave
  detectors}.
\newblock {\em Phys. Rev. D} {\bf 2008}, {\em 77},~021502,
  \href{http://xxx.lanl.gov/abs/0709.1915}{{\normalfont
  [arXiv:astro-ph/0709.1915]}}.
\newblock
  doi:{\changeurlcolor{black}\href{https://doi.org/10.1103/PhysRevD.77.021502}{\detokenize{10.1103/PhysRevD.77.021502}}}.

\bibitem[{Damour} and {Nagar}(2010)]{damour10}
{Damour}, T.; {Nagar}, A.
\newblock {Effective one body description of tidal effects in inspiralling
  compact binaries}.
\newblock {\em Phys. Rev. D} {\bf 2010}, {\em 81},~084016,
  \href{http://xxx.lanl.gov/abs/0911.5041}{{\normalfont
  [arXiv:gr-qc/0911.5041]}}.
\newblock
  doi:{\changeurlcolor{black}\href{https://doi.org/10.1103/PhysRevD.81.084016}{\detokenize{10.1103/PhysRevD.81.084016}}}.

\bibitem[{Bernuzzi}(2020)]{bernuzzi20a}
{Bernuzzi}, S.
\newblock {Neutron star merger remnants}.
\newblock {\em General Relativity and Gravitation} {\bf 2020}, {\em 52},~108,
  \href{http://xxx.lanl.gov/abs/2004.06419}{{\normalfont
  [arXiv:astro-ph.HE/2004.06419]}}.
\newblock
  doi:{\changeurlcolor{black}\href{https://doi.org/10.1007/s10714-020-02752-5}{\detokenize{10.1007/s10714-020-02752-5}}}.

\bibitem[{Bernuzzi} \em{et~al.}(2020){Bernuzzi}, {Breschi}, {Daszuta},
  {Endrizzi}, {Logoteta}, {Nedora}, {Perego}, {Radice}, {Schianchi}, {Zappa},
  {Bombaci}, and {Ortiz}]{bernuzzi20b}
{Bernuzzi}, S.; {Breschi}, M.; {Daszuta}, B.; {Endrizzi}, A.; {Logoteta}, D.;
  {Nedora}, V.; {Perego}, A.; {Radice}, D.; {Schianchi}, F.; {Zappa}, F.;
  {Bombaci}, I.; {Ortiz}, N.
\newblock {Accretion-induced prompt black hole formation in asymmetric neutron
  star mergers, dynamical ejecta, and kilonova signals}.
\newblock {\em MNRAS} {\bf 2020}, {\em 497},~1488--1507,
  \href{http://xxx.lanl.gov/abs/2003.06015}{{\normalfont
  [arXiv:astro-ph.HE/2003.06015]}}.
\newblock
  doi:{\changeurlcolor{black}\href{https://doi.org/10.1093/mnras/staa1860}{\detokenize{10.1093/mnras/staa1860}}}.

\bibitem[{Nakar}(2007)]{nakar07}
{Nakar}, E.
\newblock {Short-hard gamma-ray bursts}.
\newblock {\em Phys. Rep.} {\bf 2007}, {\em 442},~166--236,
  \href{http://xxx.lanl.gov/abs/arXiv:astro-ph/0701748}{{\normalfont
  [arXiv:astro-ph/0701748]}}.
\newblock
  doi:{\changeurlcolor{black}\href{https://doi.org/10.1016/j.physrep.2007.02.005}{\detokenize{10.1016/j.physrep.2007.02.005}}}.

\bibitem[{Lee} \em{et~al.}(2009){Lee}, {Ramirez-Ruiz}, and
  {L{\'o}pez-C{\'a}mara}]{lee09}
{Lee}, W.H.; {Ramirez-Ruiz}, E.; {L{\'o}pez-C{\'a}mara}, D.
\newblock {Phase Transitions and He-Synthesis-Driven Winds in Neutrino Cooled
  Accretion Disks: Prospects for Late Flares in Short Gamma-Ray Bursts}.
\newblock {\em ApJL} {\bf 2009}, {\em 699},~L93--L96.
\newblock
  doi:{\changeurlcolor{black}\href{https://doi.org/10.1088/0004-637X/699/2/L93}{\detokenize{10.1088/0004-637X/699/2/L93}}}.

\bibitem[{Kumar} and {Zhang}(2015)]{kumar15}
{Kumar}, P.; {Zhang}, B.
\newblock {The physics of gamma-ray bursts \& relativistic jets}.
\newblock {\em Phys. Rep.} {\bf 2015}, {\em 561},~1--109,
  \href{http://xxx.lanl.gov/abs/1410.0679}{{\normalfont
  [arXiv:astro-ph.HE/1410.0679]}}.
\newblock
  doi:{\changeurlcolor{black}\href{https://doi.org/10.1016/j.physrep.2014.09.008}{\detokenize{10.1016/j.physrep.2014.09.008}}}.

\bibitem[{Metzger} \em{et~al.}(2008){Metzger}, {Piro}, and
  {Quataert}]{metzger08a}
{Metzger}, B.D.; {Piro}, A.L.; {Quataert}, E.
\newblock {Time-dependent models of accretion discs formed from compact object
  mergers}.
\newblock {\em MNRAS} {\bf 2008}, {\em 390},~781--797.

\bibitem[{Beloborodov}(2008)]{beloborodov08}
{Beloborodov}, A.M.
\newblock {Hyper-accreting black holes}.
\newblock  American Institute of Physics Conference Series; {M.~Axelsson}.,
  Ed.,  2008, Vol. 1054, {\em American Institute of Physics Conference Series},
  pp. 51--70.
\newblock
  doi:{\changeurlcolor{black}\href{https://doi.org/10.1063/1.3002509}{\detokenize{10.1063/1.3002509}}}.

\bibitem[{Siegel} and {Metzger}(2017)]{siegel17a}
{Siegel}, D.M.; {Metzger}, B.D.
\newblock {Three-Dimensional General-Relativistic Magnetohydrodynamic
  Simulations of Remnant Accretion Disks from Neutron Star Mergers: Outflows
  and r -Process Nucleosynthesis}.
\newblock {\em Physical Review Letters} {\bf 2017}, {\em 119},~231102,
  \href{http://xxx.lanl.gov/abs/1705.05473}{{\normalfont
  [arXiv:astro-ph.HE/1705.05473]}}.
\newblock
  doi:{\changeurlcolor{black}\href{https://doi.org/10.1103/PhysRevLett.119.231102}{\detokenize{10.1103/PhysRevLett.119.231102}}}.

\bibitem[{Siegel} and {Metzger}(2018)]{siegel18}
{Siegel}, D.M.; {Metzger}, B.D.
\newblock {Three-dimensional GRMHD Simulations of Neutrino-cooled Accretion
  Disks from Neutron Star Mergers}.
\newblock {\em ApJ} {\bf 2018}, {\em 858},~52,
  \href{http://xxx.lanl.gov/abs/1711.00868}{{\normalfont
  [arXiv:astro-ph.HE/1711.00868]}}.
\newblock
  doi:{\changeurlcolor{black}\href{https://doi.org/10.3847/1538-4357/aabaec}{\detokenize{10.3847/1538-4357/aabaec}}}.

\bibitem[{Miller} \em{et~al.}(2019){Miller}, {Ryan}, {Dolence}, {Burrows},
  {Fontes}, {Fryer}, {Korobkin}, {Lippuner}, {Mumpower}, and
  {Wollaeger}]{miller19}
{Miller}, J.M.; {Ryan}, B.R.; {Dolence}, J.C.; {Burrows}, A.; {Fontes}, C.J.;
  {Fryer}, C.L.; {Korobkin}, O.; {Lippuner}, J.; {Mumpower}, M.R.; {Wollaeger},
  R.T.
\newblock {Full transport model of GW170817-like disk produces a blue
  kilonova}.
\newblock {\em \prd} {\bf 2019}, {\em 100},~023008,
  \href{http://xxx.lanl.gov/abs/1905.07477}{{\normalfont
  [arXiv:astro-ph.HE/1905.07477]}}.
\newblock
  doi:{\changeurlcolor{black}\href{https://doi.org/10.1103/PhysRevD.100.023008}{\detokenize{10.1103/PhysRevD.100.023008}}}.

\bibitem[{Fernandez} \em{et~al.}(2019){Fernandez}, {Tchekhovskoy}, {Quataert},
  {Foucart}, and {Kasen}]{fernandez19}
{Fernandez}, R.; {Tchekhovskoy}, A.; {Quataert}, E.; {Foucart}, F.; {Kasen}, D.
\newblock {Long-term GRMHD simulations of neutron star merger accretion discs:
  implications for electromagnetic counterparts}.
\newblock {\em MNRAS} {\bf 2019}, {\em 482},~3373--3393,
  \href{http://xxx.lanl.gov/abs/1808.00461}{{\normalfont
  [arXiv:astro-ph.HE/1808.00461]}}.
\newblock
  doi:{\changeurlcolor{black}\href{https://doi.org/10.1093/mnras/sty2932}{\detokenize{10.1093/mnras/sty2932}}}.

\bibitem[{Kasen} \em{et~al.}(2017){Kasen}, {Metzger}, {Barnes}, {Quataert}, and
  {Ramirez-Ruiz}]{kasen17}
{Kasen}, D.; {Metzger}, B.; {Barnes}, J.; {Quataert}, E.; {Ramirez-Ruiz}, E.
\newblock {Origin of the heavy elements in binary neutron-star mergers from a
  gravitational-wave event}.
\newblock {\em Nature} {\bf 2017}, {\em 551},~80--84,
  \href{http://xxx.lanl.gov/abs/1710.05463}{{\normalfont
  [arXiv:astro-ph.HE/1710.05463]}}.
\newblock
  doi:{\changeurlcolor{black}\href{https://doi.org/10.1038/nature24453}{\detokenize{10.1038/nature24453}}}.

\bibitem[{Cowperthwaite} \em{et~al.}(2017){Cowperthwaite}, {Berger}, {Villar},
  and {Metzger}]{cowperthwaite17}
{Cowperthwaite}, P.S.; {Berger}, E.; {Villar}, V.A.; {Metzger}, B.D.
\newblock {The Electromagnetic Counterpart of the Binary Neutron Star Merger
  LIGO/Virgo GW170817. II. UV, Optical, and Near-infrared Light Curves and
  Comparison to Kilonova Models}.
\newblock {\em ApJL} {\bf 2017}, {\em 848},~L17,
  \href{http://xxx.lanl.gov/abs/1710.05840}{{\normalfont
  [arXiv:astro-ph.HE/1710.05840]}}.
\newblock
  doi:{\changeurlcolor{black}\href{https://doi.org/10.3847/2041-8213/aa8fc7}{\detokenize{10.3847/2041-8213/aa8fc7}}}.

\bibitem[{Evans} \em{et~al.}(2017){Evans}, {Cenko}, {Kennea}, {Emery}, {Kuin},
  {Korobkin}, {Wollaeger}, {Fryer}, {Madsen}, {Harrison}, {Xu}, {Nakar},
  {Hotokezaka}, {Lien}, {Campana}, {Oates}, {Troja}, {Breeveld}, {Marshall},
  {Barthelmy}, {Beardmore}, {Burrows}, {Cusumano}, {D'Ai}, {D'Avanzo},
  {D'Elia}, {de Pasquale}, {Even}, {Fontes}, {Forster}, {Garcia}, {Giommi},
  {Grefenstette}, {Gronwall}, {Hartmann}, {Heida}, {Hungerford}, {Kasliwal},
  {Krimm}, {Levan}, {Malesani}, {Melandri}, {Miyasaka}, {Nousek}, {O'Brien},
  {Osborne}, {Pagani}, {Page}, {Palmer}, {Perri}, {Pike}, {Racusin}, {Rosswog},
  {Siegel}, {Sakamoto}, {Sbarufatti}, {Tagliaferri}, {Tanvir}, and
  {Tohuvavohu}]{evans17}
{Evans}, P.A.; {Cenko}, S.B.; {Kennea}, J.A.; {Emery}, S.W.K.; {Kuin}, N.P.M.;
  {Korobkin}, O.; {Wollaeger}, R.T.; {Fryer}, C.L.; {Madsen}, K.K.; {Harrison},
  F.A.; {Xu}, Y.; {Nakar}, E.; {Hotokezaka}, K.; {Lien}, A.; {Campana}, S.;
  {Oates}, S.R.; {Troja}, E.; {Breeveld}, A.A.; {Marshall}, F.E.; {Barthelmy},
  S.D.; {Beardmore}, A.P.; {Burrows}, D.N.; {Cusumano}, G.; {D'Ai}, A.;
  {D'Avanzo}, P.; {D'Elia}, V.; {de Pasquale}, M.; {Even}, W.P.; {Fontes},
  C.J.; {Forster}, K.; {Garcia}, J.; {Giommi}, P.; {Grefenstette}, B.;
  {Gronwall}, C.; {Hartmann}, D.H.; {Heida}, M.; {Hungerford}, A.L.;
  {Kasliwal}, M.M.; {Krimm}, H.A.; {Levan}, A.J.; {Malesani}, D.; {Melandri},
  A.; {Miyasaka}, H.; {Nousek}, J.A.; {O'Brien}, P.T.; {Osborne}, J.P.;
  {Pagani}, C.; {Page}, K.L.; {Palmer}, D.M.; {Perri}, M.; {Pike}, S.;
  {Racusin}, J.L.; {Rosswog}, S.; {Siegel}, M.H.; {Sakamoto}, T.; {Sbarufatti},
  B.; {Tagliaferri}, G.; {Tanvir}, N.R.; {Tohuvavohu}, A.
\newblock {Swift and NuSTAR observations of GW170817: Detection of a blue
  kilonova}.
\newblock {\em Science} {\bf 2017}, {\em 358},~1565--1570,
  \href{http://xxx.lanl.gov/abs/1710.05437}{{\normalfont
  [arXiv:astro-ph.HE/1710.05437]}}.
\newblock
  doi:{\changeurlcolor{black}\href{https://doi.org/10.1126/science.aap9580}{\detokenize{10.1126/science.aap9580}}}.

\bibitem[{Villar} \em{et~al.}(2017){Villar}, {Guillochon}, {Berger}, {Metzger},
  {Cowperthwaite}, {Nicholl}, {Alexander}, {Blanchard}, {Chornock},
  {Eftekhari}, {Fong}, {Margutti}, and {Williams}]{villar17}
{Villar}, V.A.; {Guillochon}, J.; {Berger}, E.; {Metzger}, B.D.;
  {Cowperthwaite}, P.S.; {Nicholl}, M.; {Alexander}, K.D.; {Blanchard}, P.K.;
  {Chornock}, R.; {Eftekhari}, T.; {Fong}, W.; {Margutti}, R.; {Williams},
  P.K.G.
\newblock {The Combined Ultraviolet, Optical, and Near-infrared Light Curves of
  the Kilonova Associated with the Binary Neutron Star Merger GW170817: Unified
  Data Set, Analytic Models, and Physical Implications}.
\newblock {\em ApJL} {\bf 2017}, {\em 851},~L21,
  \href{http://xxx.lanl.gov/abs/1710.11576}{{\normalfont
  [arXiv:astro-ph.HE/1710.11576]}}.
\newblock
  doi:{\changeurlcolor{black}\href{https://doi.org/10.3847/2041-8213/aa9c84}{\detokenize{10.3847/2041-8213/aa9c84}}}.

\bibitem[{Kasliwal} \em{et~al.}(2017){Kasliwal}, {Nakar}, {Singer}, {Kaplan},
  and et~al.]{kasliwal17}
{Kasliwal}, M.M.; {Nakar}, E.; {Singer}, L.P.; {Kaplan}, D.L.; et~al..
\newblock {Illuminating gravitational waves: A concordant picture of photons
  from a neutron star merger}.
\newblock {\em Science} {\bf 2017}, {\em 358},~1559--1565,
  \href{http://xxx.lanl.gov/abs/1710.05436}{{\normalfont
  [arXiv:astro-ph.HE/1710.05436]}}.
\newblock
  doi:{\changeurlcolor{black}\href{https://doi.org/10.1126/science.aap9455}{\detokenize{10.1126/science.aap9455}}}.

\bibitem[{Tanvir} \em{et~al.}(2017){Tanvir}, {Levan},
  {Gonz{\'a}lez-Fern{\'a}ndez}, {Korobkin}, {Mandel}, {Rosswog}, {Hjorth},
  {D'Avanzo}, {Fruchter}, {Fryer}, {Kangas}, {Milvang-Jensen}, {Rosetti},
  {Steeghs}, {Wollaeger}, {Cano}, {Copperwheat}, {Covino}, {D'Elia}, {de Ugarte
  Postigo}, {Evans}, {Even}, {Fairhurst}, {Figuera Jaimes}, {Fontes}, {Fujii},
  {Fynbo}, {Gompertz}, {Greiner}, {Hodosan}, {Irwin}, {Jakobsson},
  {J{\o}rgensen}, {Kann}, {Lyman}, {Malesani}, {McMahon}, {Melandri},
  {O'Brien}, {Osborne}, {Palazzi}, {Perley}, {Pian}, {Piranomonte}, {Rabus},
  {Rol}, {Rowlinson}, {Schulze}, {Sutton}, {Th{\"o}ne}, {Ulaczyk}, {Watson},
  {Wiersema}, and {Wijers}]{tanvir17}
{Tanvir}, N.R.; {Levan}, A.J.; {Gonz{\'a}lez-Fern{\'a}ndez}, C.; {Korobkin},
  O.; {Mandel}, I.; {Rosswog}, S.; {Hjorth}, J.; {D'Avanzo}, P.; {Fruchter},
  A.S.; {Fryer}, C.L.; {Kangas}, T.; {Milvang-Jensen}, B.; {Rosetti}, S.;
  {Steeghs}, D.; {Wollaeger}, R.T.; {Cano}, Z.; {Copperwheat}, C.M.; {Covino},
  S.; {D'Elia}, V.; {de Ugarte Postigo}, A.; {Evans}, P.A.; {Even}, W.P.;
  {Fairhurst}, S.; {Figuera Jaimes}, R.; {Fontes}, C.J.; {Fujii}, Y.I.;
  {Fynbo}, J.P.U.; {Gompertz}, B.P.; {Greiner}, J.; {Hodosan}, G.; {Irwin},
  M.J.; {Jakobsson}, P.; {J{\o}rgensen}, U.G.; {Kann}, D.A.; {Lyman}, J.D.;
  {Malesani}, D.; {McMahon}, R.G.; {Melandri}, A.; {O'Brien}, P.T.; {Osborne},
  J.P.; {Palazzi}, E.; {Perley}, D.A.; {Pian}, E.; {Piranomonte}, S.; {Rabus},
  M.; {Rol}, E.; {Rowlinson}, A.; {Schulze}, S.; {Sutton}, P.; {Th{\"o}ne},
  C.C.; {Ulaczyk}, K.; {Watson}, D.; {Wiersema}, K.; {Wijers}, R.A.M.J.
\newblock {The Emergence of a Lanthanide-rich Kilonova Following the Merger of
  Two Neutron Stars}.
\newblock {\em ApJL} {\bf 2017}, {\em 848},~L27,
  \href{http://xxx.lanl.gov/abs/1710.05455}{{\normalfont
  [arXiv:astro-ph.HE/1710.05455]}}.
\newblock
  doi:{\changeurlcolor{black}\href{https://doi.org/10.3847/2041-8213/aa90b6}{\detokenize{10.3847/2041-8213/aa90b6}}}.

\bibitem[{Rosswog} \em{et~al.}(2018){Rosswog}, {Sollerman}, {Feindt}, {Goobar},
  {Korobkin}, {Wollaeger}, {Fremling}, and {Kasliwal}]{rosswog18a}
{Rosswog}, S.; {Sollerman}, J.; {Feindt}, U.; {Goobar}, A.; {Korobkin}, O.;
  {Wollaeger}, R.; {Fremling}, C.; {Kasliwal}, M.M.
\newblock {The first direct double neutron star merger detection: Implications
  for cosmic nucleosynthesis}.
\newblock {\em A\&A} {\bf 2018}, {\em 615},~A132,
  \href{http://xxx.lanl.gov/abs/1710.05445}{{\normalfont
  [arXiv:astro-ph.HE/1710.05445]}}.
\newblock
  doi:{\changeurlcolor{black}\href{https://doi.org/10.1051/0004-6361/201732117}{\detokenize{10.1051/0004-6361/201732117}}}.

\bibitem[{Raithel} \em{et~al.}(2019){Raithel}, {{\"O}zel}, and
  {Psaltis}]{raithel19}
{Raithel}, C.A.; {{\"O}zel}, F.; {Psaltis}, D.
\newblock {Finite-temperature Extension for Cold Neutron Star Equations of
  State}.
\newblock {\em ApJ} {\bf 2019}, {\em 875},~12,
  \href{http://xxx.lanl.gov/abs/1902.10735}{{\normalfont
  [arXiv:astro-ph.HE/1902.10735]}}.
\newblock
  doi:{\changeurlcolor{black}\href{https://doi.org/10.3847/1538-4357/ab08ea}{\detokenize{10.3847/1538-4357/ab08ea}}}.

\bibitem[{Raithel} \em{et~al.}(2021){Raithel}, {Paschalidis}, and
  {{\"O}zel}]{raithel21a}
{Raithel}, C.A.; {Paschalidis}, V.; {{\"O}zel}, F.
\newblock {Realistic finite-temperature effects in neutron star merger
  simulations}.
\newblock {\em Phys. Rev. D} {\bf 2021}, {\em 104},~063016,
  \href{http://xxx.lanl.gov/abs/2104.07226}{{\normalfont
  [arXiv:astro-ph.HE/2104.07226]}}.
\newblock
  doi:{\changeurlcolor{black}\href{https://doi.org/10.1103/PhysRevD.104.063016}{\detokenize{10.1103/PhysRevD.104.063016}}}.

\bibitem[{Bozzola}(2021)]{bozzola21}
{Bozzola}, G.
\newblock {kuibit: Analyzing Einstein Toolkit simulations with Python}.
\newblock {\em The Journal of Open Source Software} {\bf 2021}, {\em 6},~3099,
  \href{http://xxx.lanl.gov/abs/2104.06376}{{\normalfont
  [arXiv:gr-qc/2104.06376]}}.
\newblock
  doi:{\changeurlcolor{black}\href{https://doi.org/10.21105/joss.03099}{\detokenize{10.21105/joss.03099}}}.

\bibitem[Bauswein and Stergioulas(2015)]{bauswein15a}
Bauswein, A.; Stergioulas, N.
\newblock {Unified picture of the post-merger dynamics and gravitational wave
  emission in neutron star mergers}.
\newblock {\em Phys. Rev. D} {\bf 2015}, {\em 91},~124056,
  \href{http://xxx.lanl.gov/abs/1502.03176}{{\normalfont
  [arXiv:astro-ph.SR/1502.03176]}}.
\newblock
  doi:{\changeurlcolor{black}\href{https://doi.org/10.1103/PhysRevD.91.124056}{\detokenize{10.1103/PhysRevD.91.124056}}}.

\bibitem[Bernuzzi \em{et~al.}(2015)Bernuzzi, Dietrich, and Nagar]{bernuzzi15a}
Bernuzzi, S.; Dietrich, T.; Nagar, A.
\newblock {Modeling the complete gravitational wave spectrum of neutron star
  mergers}.
\newblock {\em Phys. Rev. Lett.} {\bf 2015}, {\em 115},~091101,
  \href{http://xxx.lanl.gov/abs/1504.01764}{{\normalfont
  [arXiv:gr-qc/1504.01764]}}.
\newblock
  doi:{\changeurlcolor{black}\href{https://doi.org/10.1103/PhysRevLett.115.091101}{\detokenize{10.1103/PhysRevLett.115.091101}}}.

\bibitem[Dietrich \em{et~al.}(2015)Dietrich, Bernuzzi, Ujevic, and
  Br\"ugmann]{dietrich15b}
Dietrich, T.; Bernuzzi, S.; Ujevic, M.; Br\"ugmann, B.
\newblock {Numerical relativity simulations of neutron star merger remnants
  using conservative mesh refinement}.
\newblock {\em Phys. Rev. D} {\bf 2015}, {\em 91},~124041,
  \href{http://xxx.lanl.gov/abs/1504.01266}{{\normalfont
  [arXiv:gr-qc/1504.01266]}}.
\newblock
  doi:{\changeurlcolor{black}\href{https://doi.org/10.1103/PhysRevD.91.124041}{\detokenize{10.1103/PhysRevD.91.124041}}}.

\bibitem[Bauswein \em{et~al.}(2016)Bauswein, Stergioulas, and
  Janka]{bauswein15b}
Bauswein, A.; Stergioulas, N.; Janka, H.T.
\newblock {Exploring properties of high-density matter through remnants of
  neutron-star mergers}.
\newblock {\em Eur. Phys. J. A} {\bf 2016}, {\em 52},~56,
  \href{http://xxx.lanl.gov/abs/1508.05493}{{\normalfont
  [arXiv:astro-ph.HE/1508.05493]}}.
\newblock
  doi:{\changeurlcolor{black}\href{https://doi.org/10.1140/epja/i2016-16056-7}{\detokenize{10.1140/epja/i2016-16056-7}}}.

\bibitem[Clark \em{et~al.}(2016)Clark, Bauswein, Stergioulas, and
  Shoemaker]{clark15}
Clark, J.A.; Bauswein, A.; Stergioulas, N.; Shoemaker, D.
\newblock {Observing Gravitational Waves From The Post-Merger Phase Of Binary
  Neutron Star Coalescence}.
\newblock {\em Class. Quant. Grav.} {\bf 2016}, {\em 33},~085003,
  \href{http://xxx.lanl.gov/abs/1509.08522}{{\normalfont
  [arXiv:astro-ph.HE/1509.08522]}}.
\newblock
  doi:{\changeurlcolor{black}\href{https://doi.org/10.1088/0264-9381/33/8/085003}{\detokenize{10.1088/0264-9381/33/8/085003}}}.

\bibitem[Ciolfi \em{et~al.}(2017)Ciolfi, Kastaun, Giacomazzo, Endrizzi, Siegel,
  and Perna]{ciolfi17}
Ciolfi, R.; Kastaun, W.; Giacomazzo, B.; Endrizzi, A.; Siegel, D.M.; Perna, R.
\newblock {General relativistic magnetohydrodynamic simulations of binary
  neutron star mergers forming a long-lived neutron star}.
\newblock {\em Phys. Rev. D} {\bf 2017}, {\em 95},~063016,
  \href{http://xxx.lanl.gov/abs/1701.08738}{{\normalfont
  [arXiv:astro-ph.HE/1701.08738]}}.
\newblock
  doi:{\changeurlcolor{black}\href{https://doi.org/10.1103/PhysRevD.95.063016}{\detokenize{10.1103/PhysRevD.95.063016}}}.

\bibitem[Maione \em{et~al.}(2017)Maione, De~Pietri, Feo, and
  L\"offler]{maione17}
Maione, F.; De~Pietri, R.; Feo, A.; L\"offler, F.
\newblock {Spectral analysis of gravitational waves from binary neutron star
  merger remnants}.
\newblock {\em Phys. Rev. D} {\bf 2017}, {\em 96},~063011,
  \href{http://xxx.lanl.gov/abs/1707.03368}{{\normalfont
  [arXiv:gr-qc/1707.03368]}}.
\newblock
  doi:{\changeurlcolor{black}\href{https://doi.org/10.1103/PhysRevD.96.063011}{\detokenize{10.1103/PhysRevD.96.063011}}}.

\bibitem[Sarin and Lasky(2021)]{sarin20}
Sarin, N.; Lasky, P.D.
\newblock {The evolution of binary neutron star post-merger remnants: a
  review}.
\newblock {\em Gen. Rel. Grav.} {\bf 2021}, {\em 53},~59,
  \href{http://xxx.lanl.gov/abs/2012.08172}{{\normalfont
  [arXiv:astro-ph.HE/2012.08172]}}.
\newblock
  doi:{\changeurlcolor{black}\href{https://doi.org/10.1007/s10714-021-02831-1}{\detokenize{10.1007/s10714-021-02831-1}}}.

\bibitem[Sun \em{et~al.}(2022)Sun, Ruiz, Shapiro, and Tsokaros]{sun22}
Sun, L.; Ruiz, M.; Shapiro, S.L.; Tsokaros, A.
\newblock {Jet Launching from Binary Neutron Star Mergers: Incorporating
  Neutrino Transport and Magnetic Fields} {\bf 2022}.
\newblock  \href{http://xxx.lanl.gov/abs/2202.12901}{{\normalfont
  [arXiv:astro-ph.HE/2202.12901]}}.
\newblock arXiv:2202.12901.

\bibitem[{Takami} \em{et~al.}(2014){Takami}, {Rezzolla}, and
  {Baiotti}]{takami14}
{Takami}, K.; {Rezzolla}, L.; {Baiotti}, L.
\newblock {Spectral properties of the post-merger gravitational-wave signal
  from binary neutron stars}.
\newblock {\em ArXiv e-prints} {\bf 2014},
  \href{http://xxx.lanl.gov/abs/1412.3240}{{\normalfont
  [arXiv:gr-qc/1412.3240]}}.

\bibitem[Freiburghaus \em{et~al.}(1999)Freiburghaus, Rosswog, and
  Thielemann]{freiburghaus99b}
Freiburghaus, C.; Rosswog, S.; Thielemann, F.K.
\newblock R-Process in Neutron Star Mergers.
\newblock {\em ApJ} {\bf 1999}, {\em 525},~L121.

\bibitem[{Foucart} \em{et~al.}(2021){Foucart}, {M{\"o}sta}, {Ramirez},
  {Wright}, {Darbha}, and {Kasen}]{foucart21b}
{Foucart}, F.; {M{\"o}sta}, P.; {Ramirez}, T.; {Wright}, A.J.; {Darbha}, S.;
  {Kasen}, D.
\newblock {Estimating outflow masses and velocities in merger simulations:
  Impact of r -process heating and neutrino cooling}.
\newblock {\em \prd} {\bf 2021}, {\em 104},~123010,
  \href{http://xxx.lanl.gov/abs/2109.00565}{{\normalfont
  [arXiv:astro-ph.HE/2109.00565]}}.
\newblock
  doi:{\changeurlcolor{black}\href{https://doi.org/10.1103/PhysRevD.104.123010}{\detokenize{10.1103/PhysRevD.104.123010}}}.

\bibitem[{Korobkin} \em{et~al.}(2021){Korobkin}, {Wollaeger}, {Fryer},
  {Hungerford}, {Rosswog}, {Fontes}, {Mumpower}, {Chase}, {Even}, {Miller},
  {Misch}, and {Lippuner}]{korobkin21}
{Korobkin}, O.; {Wollaeger}, R.T.; {Fryer}, C.L.; {Hungerford}, A.L.;
  {Rosswog}, S.; {Fontes}, C.J.; {Mumpower}, M.R.; {Chase}, E.A.; {Even}, W.P.;
  {Miller}, J.; {Misch}, G.W.; {Lippuner}, J.
\newblock {Axisymmetric Radiative Transfer Models of Kilonovae}.
\newblock {\em ApJ} {\bf 2021}, {\em 910},~116,
  \href{http://xxx.lanl.gov/abs/2004.00102}{{\normalfont
  [arXiv:astro-ph.HE/2004.00102]}}.
\newblock
  doi:{\changeurlcolor{black}\href{https://doi.org/10.3847/1538-4357/abe1b5}{\detokenize{10.3847/1538-4357/abe1b5}}}.

\bibitem[{Metzger} \em{et~al.}(2015){Metzger}, {Bauswein}, {Goriely}, and
  {Kasen}]{metzger15a}
{Metzger}, B.D.; {Bauswein}, A.; {Goriely}, S.; {Kasen}, D.
\newblock {Neutron-powered precursors of kilonovae}.
\newblock {\em MNRAS} {\bf 2015}, {\em 446},~1115--1120,
  \href{http://xxx.lanl.gov/abs/1409.0544}{{\normalfont
  [arXiv:astro-ph.HE/1409.0544]}}.
\newblock
  doi:{\changeurlcolor{black}\href{https://doi.org/10.1093/mnras/stu2225}{\detokenize{10.1093/mnras/stu2225}}}.

\bibitem[{Mooley} \em{et~al.}(2018){Mooley}, {Nakar}, {Hotokezaka}, {Hallinan},
  {Corsi}, {Frail}, {Horesh}, {Murphy}, {Lenc}, {Kaplan}, {de}, {Dobie},
  {Chandra}, {Deller}, {Gottlieb}, {Kasliwal}, {Kulkarni}, {Myers}, {Nissanke},
  {Piran}, {Lynch}, {Bhalerao}, {Bourke}, {Bannister}, and {Singer}]{mooley17}
{Mooley}, K.P.; {Nakar}, E.; {Hotokezaka}, K.; {Hallinan}, G.; {Corsi}, A.;
  {Frail}, D.A.; {Horesh}, A.; {Murphy}, T.; {Lenc}, E.; {Kaplan}, D.L.; {de},
  K.; {Dobie}, D.; {Chandra}, P.; {Deller}, A.; {Gottlieb}, O.; {Kasliwal},
  M.M.; {Kulkarni}, S.R.; {Myers}, S.T.; {Nissanke}, S.; {Piran}, T.; {Lynch},
  C.; {Bhalerao}, V.; {Bourke}, S.; {Bannister}, K.W.; {Singer}, L.P.
\newblock {A mildly relativistic wide-angle outflow in the neutron-star merger
  event GW170817}.
\newblock {\em Nature} {\bf 2018}, {\em 554},~207--210,
  \href{http://xxx.lanl.gov/abs/1711.11573}{{\normalfont
  [arXiv:astro-ph.HE/1711.11573]}}.
\newblock
  doi:{\changeurlcolor{black}\href{https://doi.org/10.1038/nature25452}{\detokenize{10.1038/nature25452}}}.

\bibitem[{Hotokezaka} \em{et~al.}(2018){Hotokezaka}, {Kiuchi}, {Shibata},
  {Nakar}, and {Piran}]{hotokezaka18a}
{Hotokezaka}, K.; {Kiuchi}, K.; {Shibata}, M.; {Nakar}, E.; {Piran}, T.
\newblock {Synchrotron Radiation from the Fast Tail of Dynamical Ejecta of
  Neutron Star Mergers}.
\newblock {\em ApJ} {\bf 2018}, {\em 867},~95,
  \href{http://xxx.lanl.gov/abs/1803.00599}{{\normalfont
  [arXiv:astro-ph.HE/1803.00599]}}.
\newblock
  doi:{\changeurlcolor{black}\href{https://doi.org/10.3847/1538-4357/aadf92}{\detokenize{10.3847/1538-4357/aadf92}}}.

\bibitem[{Hajela} \em{et~al.}(2022){Hajela}, {Margutti}, {Bright}, {Alexander},
  {Metzger}, {Nedora}, {Kathirgamaraju}, {Margalit}, {Radice}, {Guidorzi},
  {Berger}, {MacFadyen}, {Giannios}, {Chornock}, {Heywood}, {Sironi},
  {Gottlieb}, {Coppejans}, {Laskar}, {Cendes}, {Duran}, {Eftekhari}, {Fong},
  {McDowell}, {Nicholl}, {Xie}, {Zrake}, {Bernuzzi}, {Broekgaarden},
  {Kilpatrick}, {Terreran}, {Villar}, {Blanchard}, {Gomez}, {Hosseinzadeh},
  {Matthews}, and {Rastinejad}]{hajela22}
{Hajela}, A.; {Margutti}, R.; {Bright}, J.S.; {Alexander}, K.D.; {Metzger},
  B.D.; {Nedora}, V.; {Kathirgamaraju}, A.; {Margalit}, B.; {Radice}, D.;
  {Guidorzi}, C.; {Berger}, E.; {MacFadyen}, A.; {Giannios}, D.; {Chornock},
  R.; {Heywood}, I.; {Sironi}, L.; {Gottlieb}, O.; {Coppejans}, D.; {Laskar},
  T.; {Cendes}, Y.; {Duran}, R.B.; {Eftekhari}, T.; {Fong}, W.; {McDowell}, A.;
  {Nicholl}, M.; {Xie}, X.; {Zrake}, J.; {Bernuzzi}, S.; {Broekgaarden}, F.S.;
  {Kilpatrick}, C.D.; {Terreran}, G.; {Villar}, V.A.; {Blanchard}, P.K.;
  {Gomez}, S.; {Hosseinzadeh}, G.; {Matthews}, D.J.; {Rastinejad}, J.C.
\newblock {Evidence for X-Ray Emission in Excess to the Jet-afterglow Decay 3.5
  yr after the Binary Neutron Star Merger GW 170817: A New Emission Component}.
\newblock {\em ApJL} {\bf 2022}, {\em 927},~L17,
  \href{http://xxx.lanl.gov/abs/2104.02070}{{\normalfont
  [arXiv:astro-ph.HE/2104.02070]}}.
\newblock
  doi:{\changeurlcolor{black}\href{https://doi.org/10.3847/2041-8213/ac504a}{\detokenize{10.3847/2041-8213/ac504a}}}.

\bibitem[{Rosswog} \em{et~al.}(2000){Rosswog}, {Davies}, {Thielemann}, and
  {Piran}]{rosswog00}
{Rosswog}, S.; {Davies}, M.B.; {Thielemann}, F.K.; {Piran}, T.
\newblock {Merging neutron stars: asymmetric systems}.
\newblock {\em A\&A} {\bf 2000}, {\em 360},~171--184.

\bibitem[{Korobkin} \em{et~al.}(2012){Korobkin}, {Rosswog}, {Arcones}, and
  {Winteler}]{korobkin12a}
{Korobkin}, O.; {Rosswog}, S.; {Arcones}, A.; {Winteler}, C.
\newblock {On the astrophysical robustness of the neutron star merger
  r-process}.
\newblock {\em MNRAS} {\bf 2012}, {\em 426},~1940--1949.

\bibitem[{Rosswog}(2013)]{rosswog13b}
{Rosswog}, S.
\newblock {The dynamic ejecta of compact object mergers and eccentric
  collisions}.
\newblock {\em Royal Society of London Philosophical Transactions Series A}
  {\bf 2013}, {\em 371},~20272.
\newblock
  doi:{\changeurlcolor{black}\href{https://doi.org/10.1098/rsta.2012.0272}{\detokenize{10.1098/rsta.2012.0272}}}.

\bibitem[{Farouqi} \em{et~al.}(2021){Farouqi}, {Thielemann}, {Rosswog}, and
  {Kratz}]{farouqi22}
{Farouqi}, K.; {Thielemann}, F.K.; {Rosswog}, S.; {Kratz}, K.L.
\newblock {Correlations of r-Process Elements in Very Metal-Poor Stars as Clues
  to their Nucleosynthesis Sites}.
\newblock {\em arXiv e-prints} {\bf 2021}, p. arXiv:2107.03486,
  \href{http://xxx.lanl.gov/abs/2107.03486}{{\normalfont
  [arXiv:astro-ph.SR/2107.03486]}}.

\bibitem[{Price}(2007)]{price07d}
{Price}, D.J.
\newblock {splash: An Interactive Visualisation Tool for Smoothed Particle
  Hydrodynamics Simulations}.
\newblock {\em Publications of the Astronomical Society of Australia} {\bf
  2007}, {\em 24},~159--173,
  \href{http://xxx.lanl.gov/abs/arXiv:0709.0832}{{\normalfont
  [arXiv:0709.0832]}}.
\newblock
  doi:{\changeurlcolor{black}\href{https://doi.org/10.1071/AS07022}{\detokenize{10.1071/AS07022}}}.

\bibitem[{Ridders}(1982)]{ridders82}
{Ridders}, C.
\newblock {\em Advances in Engineering Software} {\bf 1982}, {\em 4},~75.

\bibitem[Press \em{et~al.}(1992)Press, Flannery, Teukolsky, and
  Vetterling]{press92}
Press, W.H.; Flannery, B.P.; Teukolsky, S.A.; Vetterling, W.T.
\newblock {\em {N}umerical {R}ecipes}; Cambridge University Press: New York,
  1992.

\end{thebibliography}

\end{document}